# THE FUTURE OF THE ARECIBO OBSERVATORY: THE NEXT GENERATION ARECIBO TELESCOPE

White Paper, ver 2.0, 02-01-2021

**Contact Author: D. Anish Roshi[1], aroshi@naic.edu**

**Authors: D. Anish Roshi[1] , N. Aponte[1], E. Araya[2], H. Arce[3], L. A. Baker[7], W. Baan[35], T. M. Becker[4], J. K. Breakall[34], R. G. Brown[5], C. G. M. Brum[1], M. Busch[6], D. B. Campbell[7], T. Cohen[24], F. Cordova[1], J. S. Deneva[8], M. Devogèle[1], T. Dolch[30], F. O. Fernandez-Rodriguez[1], T. Ghosh[9], P. F. Goldsmith[10], L.I. Gurvits[14,27], M. Haynes[7], C. Heiles[11], J. W. T. Hessel [35, 38], D. Hickson[1], B. Isham[12], R. B. Kerr[13], J. Kelly[28], J. J. Kiriazes[5], J. Lautenbach[1], M. Lebron[15], N. Lewandowska[16], L. Magnani[17], P. K. Manoharan[1], J. L. Margot [37], S. E. Marshall[1], A. K. McGilvray[1], A. Mendez[36], R. Minchin[18], V. Negron[1], M. C. Nolan[19], L. Olmi[26], F. Paganelli[9], N. T. Palliyaguru[20], C. A. Pantoja[15], Z. Paragi[27], S. C. Parshley[7], J. E. G. Peek[6,21], B. B. P. Perera[1], P. Perillat[1], N. Pinilla-Alonso[22,1], L. Quintero[1], H. Radovan[25], S. Raizada[1], T. Robishaw[23], M. Route[31], C. J. Salter[9,1], A. Santoni[1], P. Santos[1], S. Sau[1], D. Selvaraj[1], A. J. Smith[1], M. Sulzer[1], S. Vaddi[1], F. Vargas[33], F. C. F. Venditti[1], A. Venkataraman[1], H. Verkouter[27], A. K. Virkki[1], A. Vishwas[7], S. Weinreb[32], D. Werthimer[11], A. Wolszczan[29] and L. F. Zambrano-Marin[1].**

***Affiliations are listed after the acknowledgements, immediately before the appendices.***

Please click here to endorse the contents of this white paper.

# Table of Contents

## Executive Summary

The Arecibo Observatory (AO) hosted the most powerful radar system and the most sensitive radio telescope in the world until the unexpected collapse of the 1000-ft “legacy” AO telescope (LAT) on December 1, 2020. For 57 years, the facility uniquely excelled in three separate, major scientific areas: planetary science, space and atmospheric sciences, and astronomy. Through its final day of operation, the LAT continued to produce new, groundbreaking science, adding to its long history of extraordinary achievements, including a Nobel Prize in Physics. Its collapse has produced a significant void in these scientific fields, which echoed across the extensive, worldwide scientific community. It also produced a deeply-felt cultural, socioeconomic, and educational loss for Puerto Ricans, and a tragic deprivation of opportunity, inspiration, and training for Science, Technology, Engineering, and Mathematics (STEM) students in Puerto Rico and across the U.S., all of whom represent the next generation of America’s scientists and engineers.

In the tremendous wake of the LAT, we envision a new, unparalleled facility, one which will push forward the boundaries of the planetary, atmospheric, and radio astronomical sciences for decades to come. A future multidisciplinary facility at the site should enable cutting-edge capabilities for all three of the science branches that form the cornerstones of AO exploration. To facilitate the novel, consequential science goals described in this document, the new facility must meet the capability requirements described below, which ultimately drive our telescope concept design.

**Planetary Science: 5 MW of continuous wave transmitting power at 2 - 6 GHz, 1-2 arcmin beamwidth at these frequencies, and increased sky coverage.**

**Atmospheric Science: 0° to 45° sky coverage from zenith to observe both parallel and perpendicular directions to the geomagnetic field, 10 MW peak transmitting power at 430 MHz (also at 220 MHz under consideration) and excellent surface brightness sensitivity.**

**Astronomical Science: Excellent sensitivity over 200 MHz to 30 GHz frequency range, increased sky coverage and telescope pointing up to 48° from zenith to observe the Galactic Center.**

In the following sections of this summary, we describe the key scientific objectives and novel capabilities that the new facility will offer to the three science areas and space weather forecasting, a unique new interdisciplinary application.

## Planetary Radar Science

A key role of the LAT as the host to the world's most powerful radar system was to characterize the physical and dynamical properties of near-Earth objects (NEOs), in support of NASA's Planetary Defense Coordination Office and in line with national interest and security. In recent years, AO observed hundreds of NEOs as a part of NASA's mandate by the US Congress [George E. Brown, Jr. [ADD: Near-Earth Object Survey] Act (Public Law 109-155 Sec. 321)] to detect, track, catalogue, and characterize 90% of all NEOs larger than 140 meters in size. Post-discovery tracking of NEOs with radar is an unparalleled technique for accurately determining their future trajectory and assessing whether they pose a real impact threat to Earth. These radar measurements secure the position and velocity of NEOs with a precision of tens of meters and millimeters per second, respectively. The LAT radar was also used to map the surfaces of Mercury, Venus, Mars, and the Moon, *supporting human and robotic exploration of the Moon, Mars, and near-Earth asteroids*. A new facility, with a more powerful radar system (5 MW at 2 to 6 GHz) and large sky coverage, will support Planetary Defense, Solar System science, and Space Situational Awareness by providing the following capabilities:

| Planetary Defense and Solar System Exploration | | |
|---|---|---|
| Post-discovery characterization and orbit determination of up to 90% of possible asteroid impactors | Study the surface and sub-surface of ocean worlds around Jupiter, Saturn and other Solar System objects | Observe asteroids in the outer regions of the main-belt and beyond |
| Space Situational Awareness (SSA) to categorize space debris down to mm-size in LEO, and smaller than one meter in GEO and cislunar space | Support NASA Human Exploration program by characterizing spacecraft landing sites and identifying potential hazards at low cost | Support and extend the science return of missions including NASA's DART, Janus, Europa Clipper, and Dragonfly missions; and ESA's JUICE mission |

## Space and Atmospheric Sciences

Space and Atmospherics Sciences (SAS) at AO has traditionally utilized multiple approaches to atmospheric research. The LAT's Incoherent Scatter Radar (ISR), the Light Detection and Ranging (Lidar) facility, the onsite and remote passive optical facilities, and the High Frequency facility formed the cornerstones of SAS research at AO. The powerful LAT's ISR was the only instrument of its kind and was capable of profiling ionospheric parameters beyond 2000 km of the Earth's atmosphere. The high resolution, range-resolved observations of electron concentrations, temperatures, ion compositions, and inference of electric fields in the ionosphere are important for the investigations of the coupling processes between different atmospheric regions, influence of solar and space weather disturbances on the Earth's environment, and fundamental plasma processes, since the ionosphere acts as a natural plasma laboratory. The LAT's ISR provided unique contributions in the space sciences due to its high sensitivity and power. However, a major drawback was its limited beam steering capabilities, which will be overcome with the proposed new facility. Increased sky coverage ($\geq 45°$ zenith coverage), and more power ($\gtrsim$10 MW at 430 MHz; a 220 MHz radar is also under consideration) open up new possibilities that will lead to innovative research and discoveries in the following topics.

| Advances in Space and Atmospheric Sciences | | |
|---|---|---|
| Investigate global climate change and its influence on the upper atmosphere | Unravel the mysterious causes of short-period perturbations in the ionosphere | Understand interactions in Earth's atmosphere in the northern and southern hemispheres |
| Investigate coupling between Earth's atmospheric layers to improve satellite navigation, radio wave propagation, and weather forecasting models | Disambiguate between the influence of meteorological and space weather on the neutral and ionized coupling phenomena in Earth's atmosphere | Understand the neutral and ionized atmospheric behavior by combined active and passive observation |

## Radio Astronomy

LAT's unique capabilities enabled several key discoveries in radio astronomy. The loss of the instrument was felt most keenly by pulsar, galactic and extragalactic researchers. The new facility should enable complementary observations with other existing and upcoming radio facilities. For example, the new facility must provide a substantial increase in sensitivity for Very Long Baseline Interferometry, of which the LAT was a contributing instrument whenever higher sensitivity was required. In addition, wider sky coverage, greater collecting area, increased frequency coverage, and a larger field-of-view (FoV) will substantially increase the research potential in a wide range of fields, some of which are highlighted below.

| New Frontiers in Radio Astronomy | | |
|---|---|---|
| Test General Relativity with Galactic Center pulsars | Illuminate underlying physics of pulsars, the emission mechanism, and propagation of radio waves in the interstellar medium | Gain new insights into the causes and physical processes of Fast Radio Bursts |
| Constrain cosmological theories for Dark Matter in the local Universe | Search for Exoplanets and Earth-like Worlds including studies on habitability and magnetic fields | Measure the distribution of matter to moderate redshifts to constrain Dark Energy |
| Probe Extreme Astrophysical Regimes with Very Long Baseline Interferometry | Detect the fingerprints of prebiotic molecules in our Galaxy and beyond | Detect and study Gravitational Waves using pulsar timing |
| Explore the star formation history of the Universe by observing $^{12}$CO emission from massive galaxies at redshift > 3 | Study the formation of massive stars through ammonia observations | Search for Technosignatures from advanced life forms |

## Interdisciplinary science - Space Weather Studies

The US "space weather preparedness" bill [116th Congress Public Law 181 (10/21/2020)] emphasizes the importance of space weather research and forecasting efforts. It is important to efficiently track and understand the propagation and dynamics of solar storms to improve space weather forecasting and to provide sufficient warnings for the safety of the technological systems and humans in space. The new capabilities for interplanetary space observations enabled by extended FoV coverage will facilitate solar wind measurements that probe the dynamics of space weather between the Sun and Earth at several points inaccessible to current space missions, with the goal of improving the lead time and advanced warning capabilities for space weather events.

| Forecasting Space Weather | | |
|---|---|---|
| Protect humans in space and ground and space-based technology by tracking solar storms and predicting their arrival at Earth | Study the effects of space weather on Earth's atmosphere and the near-Sun environment | Perform high frequency and spatial resolution observations of solar radio bursts associated with powerful coronal mass ejections |

## The Concept of a Next Generation Arecibo Telescope

In order to accomplish the overarching scientific goals stated above, we present a concept for the Next Generation Arecibo Telescope (NGAT) - an innovative combination of a compact, phased array of dishes on a steerable plate-like structure. ***Compared to the LAT, the NGAT will provide 500 times wider field of view, 2.3 times larger declination coverage, 3 times more frequency coverage, nearly double the sensitivity in receiving radio astronomy signals, and more than four times greater transmitting power required for both Planetary and Atmospheric investigations.*** We summarize the new capabilities and direct applications of this facility in Table 1. The new telescope will coexist with an extended High Frequency (HF) facility, and a diverse set of radio and optical instrumentation that continue to operate at AO and at the Remote Optical Facility (ROF). The largely new proposed concept for a radio science instrument requires extensive engineering studies that will be the next step to ensure the new facility achieves the driving scientific requirements for the aforementioned science objectives.

**Table 1: The principally new capabilities and significant technical improvements of the proposed concept and their impacts on the science studies.**

| | New NGAT Capability | Comparison with legacy Arecibo Telescope (LAT) | Enabled/Improved science |
|---|---|---|---|
| **Structural and Instrumental Improvements** | High sensitivity (Gain > 18 K/Jy) | 1.8 - 3.6 times more sensitive from 0.3 to 10 GHz | All fields of science enhanced by increased sensitivity, field of view, and frequency coverage |
| | Large sky coverage; zenith angle range 0°- 48° | 2.3 times declination coverage increase (5.5× increase in sky coverage) | All fields of science benefit from increased sky coverage |
| | Beam Width of each dish[1] ~6 deg. at 0.3 GHz ~3.5 arcmin at 30 GHz | ~500 times increase in FoV | All survey observations immensely benefit from increased field of view |
| | Frequency coverage from ~200 MHz to 30 GHz | A factor of 3× more frequency coverage | Enhanced spectroscopic capabilities, crucial for Space Weather studies |
| | High survey speed | Improved and New Capabilities | Pulsar, FRB, and spectroscopic surveys at various frequencies |
| | Capable of mitigating radio frequency interference (RFI) through phased nulling | Improved and New Capabilities | All observations benefit from RFI mitigation |
| | Dual observing modes as a phased array and interferometer | Improved and New Capabilities | Improves HI intensity mapping, Detecting and monitoring Coronal Mass Ejections |
| **Improved Transmitting Capabilities for Planetary Radar and Atmospheric Sciences** | 5 MW of continuous wave radar transmitter power at 2 - 6 GHz | A factor of 5× more power; maximum transmitter power was 900 kW at 2.38 MHz | Planetary defense: 90% of the virtual impactors can be tracked<br>New space situational awareness capabilities and space mission support<br>Radar of surfaces and subsurfaces of icy worlds |
| | 10 MW peak transmitter power at 430 MHz (also at 220 MHz under consideration) | A factor of 4× more power | ISR studies: Better spatial and temporal resolution to study small-scale ionospheric structures, natural or human-caused, to unprecedented levels |

[1] For the configuration given in Table 2a on page 38.

## The Necessity to Rebuild in Arecibo, Puerto Rico

We propose that NGAT be located at the Arecibo Observatory, preferably at the location of the LAT to take advantage of the existing infrastructure and the extension of the RFI active cancellation system, an active project in development at the AO location. Several other advantages for the Arecibo site include:

### Advantages of Arecibo, Puerto Rico as a site

**Scientific**

- The proximity to equatorial latitudes is ideal for observing Solar System objects.
- The location uniquely enables ISR studies both parallel and perpendicular to the Earth's magnetic field lines.
- The geographic and geomagnetic location provides unique latitude coverage which is not offered by other facilities in the world.
- It is a strategic location from which to study the effects of the South Atlantic Magnetic Anomaly (SAMA) in the Caribbean upper atmosphere as well as on the trans-ionospheric radio signals.
- The location is critical for studying acoustic and gravity waves generated by extreme weather systems approaching the U.S. and Caribbean.

**Socioeconomic**

- To serve the population of Puerto Rico by inspiring and educating new generations while contributing to the socioeconomics of the island.
- To take advantage of the existing infrastructure, which is on federal property, and has the local government support, significantly offsetting costs.
- To leverage the strategic location in the Caribbean Sea, a region with the largest traffic vessels and for which accurate geopositioning is critical, and the ISR inputs for space weather forecast models are crucial.

**Technical**

AO is located in a Radio Frequency Interference (RFI) Coordination Zone which minimizes the effects of RFI, protecting the radio bands needed for science operations.

**Legacy**

To extend and further strengthen the 'long-term legacy' ionospheric data for future climate change investigations.

## Contents of the white paper

This white paper was developed in the two months following the collapse of LAT through discussions with hundreds of scientists and engineers around the world who support the construction of a new and more powerful telescope at AO site. Our goal is to acquire vastly enhanced capabilities that will open exciting new possibilities for the future of radio science with direct applications for planetary defense and the protection of US satellites and astronauts. The remainder of this white paper is outlined as follows: the Introduction (Section 1) includes the context for the push to construct NGAT at the AO site. We discuss the Key Science Goals for planetary science, atmospheric science, and astronomy in Section 2 after first defining the new facility's projected capabilities. In Section 3 we discuss the NGAT concept. Following the main text, we discuss alternative concepts considered for the new facility in Appendix A. An important extension of the NGAT's capabilities in space and atmospheric sciences relies on relocating the High Frequency facility within the AO site, and we describe these plans in Appendix B. Appendix C describes additional science objectives the NGAT concept will enable to continue or improve, and finally, we discuss other AO science activities that interlock with NGAT in Appendices D and E. A summary of the contents of Appendix C and D is listed below. The acronyms used in the document are defined in Appendix F.

**Additional science studies that are enhanced by new NGAT capabilities**

**C.1 Additional Planetary Science Studies including**
- **Radar of comets**
- **Spectroscopic studies of comets and interstellar visitors**
- **Additional missions for spacecraft support**

**C.2 Solar Wind and Space Weather Studies including**
- **Tracking Coronal Mass Ejections (CMEs)**
- **Faraday Rotation and the internal magnetic field of CMEs**
- **Solar Radio Studies**
- **Solar Wind and Space Weather Impacts on the AIMI System**
- **In-situ Data Comparison and Cometary Plasma Tail Investigations**

**C.3 Pulsars Studies including**
- **Wide searches for**
  - new pulsars
  - Binary pulsar studies
- **Pulsar emission mechanism and individual pulses**

**C.4 VLBI Studies including**
- **General Relativity field tests**
- **Applications to Stellar Physics**
- **Observations of Extragalactic Continuum Polarization**

**C.5 Radio Astronomy at high frequencies**
- **Probing the nature of early and late-type stellar evolution from the local to the distant Universe**
- **Pulsars and Transients at high frequencies**

**C.6 A Comprehensive Snapshot of the Galactic plane**

**C.7 Further discussion of Near-Field HI 21 cm Line Cosmology**

**C.8 Detection of Cold Dark Matter and Testing the Standard Model of Particle Physics**

**C.9 Additional Space and Atmospheric Science studies**
- **Ion-Neutral Interactions in the Atmosphere**
- **Sudden Stratospheric Warming Events**
- **Plasmaspheric studies and modeling**
- **Inter-hemispheric flux of particles and its impacts on the Caribbean Sector**
- **Atmosphere-ionosphere-magnetosphere interactions (AIMI)**
- **Vertical Coupling of the Earth's atmospheric layers**
  - Science driver for the NGAT 220 MHz Coherent Radar
  - Wave energy in the F-region thermosphere
  - Climatology, morphology and equatorward propagation of MS-TIDs
  - Tropospheric Forcing on the upper atmosphere during Extreme Weather Systems
  - Aerosol and Coupling Processes in the Lower Atmosphere

## Other Science Activities at the Arecibo Observatory that interlock with NGAT

**D.1 Complementary Space and Atmospheric Science studies**
- **Ion Transport processes using Lidars and ISR**
- **Climate Studies and Forcing of the Ionosphere from below using LidarsGeocoronal hydrogen: Secular change and storm response**
- **Horizontal winds as a function of altitude: New wind measurements above the exobase**
- **The AO Remote Optical Facility (ROF) in Culebra Island**
  - High Doppler resolution measurements of vertical motion in the thermosphere
  - Field line diffusion of HF produced electrons as a function of energy
  - First Caribbean Meteor Radar and its application to enhance the atmospheric probing

**D.2 12m Telescope for radio astronomy**

**D.3 e-CALLISTO spectrometer**

**E. Planetary subsurface studies with 40-60 MHz radar observations.**

## 1.0 Introduction

The Arecibo Observatory (AO) is a multidisciplinary research and education facility that is recognized worldwide as an icon to astronomy, planetary, and atmospheric and space sciences. The AO hosts multiple radio and optical instruments onsite as well as a Remote Optical Facility (ROF) in Isla Culebra. AO's cornerstone research instrument was the William E. Gordon telescope, referred in this document as the legacy Arecibo Telescope (LAT). This telescope, completed in 1963, was a 305-meter spherical reflector dish built into a natural sinkhole of the karst topography near Arecibo, Puerto Rico. The optics to correct for spherical aberration, radar transmitters, and radio receivers to cover multiple frequency bands were housed in the Gregorian dome, suspended on a platform 150 meters above the dish.

For over 57 years the AO has led in scientific research and discoveries. Among the many profound achievements enabled by the LAT, we highlight here the discovery of binary and millisecond pulsars, the first detection of exoplanets, the determination of Mercury's rotation period, and more recently, the telescope's critical role towards the detection of long-wavelength gravitational waves, and the characterization of asteroids for NASA mission support and planetary defense purposes. To date, the data collected at AO- generated from over 1,700 observing proposals from the scientific community - has resulted in more than 3,500 scientific publications, 376 masters and PhD theses, 1,000 student STEM projects, and a number of prestigious awards including the 1993 Nobel Prize in physics, the Henry Draper Award, and the Jansky Lectureship in 2020.

On August 10, 2020, an auxiliary cable that supported the LAT's 900-ton platform experienced a failure, resulting in damage to the telescope's primary reflector dish and the Gregorian dome. In the following weeks, structural models were developed for the platform, towers, and suspension cables so that the appropriate temporary and permanent repair plans could be developed. On November 6, 2020, before the temporary repair efforts could be implemented, a main cable connected to the same tower unexpectedly snapped, possibly as a result of bearing the additional weight due to the loss of the auxiliary cable. Following the second failure, two of the three commissioned engineering reports recommended a controlled decommissioning of the telescope. Based on this report, the National Science Foundation announced on November 19, 2020 that the telescope will be decommissioned. However, on December 1, before the controlled decommissioning could be executed, the remaining cables attached to the same tower failed, causing the collapse of the platform into the 305-m receiver dish below, irreparably damaging the telescope.

In the three weeks following the collapse, AO's scientific and engineering staff and the AO users community initiated extensive discussions on the future of the observatory. The community is in overwhelming agreement that there is a need to build an enhanced, next-generation radar-radio telescope at the AO site. From these discussions, we established the set of science requirements the new facility should enable. These requirements can be summarized as (see also Executive Summary) 5 MW of continuous wave transmitter power at 2 - 6 GHz, 10 MW of peak transmitter power at 430 MHz (also at 220 MHz under consideration), zenith angle coverage 0 to 48 deg, frequency coverage 0.2 - 30 GHz and increased FoV. These requirements determine the unique specifications of the new instrument.

The telescope design concept we suggest **consists of a compact array of fixed dishes on a tiltable, plate-like structure that exceeds the collecting area of the LAT.** This concept, referred to throughout this text as **the Next Generation Arecibo Telescope (NGAT),** meets *all* of the desired specifications and **provides significant new science capabilities to all three research groups at AO.** We discuss the specifics of this unique concept in Section 3.

## 2.0 Key Science Goals

We delineate in the following section the key science goals of NGAT. These science goals were the drivers for the capabilities of the new concept described in Section 3. In order to better discuss these objectives in the context of the new, unparalleled capabilities, we first illustrate the new parameter spaces NGAT will open up. As shown in Fig. 1 (divided in four panels below), the NGAT **will have a field of view 500 times larger than the LAT and 3 times more frequency coverage. It will also have 2.3 times more declination coverage, more than 4 times the radio signal transmitting power, and nearly double the sensitivity to receive radio signals when compared with the LAT.**

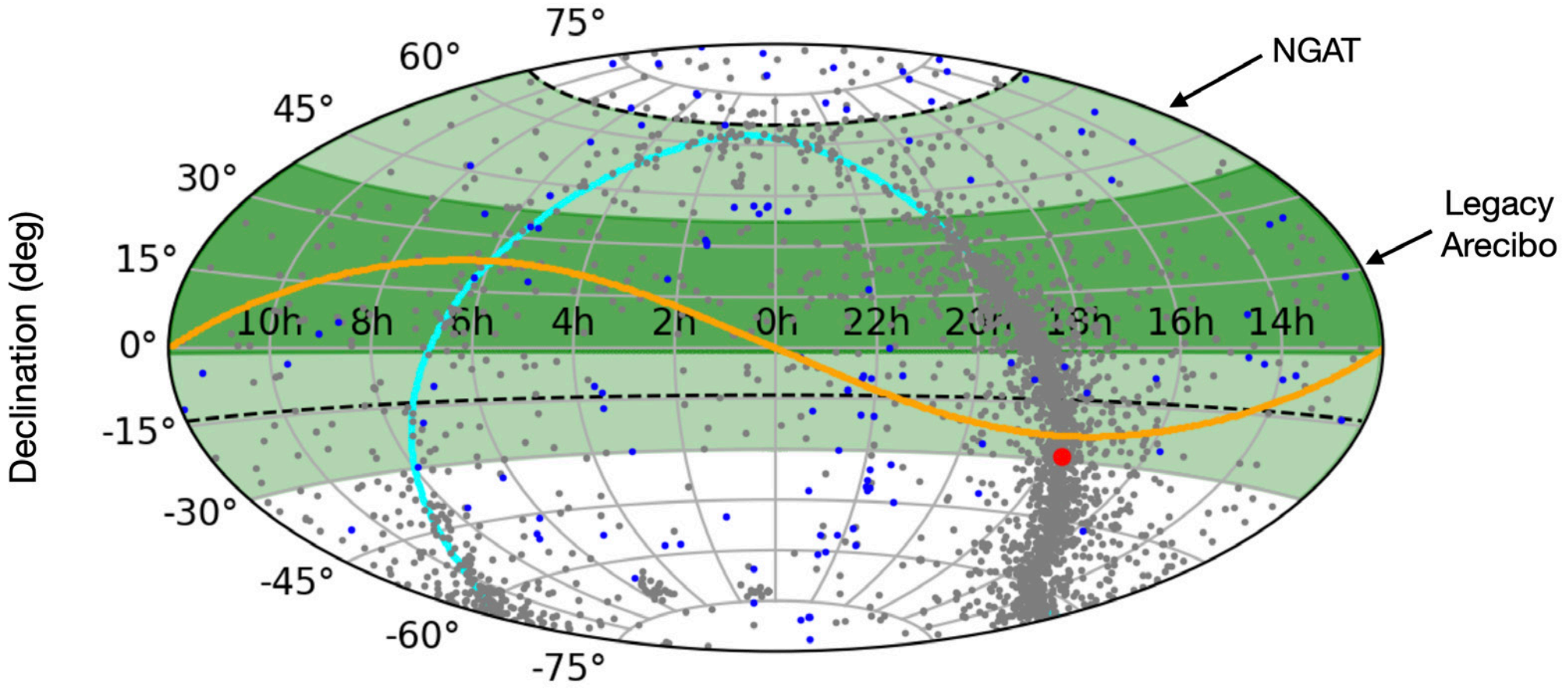

**Figure 1a.** Comparison of the sky coverage (see also Appendix C.3, Fig. 15) of NGAT with the legacy Arecibo telescope (LAT).

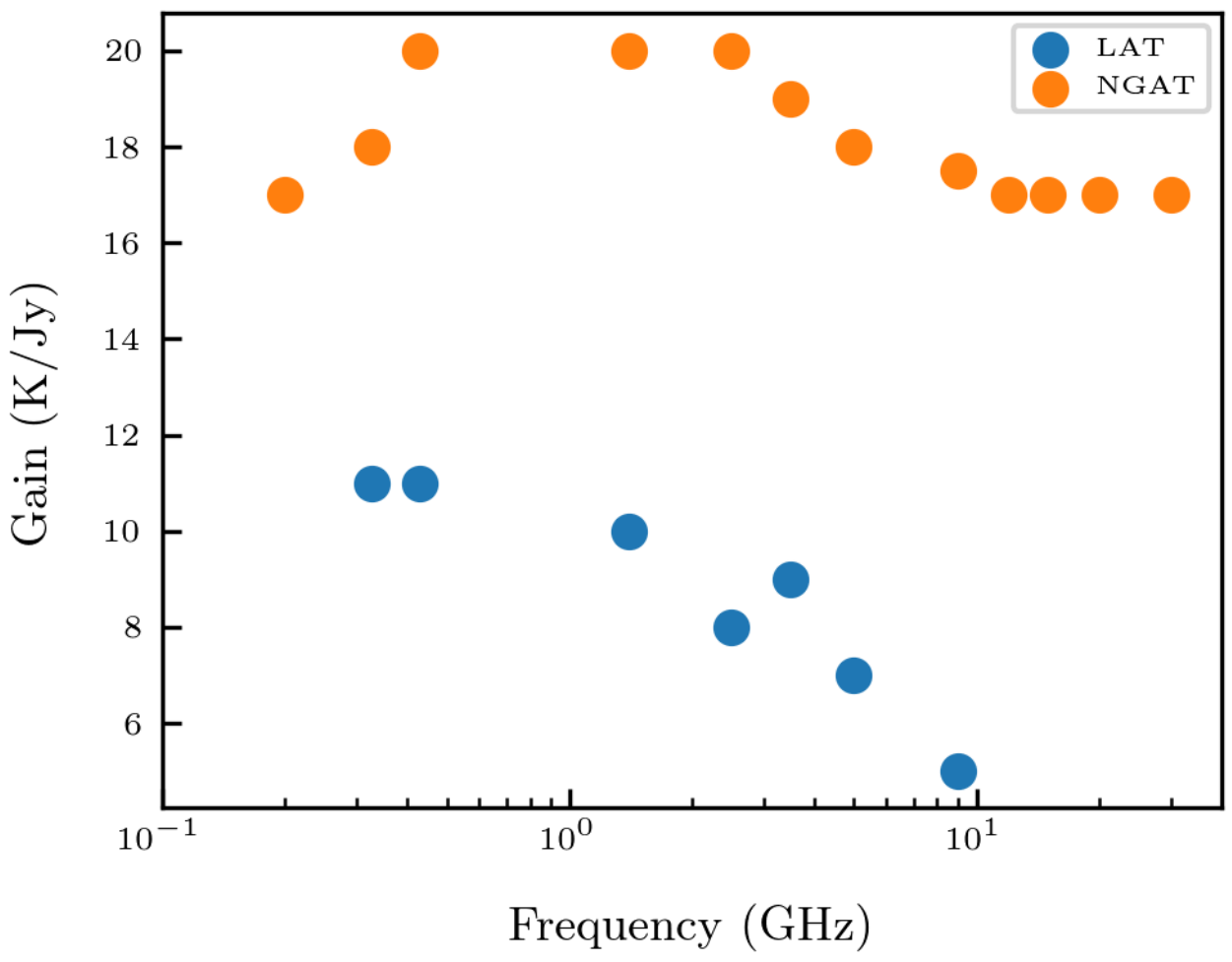

**Figure 1b.** Comparison of the telescope gain of NGAT with the LAT. The NGAT covers a frequency range from 200 MHz to 30 GHz, which is 3 times larger than that of the LAT. The system temperature above 10 GHz was estimated using the ATM model (Pardo et al. 2004; Luca Olmi, private communication).

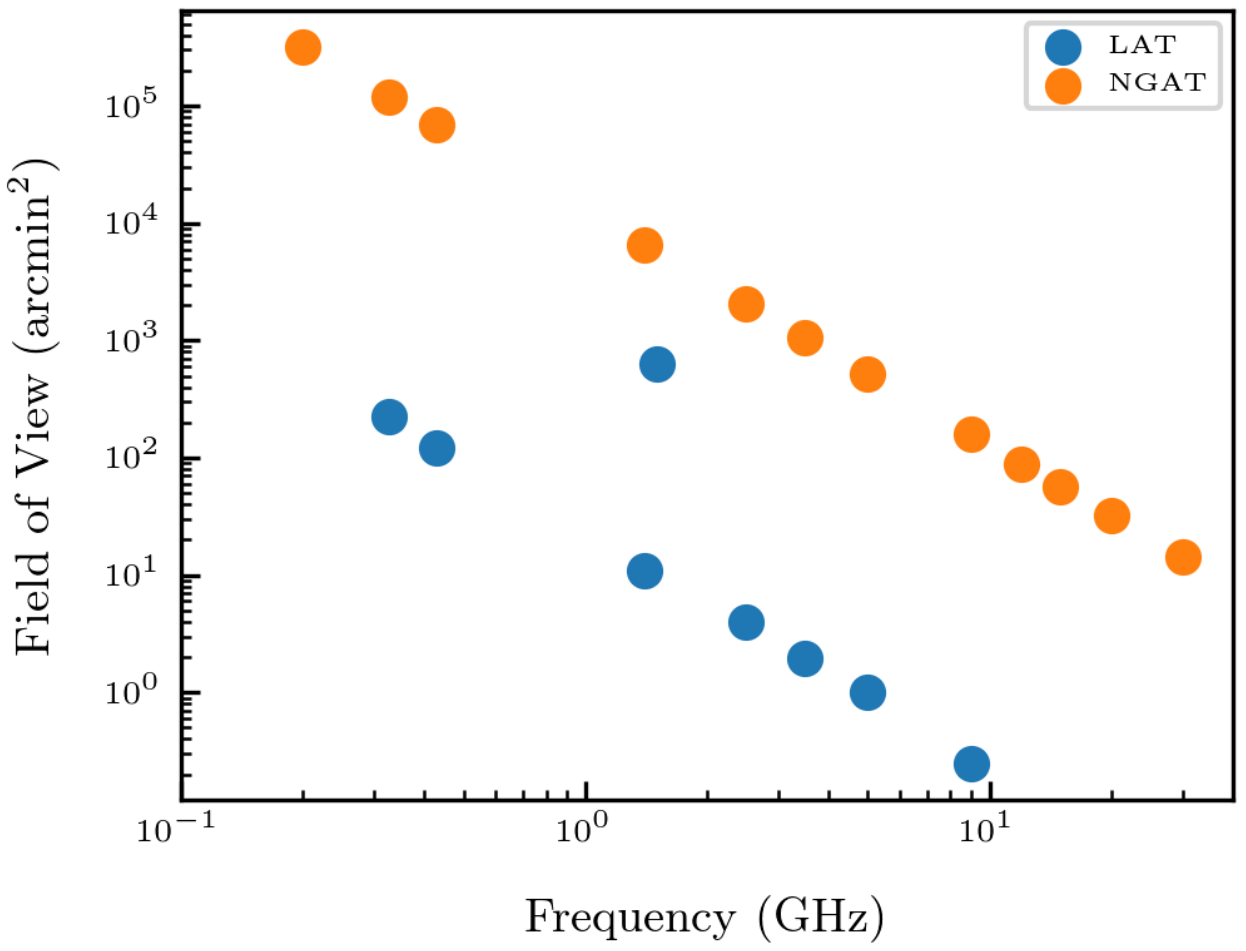

**Figure 1c.** Comparison of the field of view[2] of NGAT with the LAT.

[2] For the configuration summarized in Table 2a

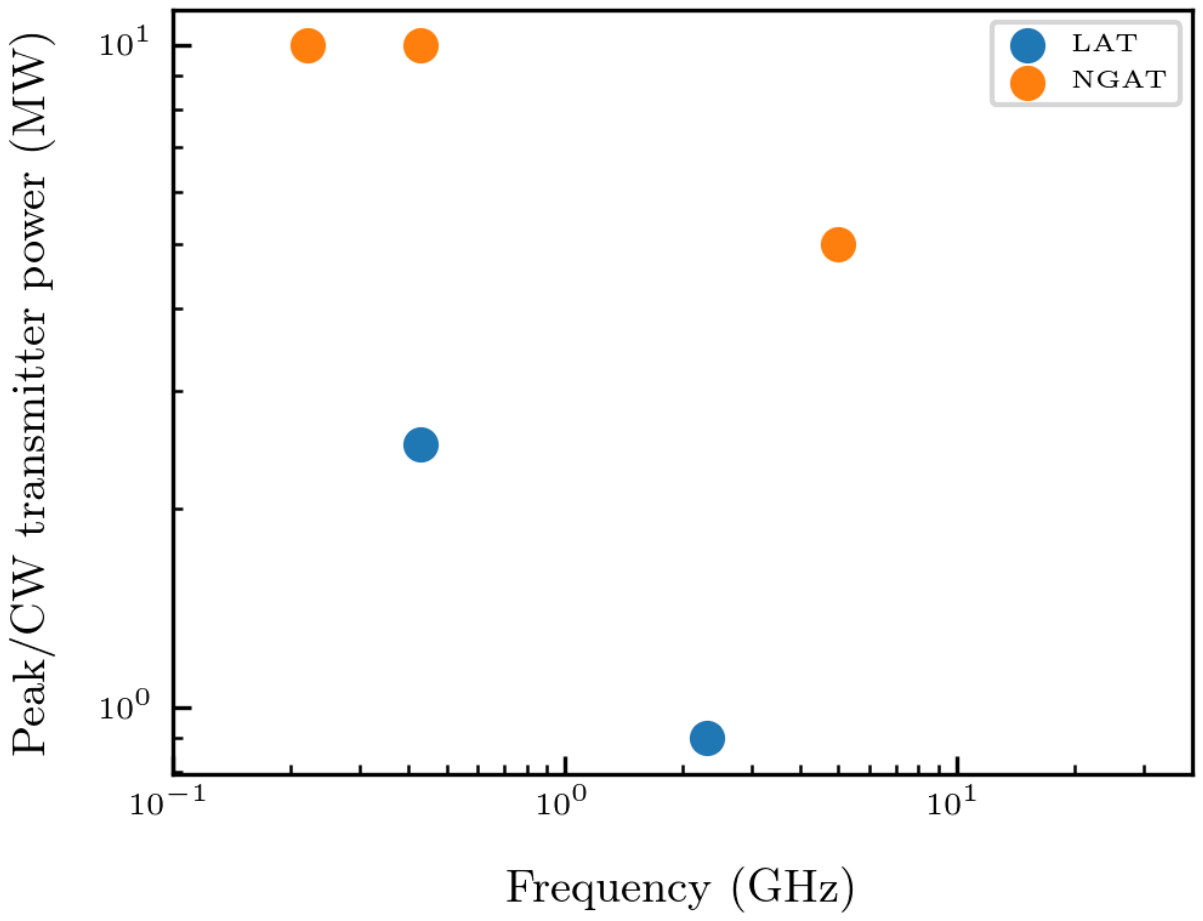

**Figure 1d.** Comparison of the transmitter power of NGAT with the LAT. NGAT will provide a significant increase in the capabilities and therefore science return. The 220 MHz transmitter is under consideration.

For additional context, we present below in Fig. 2 a comparison of the capabilities of the NGAT concept with other existing and upcoming facilities in radio astronomy, planetary radar, and incoherent scatter radars (ISRs) used in atmospheric science.

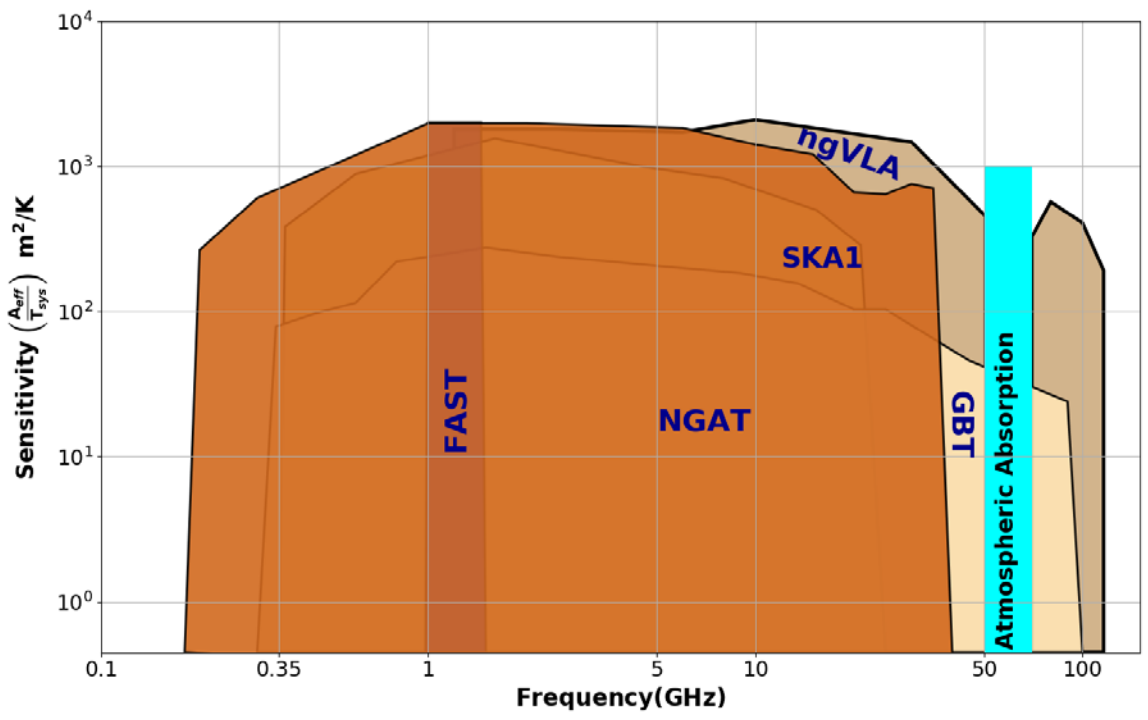

**Figure 2a.** Comparison of the sensitivity of the NGAT with ngVLA, FAST, SKA1, and GBT. We followed the metric defined in Dewdney et al. (2015) for the comparison. The system temperature above 10 GHz was estimated using the ATM model (Pardo et al. 2004; Luca Olmi, private communication).

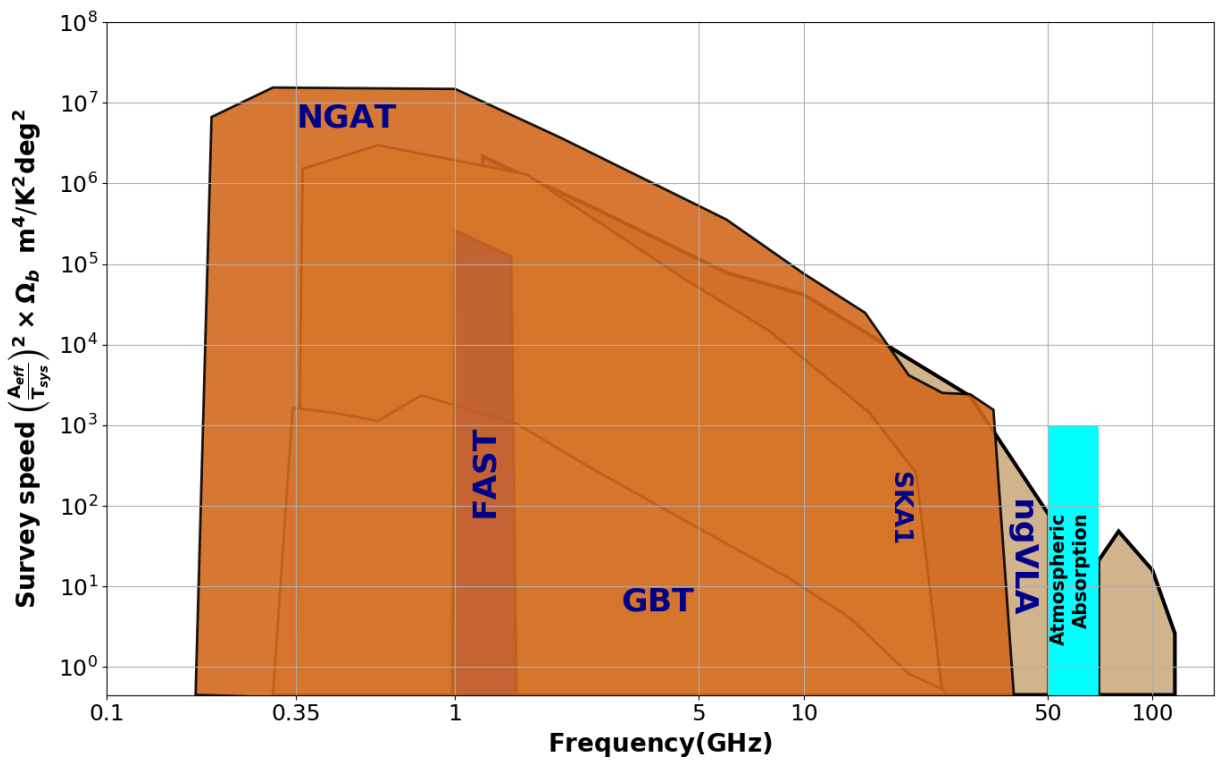

**Figure 2b.** Comparison of the survey speed of the NGAT[3] with ngVLA, FAST, SKA1, and GBT. We followed the metric defined in Dewdney et al. (2015) for the comparison. The system temperature above 10 GHz was estimated using the ATM model (Pardo et al. 2004; Luca Olmi, private communication).

[3] For the configuration summarized in Table 2a

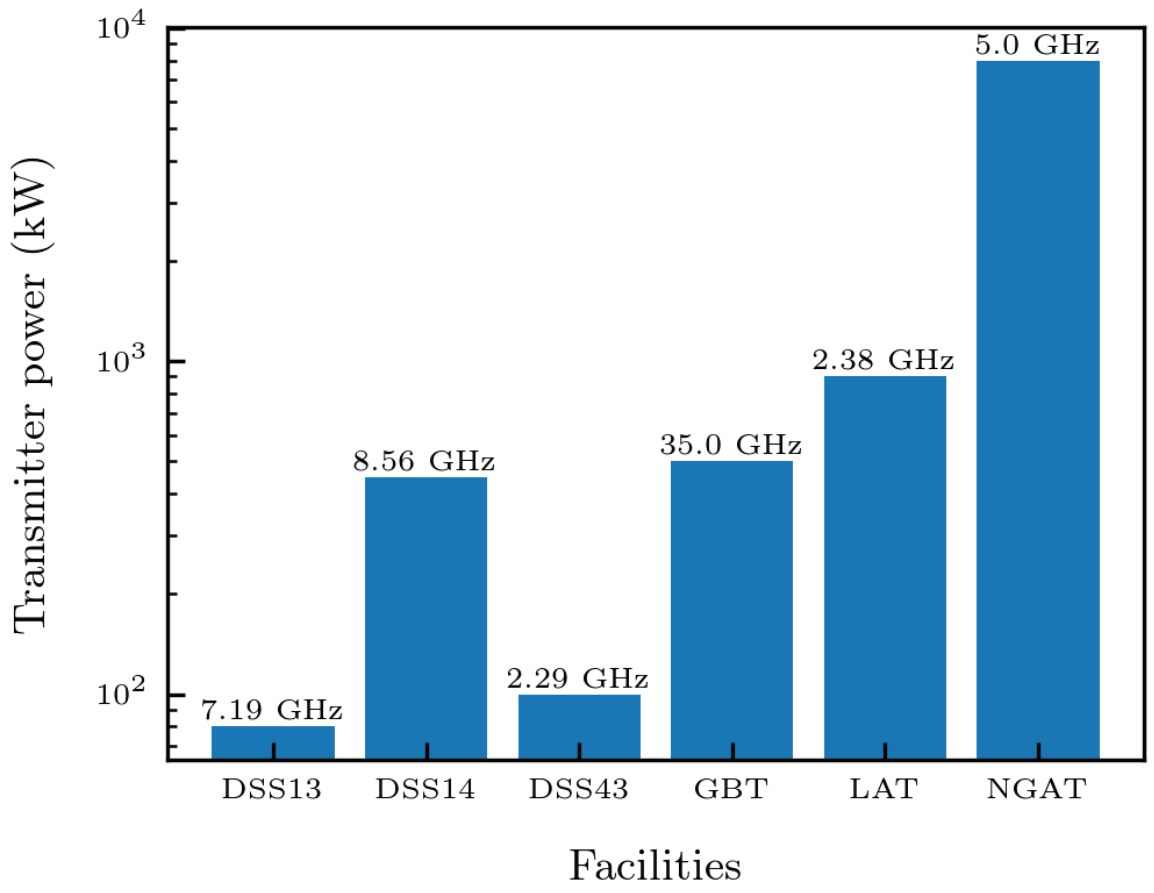

**Figure 2c.** Comparison of the transmitter power of NGAT with other planetary radar facilities.

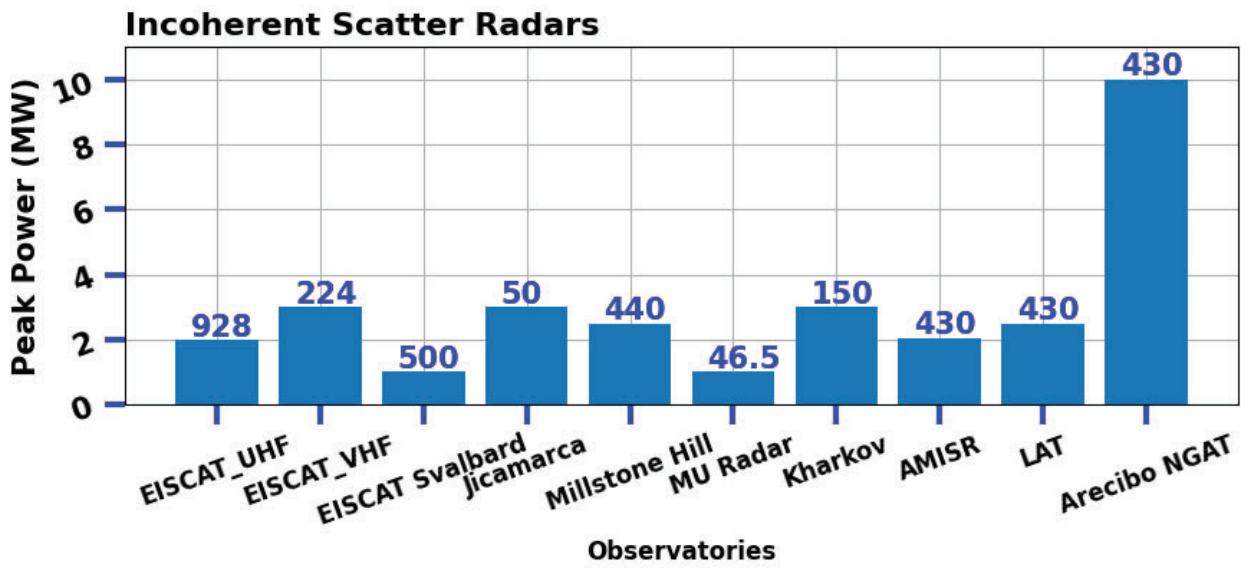

**Figure 2d.** Comparison of the transmitter power of NGAT with other ISR facilities. The frequency of operation of the facilities in MHz is marked on the top of the bar. A significant increase in NGAT capabilities compared to all existing facilities is evident from these plots.

## 2.1 Planetary Defense

Over the past two decades, the legacy AO's S-band (2380 MHz) radar system has been the most powerful and most sensitive planetary radar system in the world, observing far more targets than any other planetary radar facility. The majority of the S-band radar observation time was used for the tracking and characterization of near-Earth objects (NEOs). These efforts have been part of NASA's mandate by the US Congress to discover and characterize 90% of NEOs larger than 140 m in diameter (D) by 2020 [George E. Brown, Jr. (GEB) Act (Public Law 109-155 Sec. 321)]. According to current estimates, approximately 40% of NEOs with D ≥ 140 m have been identified, and the currently operational NEO surveys such as the Catalina Sky Survey (Larson et al. 1998), PanSTARRS (Kaiser et al. 2002), NEOWISE (Mainzer et al. 2011), ZTF (Bellm et al. 2019), and ATLAS (Tonry et al. 2018) cannot reach this goal for at least several decades due to their limited sensitivity and field of regard (Mainzer et al. 2020). The rate of NEO discoveries is expected to increase over the next decade. More capable NEO survey systems such as the Vera C. Rubin Observatory and the NEO Surveillance Mission (NEOSM) are in development. Once they come online, they are expected to achieve the goal of discovering 90% NEOs larger than 140 m within ten years of operation (Ivezić et al. 2007). NGAT will be able to follow up most of these new discoveries. NEO tracking and characterization will undoubtedly continue for decades.

In 2020, almost 3000 new NEOs were discovered[4], and ~10% of these were deemed virtual impactors (VIs)[5]. This means that they possess a nonzero probability of impacting the Earth during the next 100 years. Thus, post-discovery tracking of NEOs is crucial to reduce the uncertainties in their orbital parameters (providing information to prevent them from getting lost), and to determine whether close approaches of large NEOs pose any real threat. It should be noted that approximately 50% of newly discovered VIs cannot be removed from the list during their discovery apparition. Ruling out all of an object's potential impacts typically requires several months of optical-only observations. In contrast, the high-precision range and Doppler measurements by the Arecibo radar system provided invaluable information, and just a single radar observation is usually enough to rule out all potential impacts. In 2019, AO had an observation record of 126 NEOs, of which 78 were recently discovered objects. Very recently, in July 2020, Arecibo radar observations of NEO 2020 NK1 (in Fig. 3) ruled out possible impacts later in this century (Venditti et al. 2020).

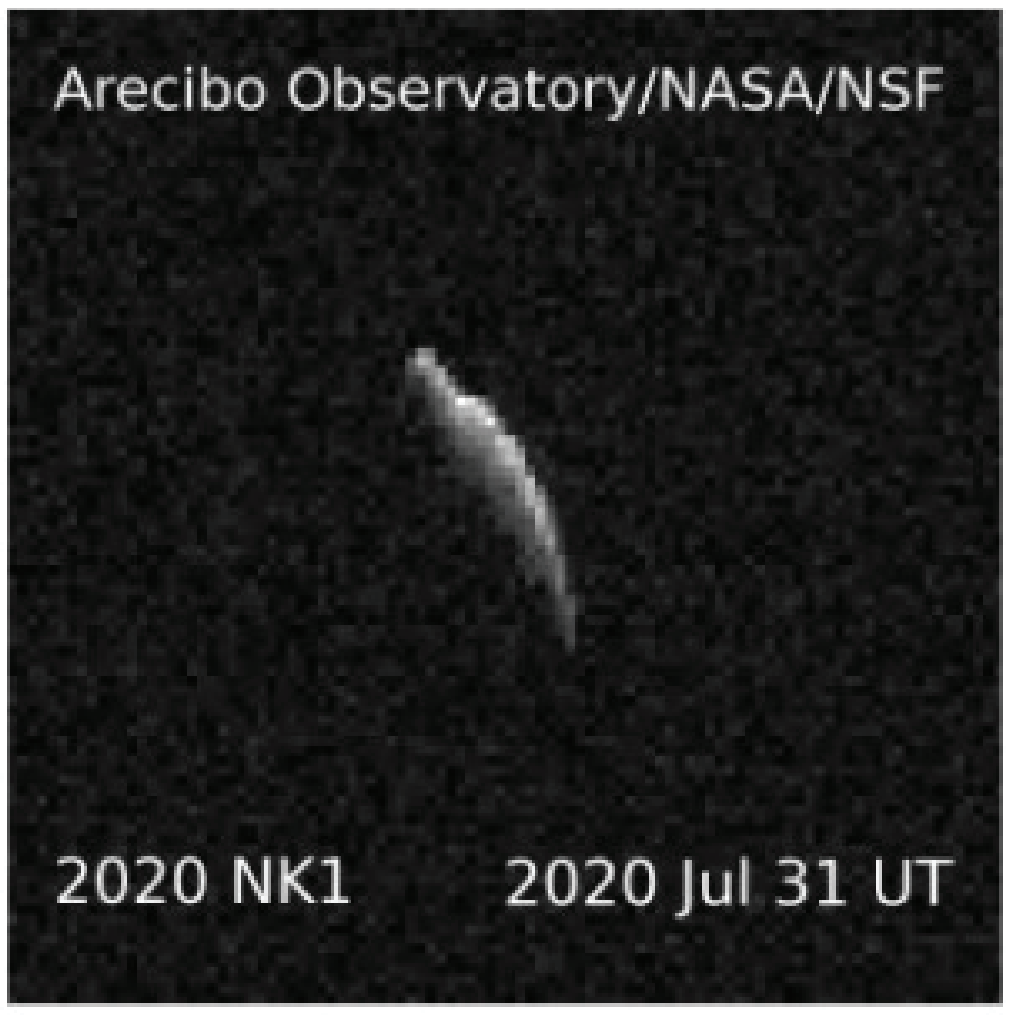

**Figure 3:** Radar delay-Doppler image of 2020 NK1 with vertical delay resolution of 0.05 µs (7.5 m) showing an elongated shape approximately 1 km (0.6 mi) along the long axis.

[4] https://cneos.jpl.nasa.gov/stats/totals.html
[5] https://cneos.jpl.nasa.gov/sentry/

**The new system proposed here will enable observations of more than 90% of the VIs larger than 20 m during their discovery apparition, compared with less than 10% enabled by the LAT**, due to the greater sensitivity (see Fig. 4) and range of sky coverage (see Fig. 5) enabled by this design. Radar observations are not only unparalleled at obtaining high-precision astrometry for accurate trajectory predictions (e.g., Giorgini et al. 2002; 2009), but also at producing physical shape models (e.g., Ostro et al. 2006), detecting satellites (e.g., Margot et al. 2002; 2015), and determining compositional information for asteroids (Benner et al. 2015). All of these are invaluable information for impact mitigation planning within the strategic Goal 2 of the National NEO Preparedness Strategy and Action Plan[6]. Goldstone Deep Space Communications Complex is currently the only facility in the world capable of providing NASA with active radar observations, which places all current radar observations at risk of a single-point failure that could follow from a major earthquake in California, or even a smaller scale technical failure such as the 18-month

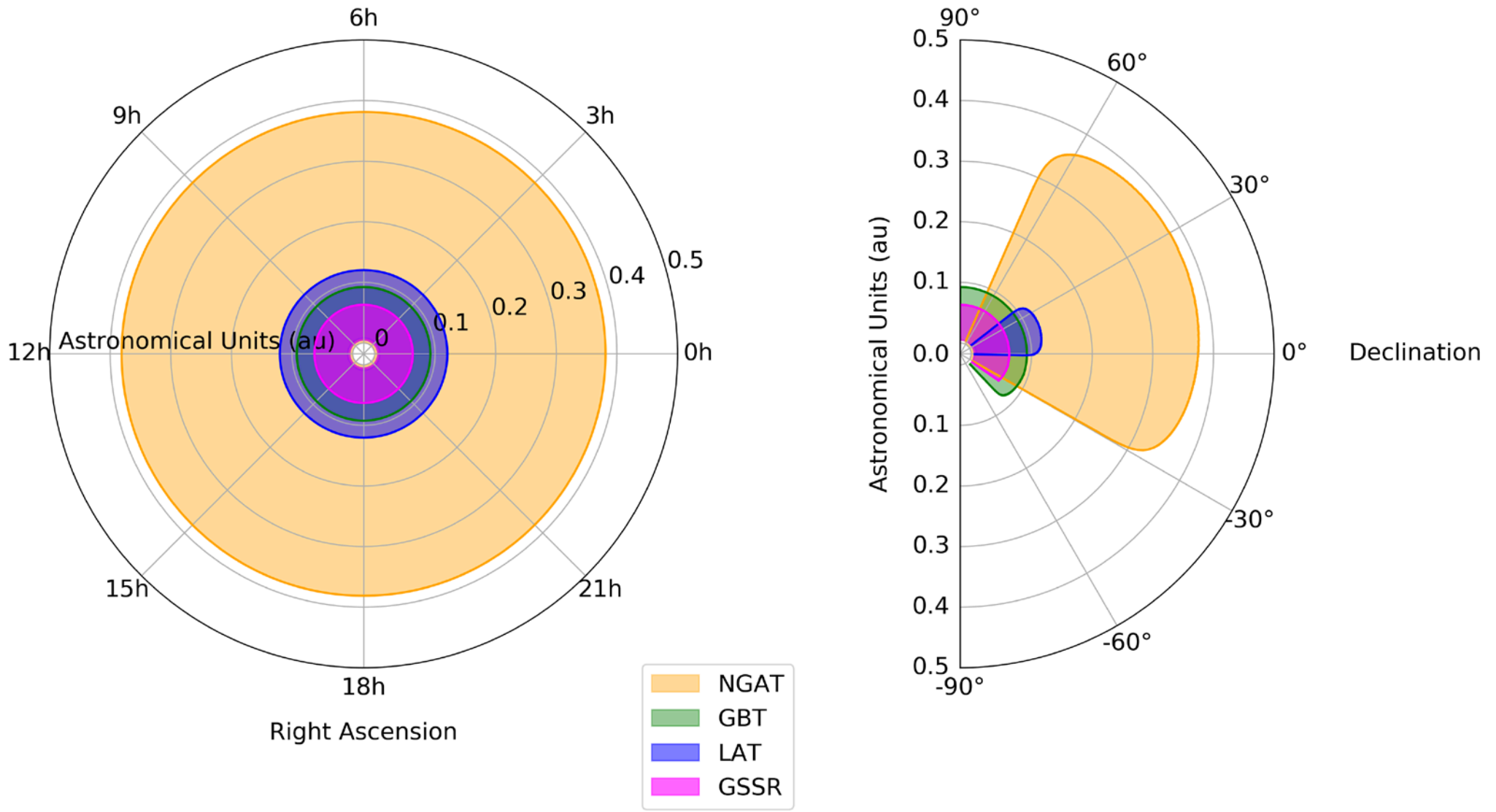

**Figure 4:** Comparison of the proposed NGAT with existing and in development (GBT) planetary radar facilities. The shaded areas correspond to regions of the celestial sphere where each telescope could detect a PHA that is 140 meters in diameter with a 2.1 hour rotation period and a radar albedo of 0.1, above the detection threshold of SNR = 6 (7.8 dB). NGAT = Next Generation Arecibo Telescope, GBT = Green Bank Telescope, LAT = Legacy Arecibo Telescope, GSSR = Goldstone Solar System Radar DSS-14.

[6] https://www.whitehouse.gov/wp-content/uploads/2018/06/National-Near-Earth-Object-Preparedness-Strategy-and-Action-Plan-23-pages-1MB.pdf

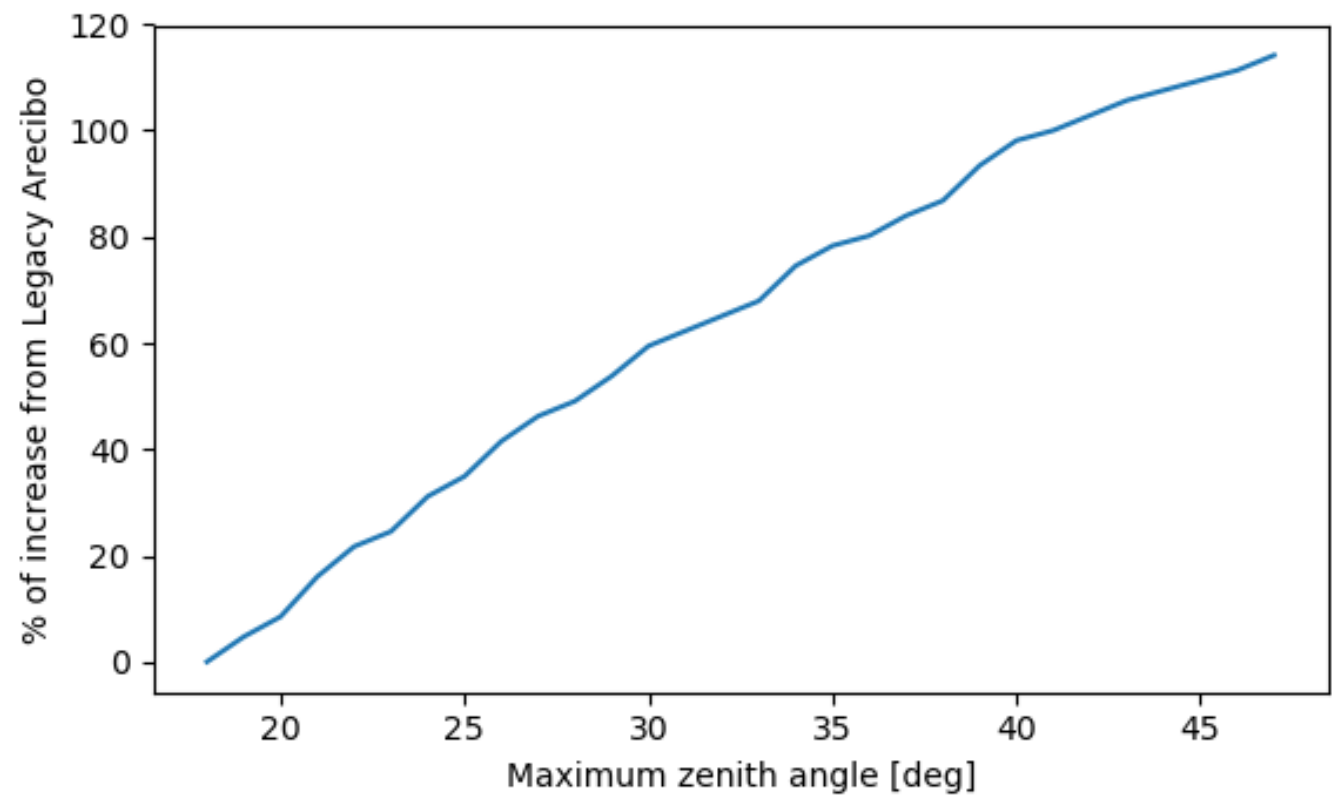

**Figure 5:** Fractional increase of the detectable number of objects as a function of the maximum zenith angle. Increasing the zenith angle from 19° to 48° increases the number of observable objects by more than 100% (factor of two), relative to the LAT. This plot only illustrates how the number of detectable asteroids increases as maximum zenith angle increases. NGAT will have an additional increase in detectable targets (relative to LAT) due to its greater transmitter power and effective area.

Goldstone downtime between May 2019 and September 2020 due to klystron issues. Also, because that facility is part of the Deep Space Network, there are time constraints for when the facility can be used for asteroid observations. (In contrast, asteroid radar observations with the LAT could be scheduled on short notice.) Furthermore, Goldstone's largest dish is 15 times less sensitive than Arecibo was, and therefore will not be able to observe as many NEOs; for example, in 2015, Arecibo observed 77 (30%) and Goldstone 32 (24%) of NEOs that were observable at each facility, respectively (Naidu et al. 2016). Compared to having optical astrometry only, having radar astrometry increases by a factor of five (on average) the time interval over which an asteroid's trajectory can be accurately predicted (Giorgini et al. 2009). In practice, in the vast majority of the cases, radar astrometry makes it possible to rule out any impact in the next few centuries. **NGAT will be able to detect six times more NEOs per year than the LAT**, enabling observations of both smaller and more distant asteroids and resulting in the accurate determination of many more asteroids' orbital parameters.

Green Bank Observatory is planning a Ka-band (35 GHz) radar system (Bonsall et al. 2019). Ka-band is more weather-dependent because atmospheric absorption is significantly more effective at Ka-band frequencies than S- and C-band (2-8 GHz) frequencies. Furthermore, GBT's radar will have a narrower beam than NGAT, which means there is less room to account for the large plane-of-sky uncertainties in some newly discovered asteroids' positions. NGAT will have a wider beam (and also more transmitter power and a greater effective area) than GBT, giving NGAT the advantage when attempting to observe asteroids whose orbital parameters have large uncertainties.

## 2.2 Exploration of Planetary Surfaces and Ocean Worlds

The first major findings of the AO in the 1960s were observations that established the rotation periods of Mercury and Venus (Pettengill & Dyce 1965; Dyce et al. 1967). Right up until its collapse, the Arecibo S-band radar system continued to provide the world's most powerful capabilities for obtaining ground-based observations of the surface of Venus to search for active volcanism (e.g., Campbell & Burns 1980; Filiberto et al. 2020), polar ice in the permanently shadowed craters of Mercury (e.g., Harmon & Slade 1992; Harmon et al. 2011a; Rivera-Valentin et al. 2020a), and lakes on the surface of Saturn's moon Titan (e.g., Campbell et al. 2003; Hofgartner et al. 2020). It has also provided radar imagery of Mars (Harmon et al. 2012; 2017), characterizing the landing site for NASA's Insight lander (Putzig et al. 2017). No other ground-based radar instrument has the capabilities needed to observe the moons or rings of Saturn. Also, **observations of the surface of Venus and Titan are better performed while using frequencies in the C-band or** **below (Neish & Carter 2014).** Goldstone's X-band and the Green Bank Telescope's planned Ka-band are absorbed by Venus' or Titan's atmosphere. Similar to the complementary Arecibo and Cassini radar observations, NASA's planned Dragonfly mission (expected to launch in 2027) will benefit from further characterization of Titan's surface with the unprecedented sensitivity of the NGAT telescope in advance of, and in conjunction with, the mission. When the Jupiter and Saturn systems were at southern declinations, there were periods of years at a time when they could not be observed with LAT, but NGAT's wider declination range will make it possible to observe them every year.

Lower frequency (longer wavelength) radar systems not only enable observations of the surfaces of celestial objects with appreciable atmospheres, but also provide data from below the regolith surfaces. This provides a more comprehensive view of the near-surface properties than observations acquired at shorter wavelengths. We can find intriguing examples as close as the Moon, such as the ongoing debate over the presence of water ice in the polar regions of the Moon (Campbell et al. 2006; Spudis et al. 2013), which could provide invaluable resources for future space exploration if verified to exist in large quantities under the lunar surface or the study of superposed ejecta layer on the Moon's southern hemisphere (Thompson et al. 2009). **In the new era of exploration of the Moon, Mars, and near-Earth asteroids, ground-based radar observations will be key for supporting the robotic and human exploration of the Solar System through landing site characterization, including identification of hazards, volatile reservoirs for in-situ resource utilization, and ultra-precise astrometric measurements** (Rivera-Valentin et al. 2020b).

Moreover, **the unprecedented power and sensitivity of the NGAT radar system will revolutionize our ability to study the surfaces and subsurfaces of the ocean worlds of our Solar System,** parallelling the advancements achieved in the 1960's and 1970's through the LAT's mapping of the rocky terrestrial planets. Comparative radar studies that probe below the surface layers of the liquid-harboring moons of Jupiter and Saturn are a new field of research to explore, not possible from any of the current (or in development) ground-based radar facilities. The NGAT will support and extend the science return for missions including NASA's Europa Clipper and Dragonfly missions and ESA's JUICE mission.

In the past decades, radar characterization from AO has supported several missions to small bodies, such as JAXA's Hayabusa mission to asteroid 25143 Itokawa (Fujiwara et al. 2006), NASA's EPOXI flyby of comet 103P/Hartley 2 (Harmon et al. 2011b), NASA's ongoing OSIRIS-REx mission to asteroid 101955 Bennu (Lauretta et al. 2017; Scheeres et al. 2019), and JAXA's upcoming DESTINY$^{+}$ mission to asteroid 3200 Phaethon (Arai et al. 2018). Radar data provided the mission planning teams with fundamental information about the target, such as high-precision astrometry, size, and shape. Furthermore, an important part of ground-based radar characterization is the ability to reveal whether an asteroid has a satellite, which helped to select targets for two of NASA's upcoming missions to binary NEOs. The first of these is the Double Asteroid Redirection Test (DART) to the binary asteroid system 65803 Didymos, which will be the first planetary defense demonstration (Cheng et al. 2018). The second, Small Innovative Missions for Planetary Exploration (SIMPLEx) mission Janus, will explore two binary NEOs, (175706) 1996 FG3 and (35107) 1991 VH, to better understand how primitive bodies form and evolve into multiple asteroid systems (Scheeres et al. 2020). The NGAT will continue this legacy of spacecraft mission support, and an additional VHF site (see Appendix E) could expand the support to possible direct measurements of internal structures of small bodies.

## 2.3 Space Debris Monitoring

With increased accessibility to Earth orbit, the amount of space debris increases, and the risks of them causing significant damage to a satellite are imminent. Space debris is a continuously growing problem in the Earth's orbit, placing satellites and the International Space Station at risk of micro-impacts. The number of pieces of space debris larger than 1 centimeter officially cataloged by the U.S. Space Surveillance Network is more than 20,000 (Anz-Meador 2020). In addition, more than 500,000 pieces of debris ≥1 cm, and approximately 100,000,000 of size ≥ 1mm are estimated to be orbiting the Earth at a speed of approximately 8 km/s. Furthermore, the collision between objects having an average impact speed of ~10 km/s generates more debris with new physical and dynamical properties (Kessler et al. 2010). Objects on the order of millimeters are harder to detect and catalog by debris surveys, but even a fragment size of millimeters could cause significant damage if it hits a vulnerable part of a satellite. Space debris characterization is an important part of Space Situational Awareness (SSA) and Space Domain Awareness (SDA) efforts, and will likely continue to be so through the next decades as multiple companies are planning to launch large constellations of satellites that will quickly increase the number of spacecraft in orbit (Virgili et al. 2016; Foreman et al. 2017). Radar observations are a powerful method for characterizing space debris due to the metallic composition that is highly reflective at microwave wavelengths. Due to decreasing detectability as distance increases, radar facilities for space debris detection and tracking are focused on low Earth orbit (LEO), and optical systems are used for geostationary orbit (GEO). Thus, there are no current radar facilities dedicated to monitoring the GEO region (a key region for numerous essential satellites). In addition, only larger debris can be tracked in GEO with existing debris monitoring capabilities. NGAT will have the ability to observe space debris smaller than one meter in GEO, due to its enhanced radar power, field of view, and fast transmit/receive switching capability. NGAT also will be able to observe more distant debris, in cislunar space, beyond the reach of existing radar facilities.

## 2.4 Incoherent Scatter Radar

One of the most important capabilities of LAT was its ability to perform the world's most sensitive Incoherent Scatter Radar (ISR) studies at 430 MHz. These active experiments measure electron density, ion and electron temperatures, ion composition, and plasma velocity in the Earth's ionosphere. It is now well documented that ionospheric and plasmaspheric density irregularities are the prime candidate for the failure of our navigation and communication systems, especially during magnetically active periods. **Understanding the turbulent regions of the ionosphere and their effects on satellite communications has important applications for the Department of Defense communications, planned networks of mobile communications satellites, and high-precision applications of global navigation satellite systems such as GPS**. In fact, the initial construction of the LAT was funded in order to monitor the ionosphere as part of the U.S. Advanced Research Projects Agency's Defender Program for ballistic missile defense.

The advantages of the LAT's ISR capabilities were primarily associated with the instrument's high sensitivity associated with the sampling speed, information storage capability, and computational power. With the increase of steering, and sky view capacity, the NGAT's ISR will enhance its spatial and temporal resolution of altitudes close to the mesosphere up to the plasmasphere, which will make possible unprecedented studies of very small scale structures of the thermosphere. The increase in transmitter power and transmitter bandwidth, both the results of the new solid state transmitter modules of NGAT, will enhance nearly all areas of Space and Atmospheric studies.

Observations of the plasma line, which traces the effects of the Sun on the Earth's ionosphere, will have as much as five times the spatial resolution, and as much as ten times the temporal resolution, depending on specific circumstances. Similarly, ion line observations in the lower thermosphere will improve by similar amounts. We will be able to observe >2000 km, the limit with the LAT's ISR system, with about 2 MW of power.

The incoherent scatter spectrum is a significant function of the angle at which the radar observes with respect to the earth's magnetic field, and so observations over the full range from parallel to perpendicular give the most information about the physics. Over Arecibo, the dip angle is about 45°, and so the full range can be obtained without looking more than about 45° from the zenith. Thus, the full angular range is achieved without greatly increasing the distance to observation regions or encountering the problems from looking to near the horizon. *The NGAT concept, which increases our sky coverage, together with the AO's location, improves our ability to study the plasma lines and vector plasma velocities, which are produced by the most fundamental behaviors of the ionosphere*. Overall, these observations are important for studies of atmospheric dynamics, driven by the neutral atmosphere, but manifesting as complicated plasma motion as a function of altitude due the magnetic field and ion neutral collisions.

NGAT will unlock new areas of plasma research never before studied. Fine structures of the ionosphere will be observable (short period waves as well as variations of electron density of small amplitude); a new understanding of the effects of solar radiation on the ionosphere will be possible; nighttime detection of plasma line would help for the understanding of the complexity of interhemispheric plasma flux; and studies of ionospheric irregularities and its impacts on radio transmission will be made. *These new areas of research will strengthen our overall understanding of the Earth's upper atmosphere and its interaction with the solar flux, which is becoming increasingly important as we continue to use radio frequencies for communication with satellites in Earth orbit.*

## 2.5 Climate Change Investigation

Climate change is the long-term alteration of the average weather pattern in a particular location or the planet as a whole. Climate change may cause weather patterns to be less predictable (short term), impacting our society. Many of the effects of climate change have been attributed to the release of greenhouse gases (e.g., carbon dioxide and methane) into the atmosphere, which ultimately results in warming of the lower atmosphere but cooling of the upper atmosphere. In the thermosphere, where the density of carbon dioxide is too low to absorb much infrared radiation coming from below, greenhouse gasses promote cooling by radiating energy acquired via collisions to space in the infrared wavelengths (NRC, 2013).

*To address the challenges of climate change, adequate science and operations support is required to enable frontier research that adds to the long-term record necessary for analyzing space climate over several solar cycles.* **This includes the knowledge of the responses of the lower and upper atmosphere to energy deposition coming from the Sun as well as galactic and extragalactic sources** (radiation; solar wind flux/density; interplanetary magnetic field; energetic particles, relativistic or not, etc). Also, the peculiarities of a given geographic location and the knowledge of the atmosphere's responses to local seasonality are critical for establishing baselines and distinguishing fluctuations related to climate change (Brum et al. 2011).

In this context, LAT's ISR has long been collecting high quality ionospheric, thermospheric, and exospheric data since the 1960s. In addition, a cluster of optical (passive and active) instruments and radio receivers have been supporting the radar observations, helping to solve numerous questions about the responses of the upper atmosphere to different forcing, local weather or from outer space. *All together, these data have resulted in the most comprehensive, long-term, ground-based database in existence with exceptional temporal and spatial resolution*.

The analysis of such data set at AO revealed unique long-term changes in the ionosphere never seen before. These include a steady lowering of the ionosphere's F-region peak altitude of about 700 m per year and a slight increase/decrease in its peak density before/after the local midnight over a period of 30 years (Santos et al. 2011, Fig. 6a). AO data also revealed a long-term trend in the meridional and zonal components of the thermospheric neutral winds which can be larger in magnitude than the seasonal, solar cycle, or geomagnetic activity influences (Fig. 6b) (e.g. Brum et al. 2012; Tepley et al. 2011, Brum et al. 2019). **These variations, which required continuous observations over the past four decades, may point to important atmospheric variations provoked by climate change, but a clear understanding of their origin has not yet been reached.**

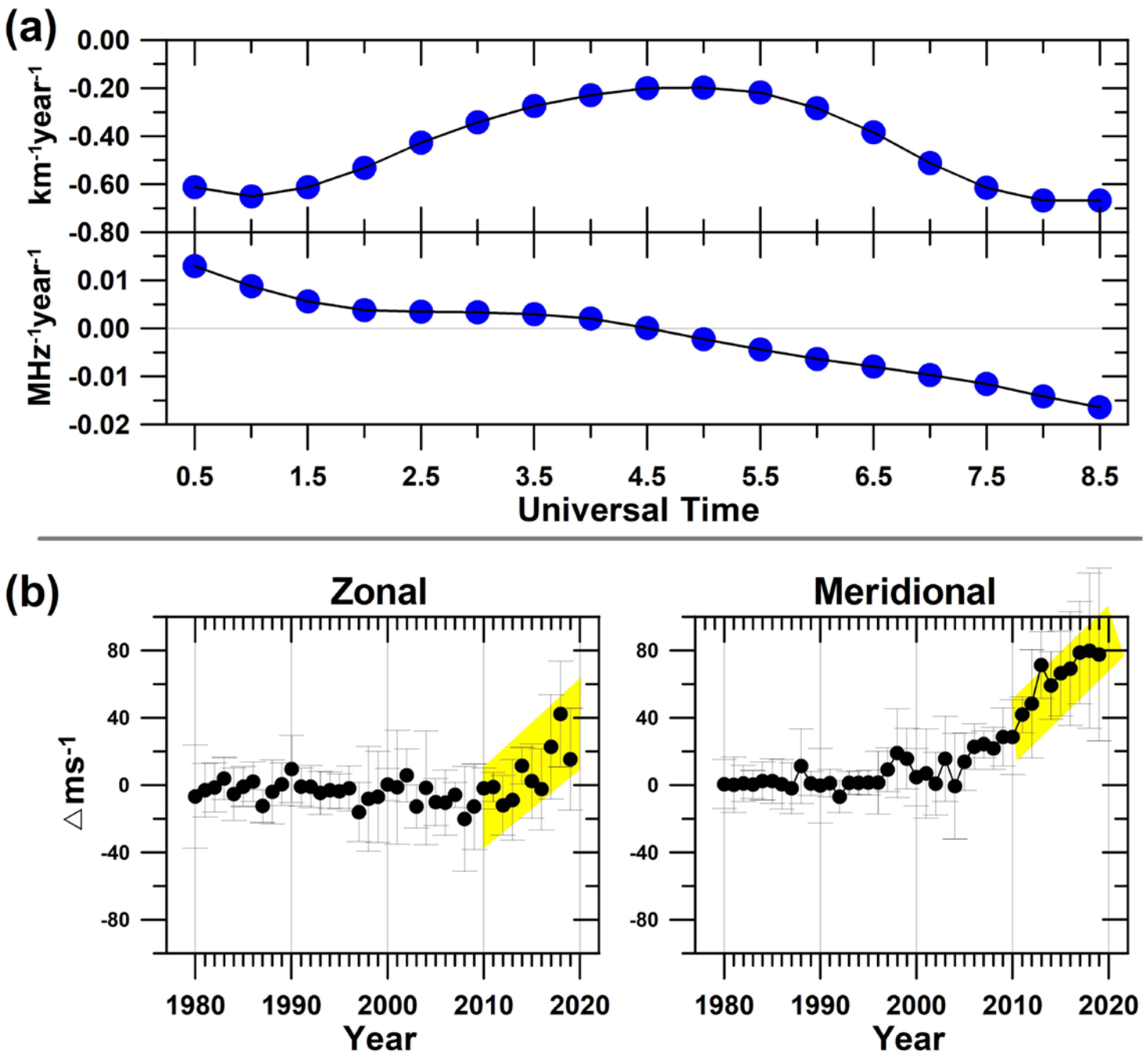

**Figure 6.** Combined long-term observations using ISR and Fabry-Perot techniques at the LAT's site; (a) The yearly rate of the long term trend of F-region ionospheric peak ISR parameters, foF2 and hmF2, in $MHz^{-1}yr^{-1}$ and $km^{-1}$ $yr^{-1}$, respectively (adapted from Santos et al. 2011). (b) Variability of the thermospheric neutral winds residual registered by Fabry-Perot at Arecibo over the last 40 years (1980 to 2019) showing a remarkable change in the rate of the components of the neutral wind.

As climate change forecasts become an urgent national enterprise, research-to-operations (and the reciprocal), data collection and dissemination should continue to be a daily routine and responsibility for the future of the Space and Atmospheric Sciences group at AO. It is only by distinguishing between the influence of meteorological and space weather on the neutral and ionized coupling phenomena in the Earth's atmosphere that we will be able to understand global climate change. Not only will the NGAT's increased sensitivity and improved ISR capabilities enhance our observations of these climate change-induced phenomena, but more critically, it will also allow us to build directly on the existing baseline derived from decades of data collected from the same location to establish long-term and periodic climate trends and changes. Furthermore, the addition of transmitting capabilities at 220 MHz will enhance NGAT's ability to provide new inputs to weather forecasting models and, in the long term, contribute to the understanding of the climate change impacts on extreme weather systems occurrence and severity.

The NGAT's observations will be augmented by the existing onsite Arecibo Optical Laboratory (AOL), lidars, radio receivers, the High Frequency (HF) Facility (see Appendix B) and the offsite Remote Optical Facility (ROF, see Appendix D) in Isla Culebra.

## 2.6 Space Weather Forecasting

The effects of "space weather" at the Earth are determined by the continuous flow of highly variable solar wind plasma, the interplanetary magnetic field (IMF) carried by the solar wind into the heliosphere, and the large-scale transient interplanetary disturbances propagating away from the Sun. Transient events such as coronal mass ejections (CMEs) and co-rotating interaction regions (CIRs) both cause moderate to severe space weather disturbances on Earth and other planets. Additionally, electromagnetic radiation produced at the time of solar flares/CMEs can also affect Earth's atmosphere and ionosphere in several ways. **The above space weather processes can have negative effects in many technologies, including satellite operations, telecommunications, navigation systems and power grids.** *For instance, the greatest solar storm event of 1-2 September 1859, the Carrington Event was the most powerful space weather event, due to which a significant part of the world's telecommunications were adversely affected and led to a real economic impact. If a similar event happened today, it could grind to a halt the high-tech infrastructure of the entire world.* **A reliable prediction of space weather is necessary to provide sufficient warning to take effective measures against its adverse effects on human technology and astronauts participating in manned space missions.**

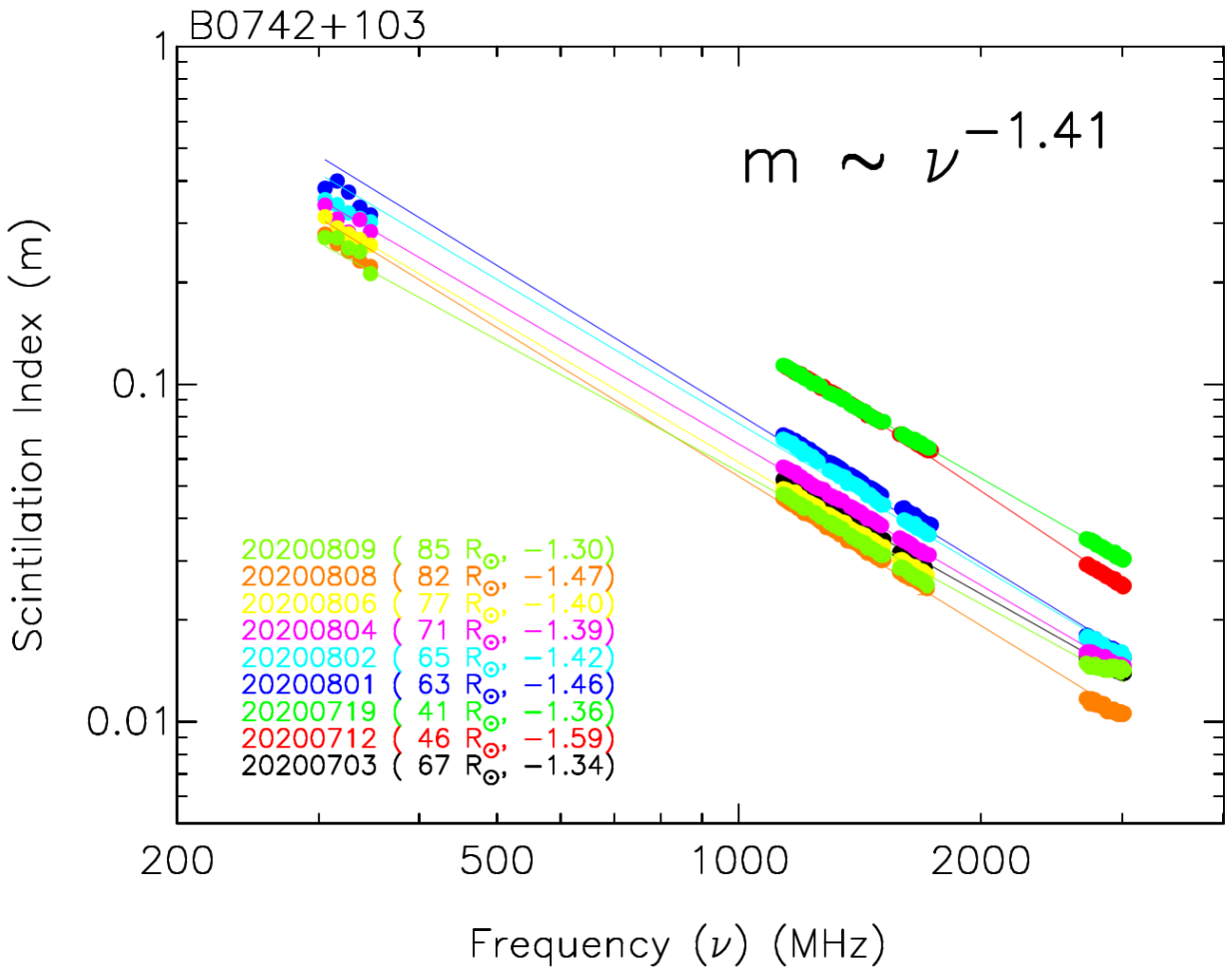

**Figure 7**. Scintillation index observed with LAT plotted as a function of observing frequency for the compact radio source B0742+103. These observations cover a period between 03 July to 09 August 2020. On some of the days, measurements have been taken at multiple frequency bands, nearly simultaneously within about 1-hour time, and these observations are joined by the best-fitted straight line. The scintillation index seems to nicely scale with the observing frequency as a power-law form. The S-band observations (>2700 MHz) can be extended to closer solar offset, <20 solar radii. For each day, the solar offset in solar radii and the slope of the best-fitted straight line are shown (Manoharan, Perillat, et al. in preparation). NGAT can provide a much better understanding of space weather studies at an improved frequency coverage.

The arrival of CMEs at the Earth has the propensity to cause intense geomagnetic storms. Several institutions and centers, like NOAA's Space Weather Prediction Center (SWPC) and the U.S. Air Force's Weather Agency (AFWA), are interested in predicting/forecasting the arrival of space weather CME events. However, predictions based on models are limited by the dynamical evolution of CMEs due to various fluid and kinetic instabilities evolving at the interface between the CME ejecta and the sheath as the CME propagates away from the Sun. On the other hand, the lead-time currently available from L1 point space mission observations is significantly less for reacting to or mitigating the effects of propagating disturbances. The efficient tracking of CMEs in between the Sun and the orbit of the Earth calls for solar wind observations at consecutive locations of the heliosphere. The radio remote-sensing technique of interplanetary scintillation (IPS) provides the properties of the solar wind in three-dimensions (e.g., Hewish et al. 1964; Coles 1978; Manoharan 1993, 2012; Tokumaru et al. 1994). Since Earth-directed CMEs could cause interplanetary disturbances, IPS observations can be used to track and forecast their arrival at the Earth. Several instruments are available to study IPS (e.g., the Ooty Radio Telescope in India, the Solar Wind Imaging Facility in Japan, and the Mexico Array Telescope). The predominant limitation constraining the usefulness of current IPS measurements and their science output is the sampling resolution of the heliosphere in space and time. Additionally, most of the current IPS arrays are operated at meter wavelengths (i.e., in the frequency range of ~100 to 327 MHz), which can be useful to probe the solar wind at distances >50 solar radii (~0.25 au) from the Sun. IPS observations over a wider frequency range are essential to probe CMEs from the near-Sun region to Earth as has been demonstrated with LAT (see Fig. 7). NGAT's wider frequency coverage (200 MHz to 30 GHz), increased sensitivity, greater sky coverage, and multi-beaming capability can revolutionize investigating the solar wind and detecting CMEs all the way from the Sun to near-Earth orbit (Manoharan 2006).

The solar magnetic field is the main cause for all kinds of solar activities. A quantitative knowledge of the solar/coronal magnetic field is essential to understand solar energetic eruptive phenomena above the photosphere. The flux density spectrum of moving type IV radio bursts at frequencies >200 MHz, along with the routinely available space-based solar images and ground-based radio images from the dedicated heliograph (e.g., Nancay Radio Heliograph), will allow us to determine the gyro-synchrotron components and their associated magnetic field within the CME (e.g., Bastian et al. 2001; Dulk & Marsh 1982; Maia et al. 2007; Simoes & Costa 2006). *The wide-frequency coverage of NGAT will be used to study the CME associated radio bursts at meter wavelengths, which can provide accurate CME magnetic field strengths along with space-based and ground-based solar images well as the characteristics of the distribution of non-thermal electrons causing the radio emission. This information is essential to understand and predict space weather.*

In summary, *for the first time, NGAT will be able to efficiently track the magnetized plasma clouds released from the Sun into the interplanetary medium and to understand the impact on Earth's atmosphere, virtually from the troposphere up to the protonosphere/plasmasphere. This will enable detection of the strong propagating disturbances during particularly powerful solar storms and, with a prediction and mitigation plan in place, it could significantly minimize their effects in communications on Earth and on humans and instrumentation in space.*

## 2.7 AO as a laboratory to test theories of Global Climate Change

The human race is becoming increasingly concerned that secular changes are occurring in the Earth's climate and that the effects of those changes may be increasingly inimical to current behaviors and lifestyles. NGAT will provide crucial, prerequisite understanding of the Geospace system and the mechanisms that mediate climate change as humanity faces this global challenge. It has become increasingly clear that the large-scale interactions between the Sun, which ultimately drives the climate throughout the inner Solar System, and the entirety of Geospace take place through relatively small-scale processes that are not yet fully understood. Without full understanding of the operation of all those small-scale processes, it is not possible to fully understand the entirety of the Geospace environment far less to predict its behavior and evolution on either short or long time scales.

The NGAT will provide the best and most detailed observational basis to describe the entire atmosphere behavior in three dimensions and over extended intervals supporting progress in elucidating and understanding all the processes which control its behavior. **Together with the HF facility (**see Appendix B**), NGAT will also have the ability to disturb pre-existing states of the atmosphere to test developing theories.** *Currently, there is no other facility in the world able to support these critical investigations in the vast volume of Geospace contained within the* *closed shield provided by the Earth's magnetic field.* Only the EISCAT3D system, currently under construction in Northern Scandinavia, will provide comparable capabilities. However, it will be dedicated to studies of the northern auroral zone.

## 2.8 Searching for Habitable Worlds

The search for exoplanets, especially Earth-like planets, is one of the most important endeavors in astronomy both now and in the coming years. Cutting-edge radio telescopes have for the first time the capability to detect and assess exoplanets by observing both bursts of circularly polarized, coherent, emission - referred to as Electron Cyclotron Maser Emission (ECME) - induced by the magnetic interaction between a star and planet (e.g. Nichols and Milan, 2016), and by observing characteristics of their atmospheres during transits. We discuss each of these possibilities in turn.

ECME emissions from a wide variety of systems are expected in the ~100 MHz to 30 GHz frequency range. These systems include:

- Very young, highly magnetized, and rapidly rotating pre-main sequence stars, often detectable in the radio (Güdel 2002);
- Hot, young planets that may possess a strong magnetic field that could be detected along with stellar radio emission at GHz frequencies (Reiners and Christensen 2010, Route & Wolszczan, 2013, Bower et al. 2016, Donati et al. 2016, and Bastian et al. 2018);
- Stellar flares may be induced by the interaction of the magnetic fields of hot Jupiter exoplanets and stellar active regions that are detectable at radio wavelengths (Cuntz et al. 2000; Route 2019)
- Brown dwarfs at temperatures of 1,000 K and below exhibit radio emission (Route and Wolszczan 2012, 2016), which implies that even the coolest dwarfs are capable of generating bursts of ECME signals;
- Close-in planets in habitable zones of sufficiently magnetized low-mass stars and brown dwarfs are potentially observable at GHz frequencies by exploiting the unipolar dynamo (or inductor circuit) mechanism (emission similar to the Jupiter satellite system; Zarka et al. 2018);
- Low-mass stars exhibit stellar flares that can be observed with dynamic radio spectroscopy to further characterize the related magnetic activity and its impact on planetary habitability, and such studies have produced interesting results (Villadsen and Hallinan, 2019).

Detection of radio emission from exoplanets would dramatically extend the experimental basis for studies of their physics and their habitability (e.g. Lazio et al. 2019), as this would enable measurements of planetary magnetic field strengths and topologies, constrain their stability, and provide insight on the nature of the dynamo mechanism. The unknown magnetic fields in these systems require an ECME search over a wide frequency range well outside the protected bands for radio astronomy. ***The NGAT will provide the community with a new tool with optimal capabilities to perform this search. The wide frequency coverage of NGAT, its sensitivity, and its ability to mitigate RFI are ideal for searching for ECME emission from exoplanets and identifying those that could harbour life.***

In addition to tracing ECME emission, the broader sky coverage of NGAT would allow for the characterization of many targets of interest discovered by NASA K2 and the Transiting Exoplanet Survey Satellite (TESS), such as systems like TRAPPIST-1 (see, e.g., Pineda and Hallinan, 2018). In particular, stellar emissions of all radio wavelengths are obstructed by a rocky planet and its atmosphere during a transit. The obstruction due to the planet's atmosphere is different at radio and optical wavelengths. Thus, the combination of optical and radio transit measurements would quantify properties of the atmospheres of rocky exoplanets. Further, radio observations can make use of the sharp features of star-spots allowing for significantly improved transit depth measurements (Pope et al. 2019). ***The sensitivity and frequency range of the NGAT will complement observations of optical transits of exoplanets providing new insights into their atmospheres and plasma environments.***

## 2.9 Detecting Gravitational Waves with Pulsar Timing Arrays

The clock-like behavior of millisecond pulsars (MSPs) makes them excellent laboratories to probe fundamental physical phenomena such as gravitational waves (GWs – Detweiler 1979, Hellings & Downs 1983). These are fluctuations in space-time caused by accelerating massive objects and a key prediction of Einstein's theory of general relativity (GR). The North American Nanohertz Observatory for Gravitational waves (NANOGrav) is a collaboration of scientists and students in the US and Canada, aiming to directly detect nanohertz GWs and uncover a new observational window in the low frequency regime of the gravitational wave spectrum (Ransom et al. 2019, Cordes et al. 2019). NANOGrav monitors regularly an array of more than 70 highly stable MSPs mainly using the largest radio telescopes in the US – the Green Bank telescope in WV and the Arecibo telescope in PR (until the collapse) – to achieve the required high timing accuracy necessary for the GW detection. **The nanohertz frequencies are accessible by only MSP timing and not through any other methods of GW detection**. The most promising sources that produce GWs in this frequency regime are supermassive black hole binaries (SMBHBs – Colpi et al. 2019, Taylor et al. 2019). They are billions of times more massive than those recently detected by LIGO, reside at the centers of galaxies, and merge when galaxies collide.

A detection of these GWs will revolutionize our understanding of galaxy formation and evolution through cosmic time. NANOGrav's primary objectives are: detect a stochastic background of nanohertz GWs; characterize this background in terms of the demographics of the expected emitting population of SMBHBs; detect individual SMBHBs, and perform multi-messenger follow-up with large synoptic photometric and spectroscopic surveys operational in the 2020s; constrain fundamental physics by probing the neutron star equation of state, cosmic strings, primordial GWs, first-order cosmological phase transitions, beyond-GR theories of gravity, and dark matter (Ransom et al. 2019).

The first NANOGrav detection is likely to be the stochastic GW background, i.e. the cumulative emission of GWs from unresolved SMBHBs throughout the universe. A stochastic signal was detected in the recent NANOGrav 12.5-yr data set that included 47 MSPs; however, this signal did not show a statistically significant "Hellings & Downs" correlation (see Fig. 8, bottom panel) to claim a detection of GW background (Arzoumanian et al. 2020). The high-quality timing data collected with LAT over 15 years until its closure supported 50% of NANOGrav's sensitivity to GWs. With more data and MSPs in the timing campaign, NANOGrav expects to detect the GW background within the next 3 – 7 years (see Fig. 8, top panel – Ransom et al. 2019). The sensitivity of NANOGrav improves over time and it is in the regime of detecting continuous GWs from individually resolved, sufficiently nearby ($z < 0.5$) and/or massive SMBHBs (chirp mass $10^9$—$10^{10}$ $M_\odot$). They expect to detect one or more individual SMBHBs as a quasi-monochromatic GW signal by 2030 (Ransom et al. 2019). The searches for GWs produced by individual SMBHBs will be cross-validated with near-future EM surveys that can register photometric (e.g. LSST) and spec-troscopic (e.g. SDSS-V) signatures of dual active galactic nuclei (AGNs), allowing for a complete multi-messenger portrait of merging SMBHs (Ransom et al. 2019, Kelley et al. 2019). However, the loss of LAT will impact and delay their efforts, since it was used to monitor approximately half of their pulsars and it obtained high signal-to-noise timing data.

As shown in Fig. 8 (bottom panel), the sampling of “Hellings & Downs” correlation becomes poor with a single telescope alone. **Therefore, NGAT will be a vital instrument for NANOGrav, and the large collecting area and advanced instruments (e.g., broadband feeds and RFI mitigation capability) will produce even better-quality high signal-to-noise MSP timing data.** Further, including more MSPs is important for Pulsar Timing Arrays (PTAs) as their sensitivity improves directly with the number of pulsars in the array (Siemens et al. 2013). Therefore, the improved sky coverage of NGAT will allow NANOGrav to observe more MSPs that were not visible before with LAT. Also, the improved NGAT sensitivity will provide pulsar surveys to discover more highly stable MSPs that can be included in NANOGrav’s timing campaign in the future (see Appendix C.4). NANOGrav also works together with the International PTA (IPTA -- Perera et al. 2019b) and *NGAT timing data would drive not only the national, but also the international effort of low frequency GW detection over this decade.*

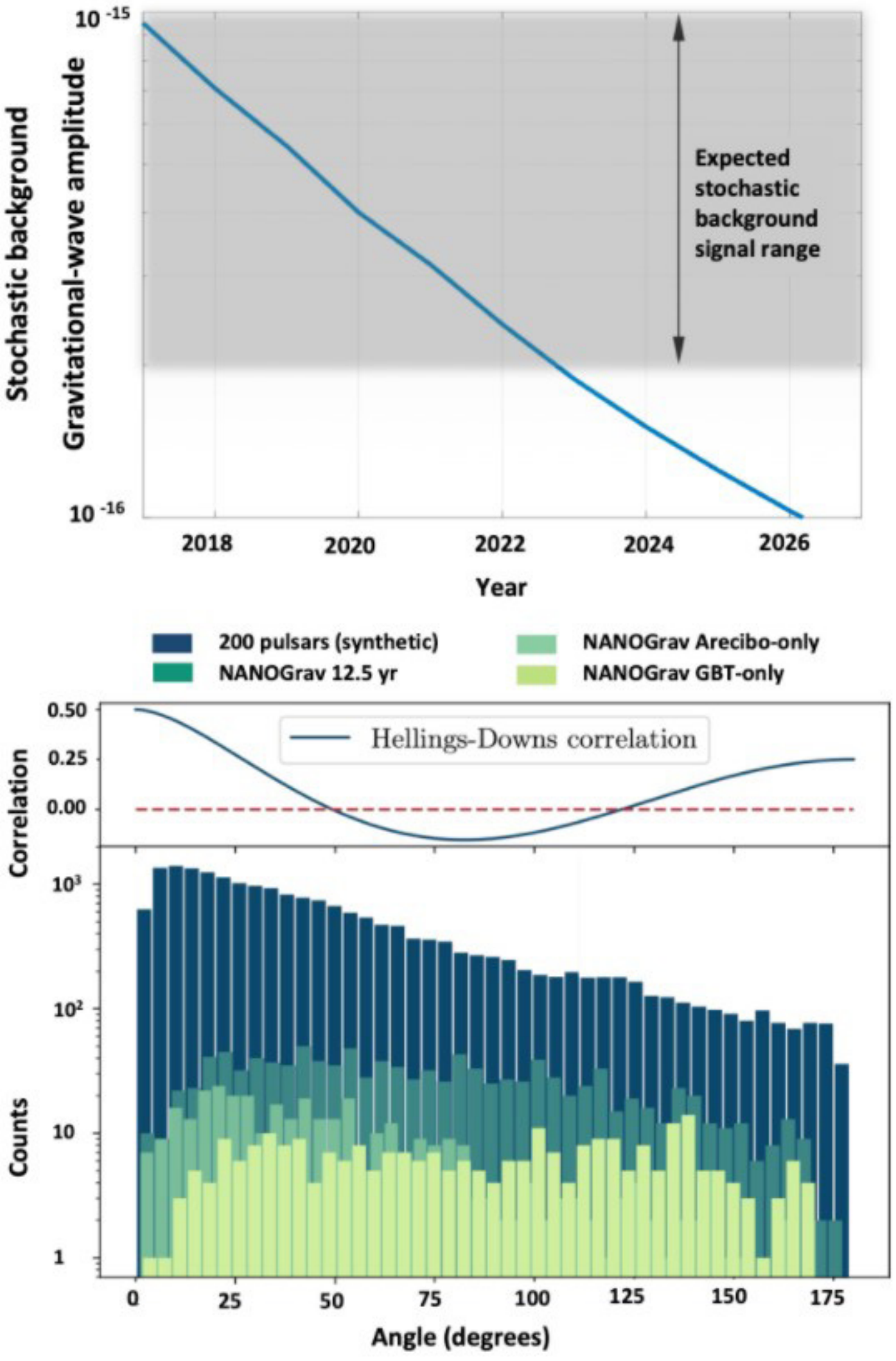

**Figure 8**. Top: NANOGrav’s detection projection for nanohertz GW background. The simulation assumed the existence of LAT and is based on forecasting of pulsar noise properties, adding future surveys, and modeling the properties of SMBHB population. Bottom: The expected “Hellings & Downs” correlation in the GW background detection as sampling with pulsar pairs by the NANOGrav. The number of pulsar pairs per angular separation bin for the 12.5-yr NANOGrav data (GBT only, Arecibo only, both telescopes), and also the sampling for a projected 200-pulsars. A single telescope alone provides a poor sampling, and a 200-pulsar array produces over an order of magnitude improvement in the number of samples per angular separation bin. The figures are taken from Ransom et al. (2019).

## 2.10 Probing innermost regions of AGN using Multi-Messenger Astronomy and VLBI

The co-evolution of galaxies and central supermassive black holes is observationally evident from several scaling relationships: black hole mass-stellar velocity dispersion of the host galaxy bulge, black hole mass-galaxy bolometric luminosity correlation, etc. (Kormendy & Ho 2013). These correlations came about due to mergers of less massive galaxies and the subsequent growth of the black holes. The correlations were observed to be tighter near black hole masses ~$10^6$ $M_{Sun}$ but no clear evidence has been observationally established at higher masses. Furthermore, our understanding of active galactic nuclei (AGN) feedback on the co-evolution is incomplete particularly in systems with SMBHBs. Enabling the NGAT as an element in Very Long Baseline Interferometer (VLBI) networks provides a very powerful tool to detect nuclear activity from massive black holes and SMBHBs, especially in regions of space where emission from other parts of the electromagnetic spectrum is absorbed by dense nuclear dust and gas. This allows us to separate emission from AGN powered by black holes and radio emission from star-formation in the early Universe. Though VLBI observations can be used to infer the presence of SMBHBs from detecting interactions with accreting gas (for example, Doppler broadened spectral lines from AGNs), these signatures could also be mimicked by single AGNs. Detecting continuous GW emission with pulsar timing arrays (see Section 2.9) and upcoming space interferometers such as LISA unambiguously identifies the presence of SMBHBs in their late stages. Thus, combined VLBI observations of such systems to study the bulge of the host galaxy and AGN properties along with GW measurements are the only way to probe the co-evolution of SMBHBs and post-merger systems (Simon & Burke-Spolaor 2016).

The majority of AGN are radio-quiet and have weak radio emission. Studies of products of the events detected via their GW emission typically require milli- and sub-milli-jansky levels of detection. It necessitates use of the most sensitive elements of VLBI network and phase referencing. The latter has made it possible to study very weak radio sources by increasing the effective coherence time from a few minutes to hours. This technique can be successfully carried out for observations at 1 GHz and above. However, at frequencies below 1 GHz the raw coherence times become very short due to ionospheric effects, requiring the phase calibrator to lie within the (voltage) primary beam of all antennas in the array. *The high sensitivity, increased field of view and multi-beam capability of NGAT will provide efficient phase-referencing with two phased beams over 200 MHz to 30 GHz, thus enabling imaging or detection of radio emission from AGNs and SMBHBs.* AO is currently a member of the European VLBI network (EVN) and the high sensitivity array (HSA), and with improved capabilities of NGAT the contributions to VLBI observations will be invaluable.

## 2.11 Prebiotic Molecules: the Precursors to Life in the Universe

High-sensitivity observations at centimeter wavelengths are a powerful tool that can result in unambiguous detections of heavy molecules, the precursors to life in the interstellar medium (ISM). Such molecules, including polycyclic aromatic hydrocarbons, can be practically impossible to identify at high frequencies due to line-forests at mm/sub-mm wavelengths, i.e., heavy molecules generally have lower abundances, and in the millimeter regime their spectral lines fall below the confusion level of the sea of lighter molecules (Heiles et al. 2019). In the 1-15 GHz range, the heavy molecules stand out clearly and unambiguously, and hence this frequency range provides an effective window into the life-bearing astrochemical processes of the ISM. Previous detections of carbon-chain molecules, such as $C_4H$ at 9.5 GHz, toward TMC-1 (Kalenskii et al. 2004), one of the closest molecular cloud complexes to Earth, exemplify the potential for such studies. In addition, numerous extragalactic detections transform our expectations of the scale over which we can hope to investigate the building blocks of life. A ground-breaking example is the detection of multiple new molecules in the Ultra Luminous Infrared Galaxy Arp 220, including pre-biotic $CH_2NH$ (methanimine) and HCOOH (a possible detection; known as formic acid) both of which play a role in the formation of glycine, the simplest amino acid (Salter et al. 2008). *NGAT's vast sensitivity, wider frequency and sky coverage (which provides greater access to the Galactic plane) and its ability to efficiently mitigate RFI (and hence discriminate between prebiotic molecular spectral lines and RFI) will enable the launch of the next generation of Galactic and extragalactic searches for prebiotic molecules.*

## 2.12 Pulsars near Sgr A*: a new test bed for General Relativity

The Galactic center (GC) region has some of the most extreme conditions and environments in the Galaxy due to high stellar and gas densities and the central black hole (Sgr A*). Pulsars in the GC are important to study GR, stellar formation and evolution, stellar dynamics, and the interstellar medium (Bower et al. 2019, Perera et al. 2019a). In particular, discovering a pulsar orbiting around Sgr A* would provide the opportunity to study and test GR and the black hole (BH) physics, measuring BH mass and the spin. Due to the high stellar formation rate and high-mass star density, we may expect to detect thousands of pulsars in the GC region (Pfahl & Loeb 2004, Rajwade et al. 2018), and the massive stars around the BH could be the progenitors for neutron stars and pulsars (Pfuhl et al. 2011). However, this large pulsar population is not accessible with current telescopes due to sensitivity limitations. The extreme propagation effects such as interstellar scattering in the central region make it difficult to discover this population. Thus, observations at frequencies >9 GHz are essential to probe the GC (Rajwade et al. 2018). Many searches have been conducted near the GC, but they discovered only six slow and isolated pulsars within 0.5° of Srg A* (Johnston et al. 2006; Deneva et al. 2009; Eatough et al. 2013). The GC magnetar (located only 0.1 pc in projection from Sgr A*) is the most interesting among these sources; however, its orbit around the central BH is not sufficiently short for relativistic effects to be observed.

The main limitation of earlier searches is the telescope sensitivity. The FAST radio telescope has a large collecting area, but its declination coverage is limited and current operating frequency is not favorable for GC searches. *NGAT, with its high sensitivity, sky and frequency coverage* will *enable us to probe the GC and discover more pulsars in the region, including a pulsar in a tight orbit around the central BH, allowing us to carry out unprecedented tests of fundamental physics and gravity theories including GR, and also discover new classes of radio sources through transient events.*

## 2.13 Elucidating the mysterious nature of Fast Radio Bursts and other energetic events in the Universe

The FoV of the telescope potentially helps the rate of the Fast Radio Bursts (FRBs) detection (e.g. the Canadian CHIME Telescope). NGAT will have a larger FoV, sky coverage, and higher sensitivity, placing it in a better configuration to detect many more FRBs than was possible in the past, including particularly distant and faint FRBs. FRB160102 has the largest measured dispersion measure (DM; 2583 pc $cm^{-3}$) to date, leading to a redshift of 3.06 (Bhandari et al. 2018). This particular source was detected with the Parkes telescope with a S/N of 16. Using a high sensitivity telescope, e.g. FAST, it is shown that the same FRB can be detected up to a redshift of 10.4 (with a S/N threshold of 10), leading to a DM of ~ 6500 pc $cm^{-3}$ (Zhang 2018). It is also shown that sensitive telescopes can search FRBs up to a redshift of ~15, setting up a DM upper limit of ~9000 pc $cm^{-3}$ (Zhang 2018). The high redshift FRBs provide unique opportunities to probe deep in the intergalactic medium which is crucial for constraining the epochs of hydrogen and helium reion-ization incited by the ignition of the earliest stars and galaxies (Keating et al. 2015, Beniamini et al. 2020). FRB detection and localization is one of the highest ranked sciences in the current era. Large collecting area is very important to detect low fluence short bursts. The highest sensitivity of LAT together with its dedicated pulsar backend made possible the detection of three out of four VLBI detected pulses and all of them have low fluences compared to the known FRB population[7] (Petroff et al. 2016), indicating that sensitive telescopes are important in detecting faint FRBs. The improved sensitivity with the NGAT will enable us to detect more faint FRBs. NGAT as a station in the VLBI network is indispensable in localizing repeating FRBs and probing spectro-temporal structure in any repeating bursts. A real-time FRB detection via a burst triggering system will be used at NGAT for commensal and other observations while recording and processing the data on petabyte storages and GPU clusters. Such observations involving NGAT have potential to reach the highest baseline sensitivity which is critical for decisive astrometric observations of FRBs (Marcote et al. 2020). With these improved capabilities, *NGAT* will *detect more FRBs compared to the legacy Arecibo, including very faint ones accompanied by low fluences and/or located further than ever before, leading to key insights into the physics behind these mysterious events, the nature of the intergalactic medium, and the history of the universe.*

NGAT will also offer significant new capabilities for the exploration of Gamma-ray Bursts (GRBs), the most energetic electromagnetic events in the universe at cosmological distances whose origin is still unclear (Chandra et al. 2016). Our understanding of GRBs has drastically improved with the detection of their afterglows at X-ray, optical, and radio wavelengths (Kann et al. 2010). Only 31% of all GRBs have been detected in radio bands due to the sensitivity limit of the current telescopes (Chandra & Frail 2012). The high sensitivity of LAT significantly contributed to the detection and investigation of the afterglows of GRBs (e.g., Taylor et al. 2004, 2005, Pihlstrom et al. 2007). It has been estimated that sensitive telescopes such as FAST can detect GRBs up to a redshift of ~15; however, its frequency coverage (< 3 GHz) does not have enough sensitivity to detect sub-luminous GRBs. With the enormous sensitivity throughout its broad frequency coverage (from 200 MHz to 30 GHz - see Fig. 2), NGAT will enable detection of all GRBs including sub-luminous events, allowing investigations into their origin and physics far better than before. Combining NGAT's large collecting area with the VLBI network will be particularly useful to precisely locate and monitor the GRB afterglow. Recent VLBI studies, for example, provided decisive physical diagnostics on the products of first detection of gravitational waves from neutron stars collision (Mooley et al. 2018, Ghirlanda et al. 2019).

## 2.14 Quantifying the local Dark Matter content with Near-Field HI 21 cm Line Cosmology

According to the ΛCDM paradigm, the local Universe should be teeming with dark matter minihalos with masses < $10^9$ $M_\odot$. While the ΛCDM model is extremely successful in describing the overall evolution of cosmic structure, observational results fail to detect its predicted large numbers of "satellites." While the number of known satellites in the Local Group has increased dramatically in recent years, the number with small circular velocities remains too low. Interactions between satellites and the hot coronae of large galaxies can explain the dearth of stars and gas in small halos, but explaining the kinematic mismatch between the smallest galaxies and their host halos is harder. In fact, the occupation of low mass (< $10^9$ $M_\odot$) halos by visible galaxies is poorly constrained, and the observed luminosity and velocity functions fall well below the predictions of

[7] http://frbcat.org

ΛCDM. Converging the paradigm with our local Universe requires a complete multiwavelength census of low mass galaxy populations and their general characteristics. A cornerstone of these studies rests on observing the HI content of the low mass populations, as it provides not only information on the HI mass but also dynamical information including recessional velocities and velocity widths. *The large field of view, sky coverage, and sensitivity of NGAT are essential in order to detect faint HI 21cm line emission from a large sample of galaxies with low halo mass.* (see also Appendix C5.5)

## 2.15 Probing Dark Energy with HI Intensity Mapping

The question of the nature of dark energy remains one of the most perplexing cosmological mysteries of our time. Modern radio telescopes have a unique opportunity to participate in the steps to uncover this mystery in the coming decades, particularly because neutral hydrogen both provides an excellent tracer of baryonic matter, and is, in principle, easily observable. Large-scale intensity mapping projects help determine not only the distribution of galaxies (and hence the distribution of dark matter) at various redshifts, but also the rate at which cosmic structures grow, providing unique constraints on dark energy. The onset of cosmic acceleration is a particularly important region to probe. However, observing the z=1 region is also prohibitively difficult at shorter wavelengths due to the opacity of Earth's atmosphere. Therefore, observing neutral hydrogen with highly sensitive radio telescopes offers an opportunity to constrain cosmological models in a very useful way.

Known as HI intensity mapping (Chang et al. 2010, Masui et al. 2013), measuring the power spectrum of intensity fluctuations in the 21 cm line can yield useful observations on the distribution of matter as well as the size of baryonic acoustic oscillations at low to moderate redshifts. In addition, by mapping the same line redshifted to different wavelengths, this technique can effectively reveal the distribution of matter in 3-dimensions and hence over large co-moving volumes. Resolving individual galaxies is not required for such large-scale measurements, and hence high sensitivity radio telescopes are uniquely poised to contribute to these observations. *NGAT's sensitivity, frequency coverage, and correlation mode of observation will provide unique capabilities that enable it to serve as a prime instrument in such studies.*

## 2.16 Searching for Advanced Life in the Universe

We now know that our Galaxy, and indeed the Universe, is awash with oases for life. As such, one of the most profound questions of this century is the search for life beyond Earth. The Search for Extraterrestrial Intelligence (SETI), more recently known as the search for 'technosignatures' is a viable method for uncovering evidence of advanced (technological) life, particularly because it probes many simultaneous targets over large volumes of space (Margot et al. 2019, Grimaldi and Marcy 2018). Radio emission at centimeter wavelengths is an especially promising tracer of advanced life, both because radio waves easily permeate the interstellar and intergalactic media and because they are easily generated, as demonstrated by our own use of radio waves on Earth (Grimaldi and Marcy 2018). As the Terrestrial Microwave Window, the portion of the radio spectrum from 1 to 10 GHz, is largely free of naturally generated cosmic and terrestrial noise sources, it is therefore ideal for technosignature searches (see, e.g., Fig. 5.2 in the Cyclops Report by Oliver et al. 1971), and this range will be fully covered by NGAT.

The relative probability of successful technosignature detections compared to other search methods depends on both the likelihood that life evolves technological capacities and the length of time over which it produces technosignatures (Grimaldi and Marcy 2018). Maximizing sky coverage, sensitivity, frequency coverage, and spectral resolution are the most effective means to increase the probability of a detection. Over the coming years, the enormous FAST telescope in China will be one of the premier instruments for technosignature searches on the planet due to its large collecting area. NGAT will have a field of view 58 times larger than the current 19 element L-band system (centered at 1.2 GHz) on the FAST telescope. *Thus NGAT (centered at 1.5 GHz) will have a Drake figure of merit ~37 times that of FAST assuming similar processing bandwidths for both telescopes.*

The detection of technological life beyond the Earth would represent the most profound discovery in the history of humankind. The significantly enhanced performance of NGAT over most available instruments in the current and next generation of technology will also constrain our estimates of the prevalence of advanced life in the Universe, an important result that reinforces the rarity of our own species. Even in the absence of a technosignature detection, null results serve as quantitative and specific evidence of the need to carefully care for our home planet and boldly plan for extending our own reach beyond the bounds of Earth.

## 2.17 Molecular Gas in the distant Universe

The star formation history of the universe is well summarized by the so-called ``Lilly-Madau" plot. This plot (not shown here), which illustrates the total star formation rate (SFR) in galaxies in a comoving volume as a function of redshift, essentially describes the evolution of star formation in galaxies across cosmic time. The SFR density increases from an early epoch ($z > 8$), peaks near $z\sim2$ and then declines by a factor of ~20 by $z=0$ (for a recent review, see Madau & Dickinson 2014). Understanding the physical process driving this evolution is fundamental to understanding galaxy evolution. At least three key quantities drive this evolution: the growth rate of dark matter halos, the gas content of galaxies (specifically the molecular gas available for star formation), and the efficiency at which molecular gas is transformed into stars.

The low energy rotational transitions of CO are good, direct tracers of the bulk of the molecular gas in galaxies which fuel the star formation (Carrilli & Walter 2013). However, detecting these lines over the cosmic history is most demanding in terms of telescope time. A fruitful strategy is to search for CO lines from the more massive galaxies ($> 10^{10}$ $M_\Theta$) such as the sub-millimeter wave galaxies (SMGs). SMGs typically have molecular gas mass ~$5\times10^{10}$ $M_\Theta$ and their SFRs are 50-1000 $M_\Theta$ $yr^{-1}$. They contribute to ~20% of the total SFR density, at an earlier epoch of the Universe, ~10 Gyr ago. They reside in massive dark matter halos ($M_{DM}\sim5\times10^{13}$ $M_\Theta$) which ultimately evolve into virialized galaxy clusters with a hot IGM (Casey et al. 2014). Their high CO luminosities and high number densities in proto-galaxy clusters make them ideal tracers of massive dark matter halos over a wide swath of cosmic time.

As an example, consider the proto-cluster SPT2394-56 ($z=4.31$), a bright ($S_{1.4mm}$=23.3 mJy) unlensed object discovered in the South pole telescope 2500 sq. deg. survey (Vieira et al. 2010). There are 14 gas-rich galaxies in this proto-cluster, and the CO(2-1) transition was detected with the ATCA (see Fig. 9) providing a molecular gas mass of ~$10^{11}$ $M_\Theta$. The 14 line-emitting objects are located in a region of ~130 kpc in size (Miller et al. 2018).

The potential impact of NGAT to survey the redshifted CO transitions from distant ($z > 3$) proto-clusters will be enormous. The interferometric mode of operation of NGAT will provide excellent spectral baselines, which is crucial for the detection of such weak lines. Further, the large field of view of NGAT will allow simultaneous searches for line emission from almost all members of proto-clusters. Such a system will allow astronomers to measure the gas content of these proto-cluster

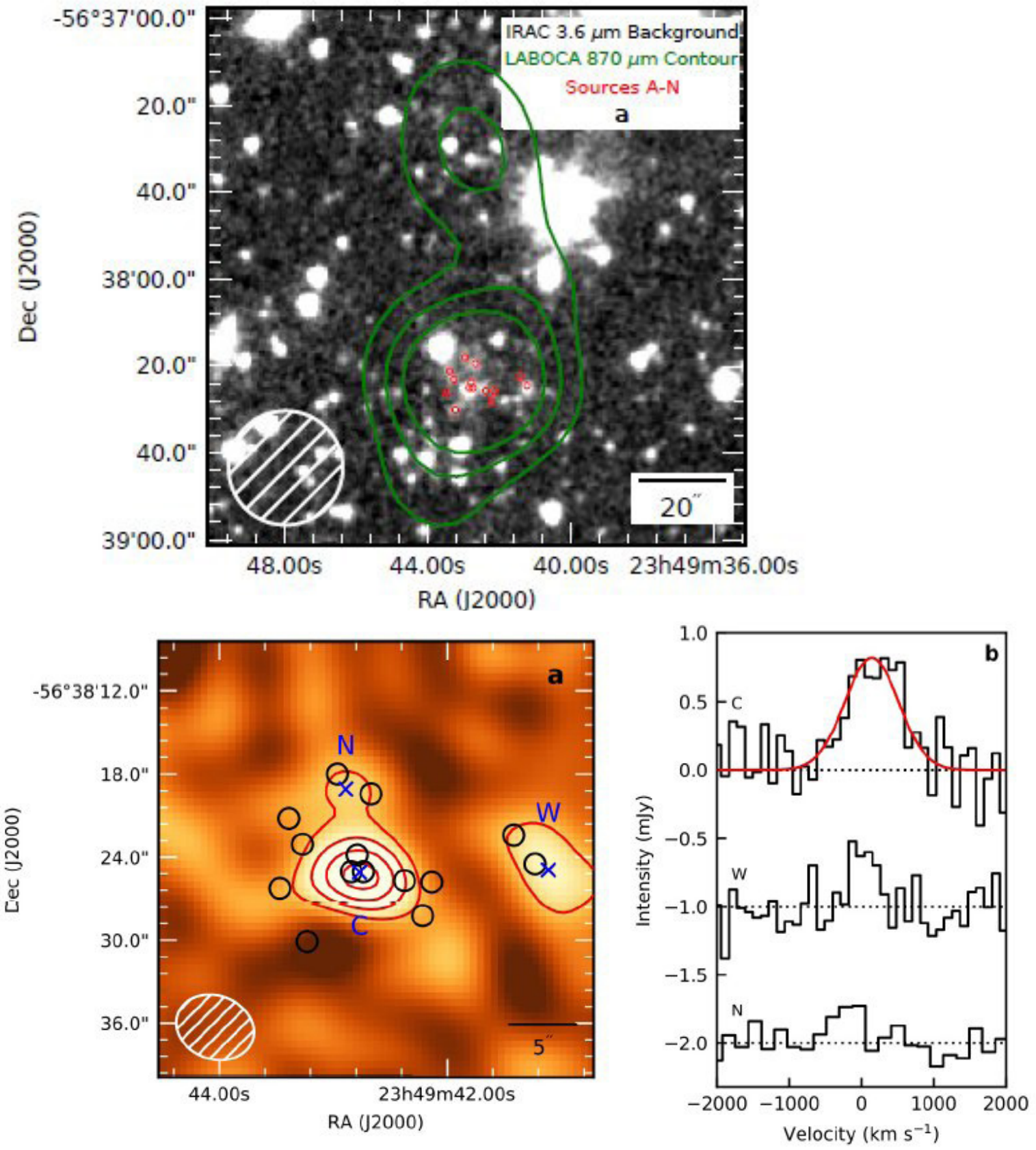

**Figure 9.** (Top): APEX 850μm image of SPT2349-56 showing the extended emission associated with the proto-cluster. (Bottom): CO(2-1) detection toward the central core of the proto-cluster by ATCA Miller et al. 2018).

members providing crucial information for the advancement of this field.

## 2.18 High Mass Star Formation in the Galaxy

High-mass stars, or O and B stars ($M > 8\ M_{\odot}$), play a major role in the energy budget of the Galaxy through their radiation, stellar wind, and eventual death in supernovae explosions. Their formation, however, is not as well understood, although it is generally known that the collapse of dense, sub-parsec sized cores in the molecular interstellar medium (ISM) produces both low and high mass stars. These dense cores as well as almost all young stars are located in large scale (extending over several pc) filamentary structures observed in the molecular phase of the galactic ISM. Although a large body of observations of structures in molecular gas exists there are several open questions (for a complete review, see Motte 2018), including: (1) How is the formation of massive stars related to the large scale structures such as the filaments? (2) What is the dominant physical process that results in the formation of filamentary structures? (3) How are dense cores formed from filaments? Are cores formed from mass accretion or instabilities in filaments? How are the high-mass star formation activities distributed over the disk of the Galaxy?

While there are many ways to study these processes in the molecular ISM, one of the most powerful means involves using spectral line observations of various transitions of ammonia, $NH_3$, near 23 GHz. These transitions require high densities ($> 10^3$ cm$^{-3}$) for excitation and have moderate optical depth. Thus, $NH_3$ observations can filter out the foreground and background contamination as they preferentially trace dense regions in molecular gas (see, e.g., Fig. 10). The line ratios also serve as a thermometer, yielding both velocity, density, and temperature information for the gas clouds. These data are crucial to understand and differentiate between the theories of high mass star formation (Motte 2018).

Surveys of $NH_3$ line emission from Galactic molecular clouds with modest angular resolution (~1 arcmin) exist but are limited to ~10 sq. deg. due to survey speed limitations (Hogge et al. 2018, Friesen et al. 2017). As high-mass stars comprise less than 1% of all stars, observing a statistically significant sample of high-mass star forming region requires probing a large number of cloud complexes. The large survey speed of NGAT and its excellent surface brightness sensitivity will make it a powerful instrument both to conduct $NH_3$ surveys over several 100 sq. deg. and to probe the faint, diffuse emission surrounding the very dense regions. These observations complement the wealth of information obtained by ongoing ALMA observations at sub-parsec scale resolution by providing information on scales $\gtrsim$ pc. In this respect NGAT will greatly transform our understanding of star formation in the Galaxy.

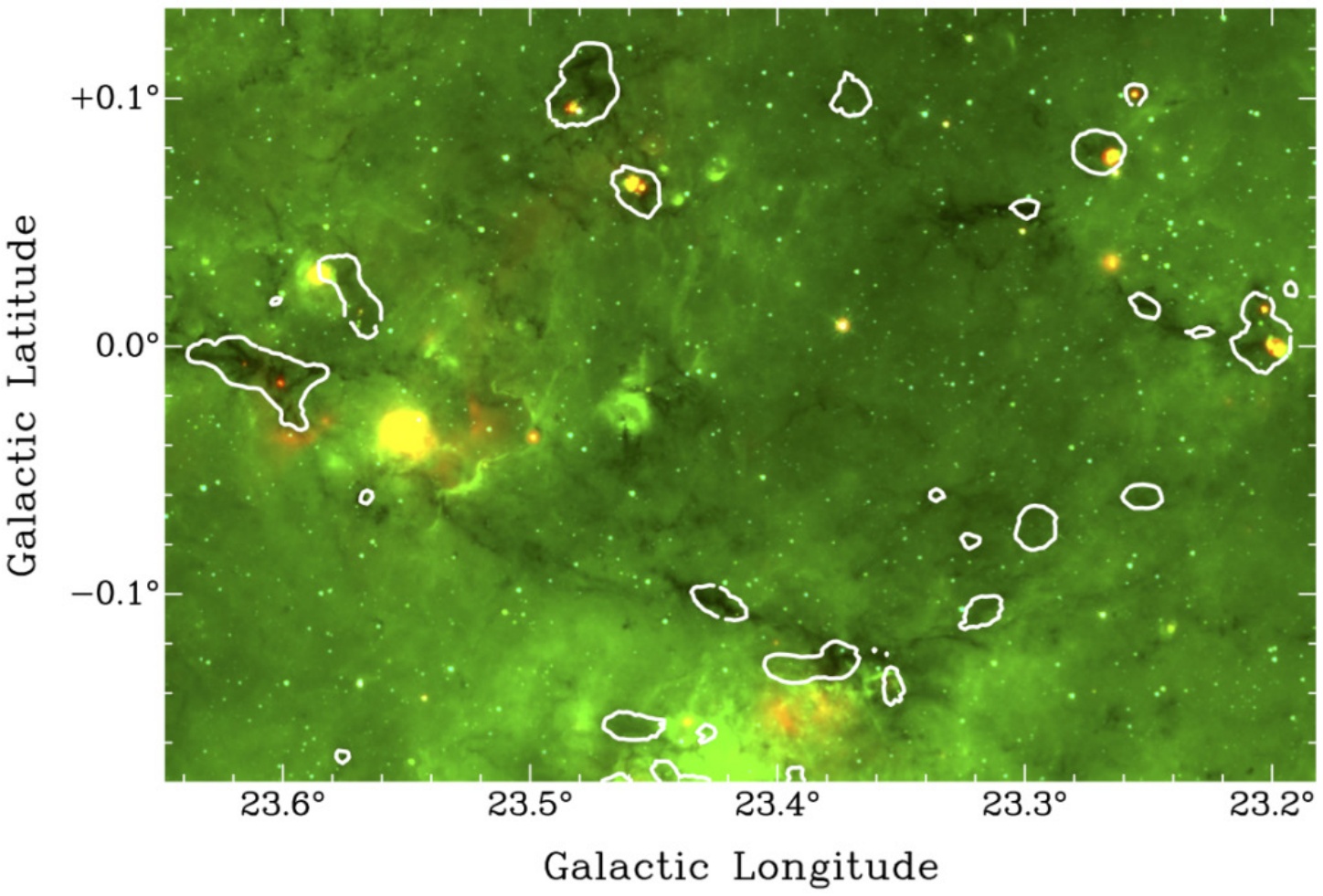

**Figure 10**. Composite image of the 3.6 μm (blue), 8 μm (green) and 24 μm emission from a star forming region from GLIMPSE and MIPSGAL superposed with RAMPS $NH_3$(1,1) integrated intensity contours in white. The $NH_3$ emission preferentially originates from dense regions in the cloud complex. (Image from RAMPS project website, Hogge et al. 2018).

[8] http://sites.bu.edu/ramps/project-summary/

# 3.0 NGAT Design Concept

## 3.1 Compact dish array on a single plane

The science requirements for the three research groups (Planetary Radar, Space and Atmospheric Science, and Radio Astronomy) at AO can be summarized as follows:

a) sky coverage 0 to 48 deg in zenith angle;
b) large collecting area;
c) excellent surface brightness sensitivity; and
d) several megawatts of transmitting power.

We considered the merits of meeting these requirements with both a single dish or a compact array with large collecting area. However, as was the case for the LAT, preliminary investigations showed that a 305-m single dish would not be able to meet the zenith angle range required. Further, there is no proven technology that ensures the possibility of transmitting several megawatts of transmitter power with a single dish. If we were to use an array of individually steerable dishes, the desired high surface brightness sensitivity would require a 'tight' packing of the dishes. Such a packing would create significant shadowing which would effectively render such an array not useful for observations. Brief descriptions of the single dish and the array of dishes considered are given in Appendix A.

**The limitations of the single dish and of individually pointable dishes led us to the concept of the 'compact dish array on a single plane'.** A schematic of this idea is shown in Fig. 11. An array of dishes are packed so as to be on the same plane, either on one very large plate-like platform or on smaller plates each with its own platform. Sky coverage is achieved by tilting the packed array so that it can point to different zenith angles.

It would be ideal if the entire array of dishes could be placed on one single movable structure that could be tilted; however, if this is not mechanically feasible, the array support structure could be divided into smaller, individually movable segments. Compact arrays mounted in a single plane are in existence, in much smaller sizes (see Fig. 12).

Detailed engineering and design studies will be undertaken. We discuss below the major design considerations.

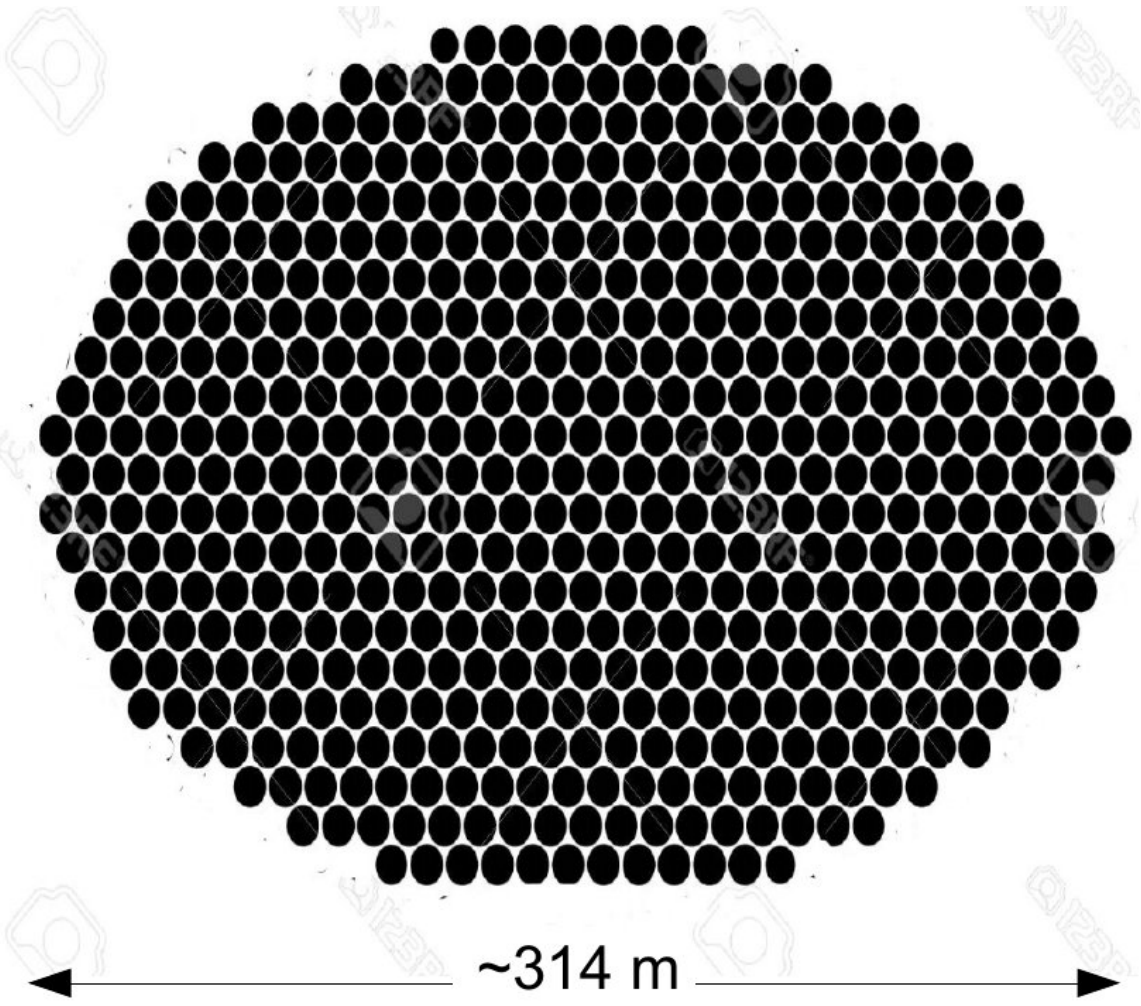

Fig. 11a

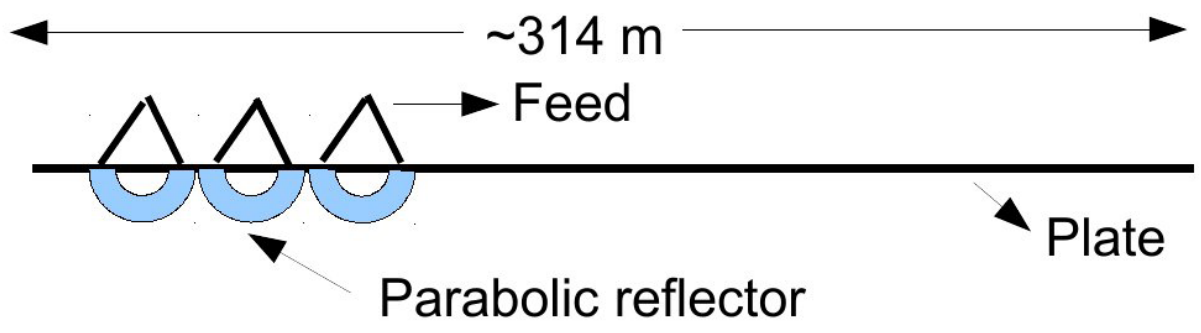

Fig. 11b

**Figure 11:** Schematic showing the concept of the NGAT. In the example configuration discussed in Table 2a, there are 1,112 parabolic reflectors each of size 9 m. These dishes are to be fixed on a structure in a single plane within a circle of diameter 314 m. The view of such an arrangement from the zenith is shown in Fig. 11a and the side view is shown in Fig. 11b. Each parabolic reflector is fed with prime focus receivers (Fig. 11b). In this diagram, we assume that the parabolic reflectors are symmetric, but off-axis reflectors are another possibility.

**3.1.1 Dish size, shape and packing:** Maximally packing the dishes on the plate/segment is important so that the largest possible total collecting area is realized. The shape of the individual dish could also be hexagonal if it increases the net collecting area compared to a circular dish. Individual dish parameters also need to be optimized in order to maximize the sensitivity over 200 MHz to 30 GHz, minimize grating lobe amplitude, reduce cost and provide maintenance access to the feed and front-end electronics. Two example configurations are shown in Fig. 13 and a summary of the configuration and capabilities provided by them is given in Table 2. The 9 m size provides only 6 wavelengths across the dish when operating at 200 MHz, which certainly is not an optimal solution for the lowest frequency but could be a cost effective way to make reflectors with good surface RMS values for 30 GHz operation. Note that the VLA 25 m antennas were used for imaging at 75 MHz.

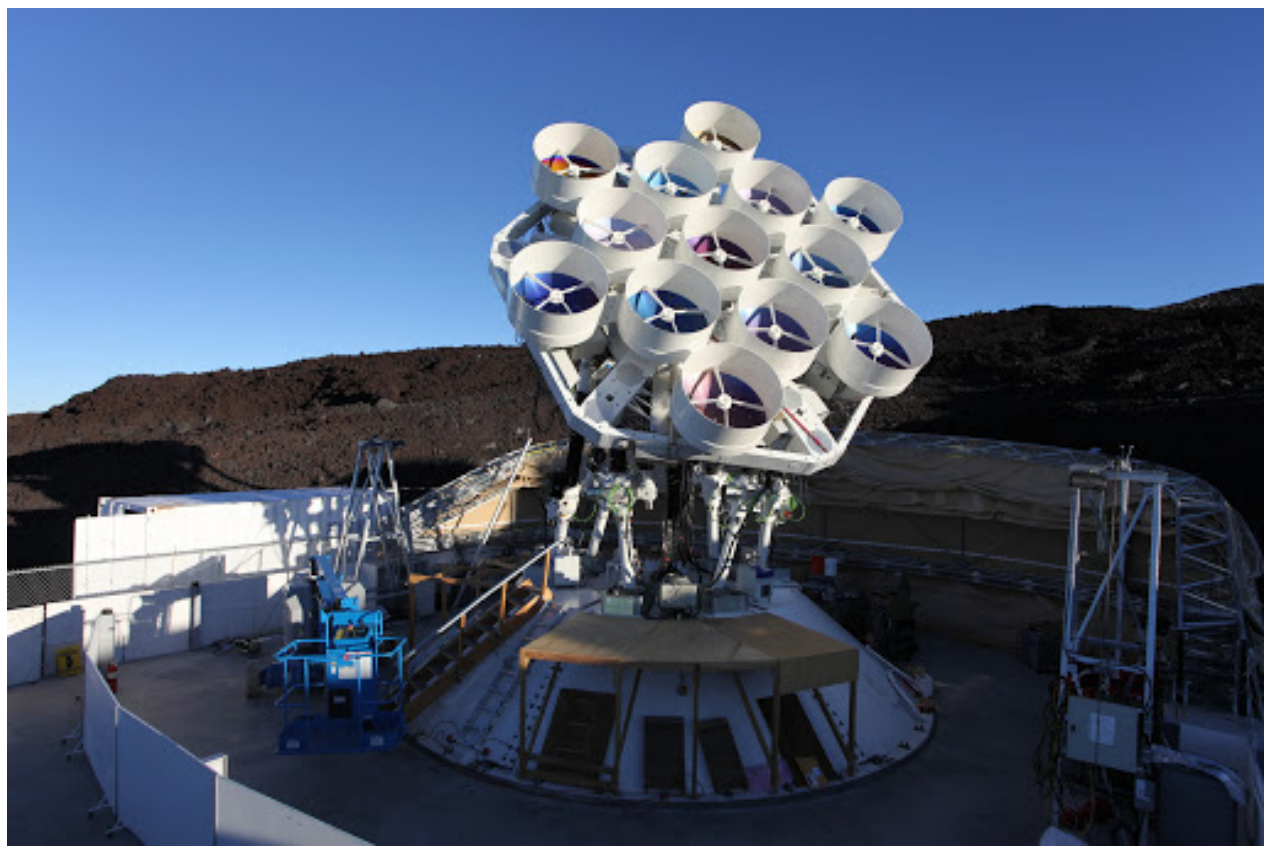

Fig. 12a

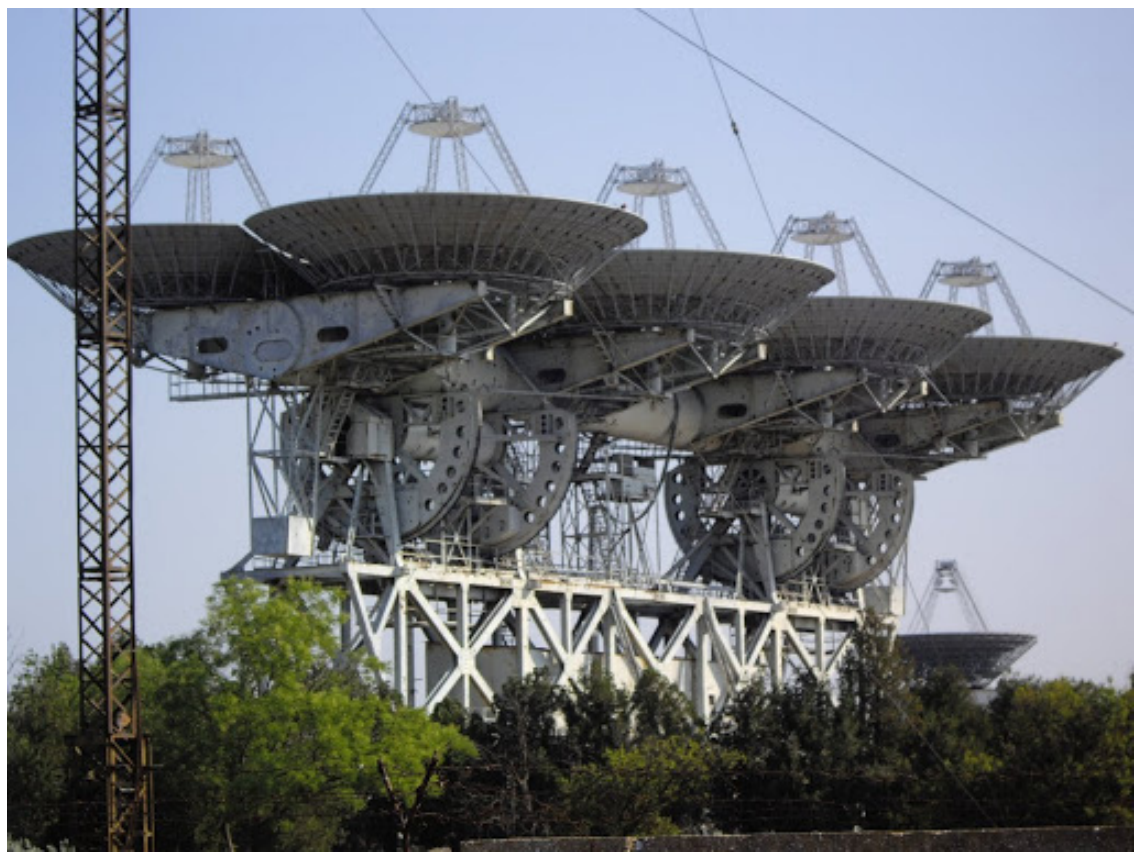

Fig. 12b

**Figure 12:** A much smaller version of such a telescope, the AMiBA[9], is being used for radio astronomy observations, and is shown in Fig 12a. A similar configuration was used by Pluton[10] deep space communications and planetary radar in Crimea, which consists of 8 reflectors of 16 m diameter (Fig. 12b).

[9] https://en.wikipedia.org/wiki/AMiBA

[10] https://en.wikipedia.org/wiki/Pluton_(complex)

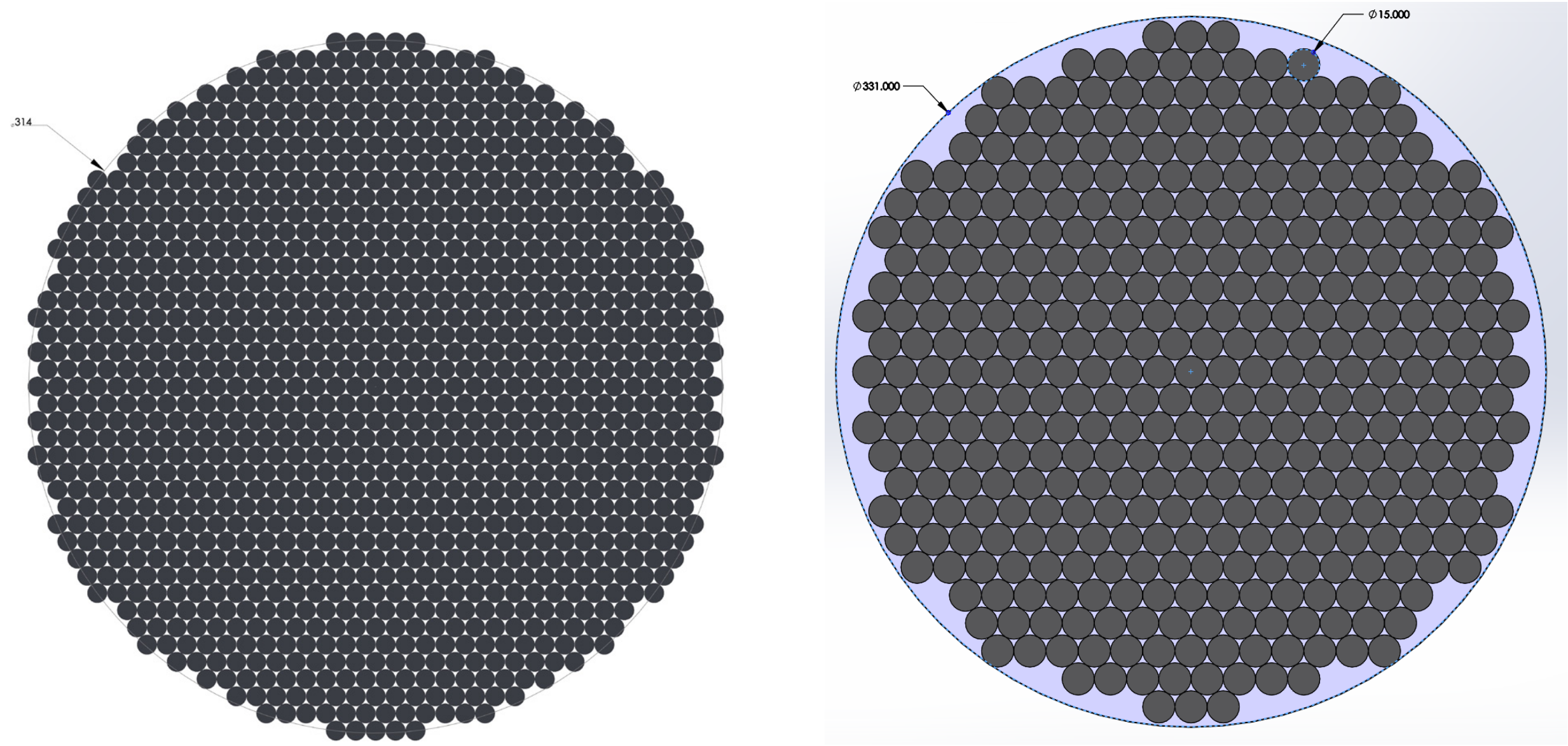

**Figure 13:** An array of 1,112 dishes of 9 m in diameter within a ~314 m diameter circle is shown on the left. The number of dishes could be reduced by increasing the dish diameter, as shown in the right figure where an array of 400 dishes of 15 m diameter within a 331 m diameter circle. In both cases, the array will provide an equivalent collecting area of a 300 m diameter dish. It is evident that the packing efficiency is not optimum when the diameter of the dish is increased. Dishes with hexagonal shape could be considered during the engineering study to investigate whether higher collecting area can be achieved with less number of dishes.

**3.1.2 Dish optics:** The reflectors could be symmetric or off-axis parabolic. The optics could be simplified with prime focus operation. With a prime focus configuration, the focal-length over diameter (F/D) ratio could be varied to achieve optimal performance in the 200 MHz to 30 GHz frequency range. A combination of primary and secondary reflectors could also be considered if costing is viable.

**3.1.3 'Plate' configuration and pointing:** The plane of the entire array needs to be moved to point at different zenith angles. The entire array could be placed on either one single large movable platform or be segmented and placed in smaller structures with each one movable for pointing individually, if deemed structurally necessary. Inevitably, segmenting the plate would result in loss of collecting area at large zenith angles so minimal segmenting would be ideal. To reduce effects due to shadowing at zenith angles away from zero, the segments should be as large as possible and they must be placed as near as possible. Segment shape could be a polygon, if it helps closely pack the array. Each segment could be moved using a set of jack mechanisms for pointing at different sky positions. **Another benefit of this segmented design is that we could begin with a single segment, and build up capability with time.**

**3.1.4 Radio receivers:** We envisage that 200 MHz to 30 GHz frequency range could be cov-ered using 4 broadband (~1:4 frequency ratio) feeds for reception. The feeds will be optimized to provide the maximum gain over system temperature in the frequency ranges 200 MHz to 800 MHz, 0.7 GHz to 2.5 GHz, 2.0 to 8 GHz and 8 to 30 GHz. To reduce the number of cryostats and required maintenance, room temperature low-noise amplifiers (LNAs) with Peltier cooling can be used for frequencies below 2.5 GHz (Schleeh et al. 2016; S. Weinreb, Caltech, private communication). Arecibo is located in an RFI coordination zone. Room temperature LNAs are expected to have enough dynamic range to accommodate the power in the RFI, but the bandwidth of the successive receiver stages need to be reduced to filter out strong RFI signals. The bandwidth of each receiver is to be determined based on various considerations such as RFI power, achievable noise performance, etc. The 2.0 to 8 GHz and 8.0 to 30 GHz receivers need to be cryogenically cooled. The 2.0-8.0 GHz receiver will be used by planetary radar for signal reception. This radar receiver needs to be cooled to maximize the performance of the planetary radar facility. Optimization of the power consumption of the cryostat can be achieved by reducing the cooling load (for example, by not cooling the horn), using high-efficiency cryo-coolers[11] and also exploring the operation of LNAs at 60 to 80 K temperature. This optimization is essential to keep the operating cost of the telescope low.

**3.1.5 Radio Transmitters:** A key objective of the new telescope is to rebuild the world's most powerful and sensitive planetary and atmospheric radar systems. The new radar systems could take advantage of new advances in microwave and radio transmitter technology, using a large number of low-power radar transmitters rather than one high-power transmitter. The great disadvantage of high-power klystrons has been their high cost and long manufacturing time. In addition, a single transmitter will face more serious power density issues such as arcing. When using cumulative power from several transmitter units with relatively less power, arcing can be avoided. Use of a large number of low-power (5 - 25 kW), high efficiency transmitters reduces operational risks and cost, although the investment in the beginning is greater than using 1-2 high-power transmitters.

We plan to divide the transmitters into two groups. The 430 MHz ISR will use a 200 MHz to 800 MHz feed and a broadband (200 to 500 MHz) power amplifier. We envisage that each dish could be equipped with a 10 to 25 kW amplifier (depending on the telescope configuration; see Table 2), thus providing a peak power from the array of ~10 MW. The second transmitter, to be used for planetary defense purposes, will consist of a 2 to 8 GHz feed, and a narrow-band (centered between 2 and 6 GHz; yet to be decided), high efficiency power amplifier. We envisage the amplifiers could provide a total power from the array of ~5 MW in continuous wave operation using a set of 5 - 12 kW power amplifiers depending on the telescope configuration.

The planned transmitter at 430/220 MHz, which is primarily for atmospheric studies, can also be used for planetary exploration. For the planetary radar, frequency range 2 to 6 GHz is considered optimal in terms of weather factors and beam width. The specific frequency could be determined near the time of construction, as WiFi systems also use frequencies in this range. Specific bands will possibly be assigned for new systems, and the selected frequency band should not overlap with those that could cause significant RFI. The ideal solution would be a dual-band or a wide-band transmitter, if one were available at the time of construction at a reasonable cost, as it would allow both flexibility and easy access to multiwavelength data. The design summary table lists options for optimal powers per frequency; however, only one will likely be selected if high efficiency dual-band or wide-band transmitters are not available. A highly efficient transmitter is essential to keep the operational cost low.

[11] https://www.northropgrumman.com/wp-content/uploads/HighEfficiencyCryocoolerPerformanceICC18.pdf

Greater total power requires more power generation for the facility, which has been restricted by the pollution levels of existing generators. More environmentally-friendly power generation for the new facility with newly emerging technologies should be investigated.

**3.1.6 Frontend instrumentation and configuration:** The front-end structure depends on the optics of the dish. Broadly speaking, each dish will have a turret to house the 4 receiving and the 2 transmitting feeds. The receiver and transmitter for the higher frequencies will be located nearby in the turret so that they could be switched between transmit/receive applications (e.g., planetary defense application). The high-power amplifiers for the transmitter need to be located close to the dish. They could be placed below the dish, attached to the support structure of the telescope. Wave guides equipped with dehydrators can carry the output of the power amplifier to the feed. The approximate estimate of the length of the wave guides is about 10 m. For the 430/220 MHz transmitter, we will use high-power heliax cable to connect the feed to the power amplifier. The power amplifiers require cooling. Liquid cooling is used in many commercial amplifiers, but to reduce the weight on the structure, we explore the possibility of using air cooling.

**3.1.7 Signal conditioning and processing:** The signals will be digitized close to the frontend for the receivers using 14-bit Analog to Digital Converters. The digitized signals must be passed through fiber optics to the backend room after minimum signal processing. All further processing (delay compensation, beamforming, correlation) will be done in the backend room, to be located in RFI shielded rooms near the telescope control room. Telemetry signals to control the turret, receiver monitoring, etc. will also be transported through the fiber optic link. The waveform of the modulated signals to be transmitted will also be sent digitally to the telescope. We envisage that all the amplifiers for the transmitter could be located close to the frontend. A separate network of fiber optic cables will carry the telemetry signal for telescope pointing.

**3.1.8 Grating Lobe Suppression:** The spatial frequencies are sampled at half the Nyquist rate in the array configuration, which implies the far-field pattern will have grating lobes. A variety of techniques exist to suppress grating lobes/sidelobes inherent to a close-packed multi-dish design, such as: array thinning, tapering, non-uniform dish placement, dish size variations, and annealing (Steinberg 1983). Tradeoffs of the array sensitivity (including aperture efficiency) versus sidelobes will be assessed during the engineering study. Our preliminary simulation study[12] shows that non-uniform dish size or keeping the edge taper of dish illumination < 10 dB are promising ways to suppress grating lobes without reducing the net collecting area.

**3.1.9 Phasing the array:** The phasing of the array in receiving mode can be done using the techniques used in interferometers by observing radio sources. However, phasing the transmitter requires additional calibration techniques. Methods to phase arrays with a large number of transmitters have been investigated in the context of solar power satellite antennas in the microwave regime (Chernoff 1979). In recent years, significant progress has been made in phasing individually steerable dish arrays for planetary radar applications near 7 GHz (Geldzahler et al. 2017, D'Addario et al. 2009). We plan to build upon these techniques to phase the proposed transmitter array.

[12] http://www.naic.edu/~aroshi/misc/GratingLobeSuppression.pdf

**3.1.10 Structure and structural deformation:** The telescope structure has to support the weight of the reflector and the electronic instrumentation. This weight depends on the number of reflectors required. We approximately estimate the weight of the reflector and electronic components by considering carbon composite reflectors (Lacy, G., NRC, Canada, private communication), SKA estimates on the front-end electronics weight (Swart & Dewdney 2020), and commercial brass transmitter waveguide and power amplifiers to be 1800 or 1300 tons for dish sizes in the range 9 to 15 m, respectively. Thus, if the entire array were on a single support structure it would weigh about 1800 tons. If the array is segmented, then the weight of the reflector and equipment to be supported will reduce accordingly.

The structure will deform during tracking due to gravity, thermal expansion, and other perturbations (wind, etc.). Assuming a 9 m antenna size, we estimate a tolerance for the deformation at 30 GHz to be ~0.8 cm for an array size of ~314 m. The tolerance requirement will be lower if the array is segmented as it is proportional to the size of the array. The large tolerance is due to the larger primary beam of each individual dish and hence depends on the assumed dish size. For phasing the array, however, we need to know the deformation length in the direction of the source to a fraction of wavelength (i.e. ~1 mm at 30 GHz). A finite element model (FEM) of the structure would accurately provide this pathlength as demonstrated by the operation of the Green Bank telescope (GBT). The panels of the GBT are adjusted using FEM model predictions to achieve a high aperture efficiency up to ~28 GHz. Whether we could use the FEM model prediction to aid array phasing depends on the amplitude of the fluctuations of the structure deformation due to external perturbations (like wind) as well as how accurately the FEM can predict the effects due to a temperature gradient. Feedback from a laser metrology system will be required to phase up the array if the latter two effects dominate the structure deformation.

**3.1.11 Locating the telescope in the sinkhole:** The sinkhole in which the LAT was located is surrounded by karst hills which provide good RFI shielding. We would like to take advantage of the RFI shielding property of the sinkhole. However, the suitability of the site for this telescope needs to be assessed. If the sinkhole is not found to be suitable, it is certainly our preference (as described in the Executive Summary) that another, suitable location is identified within the Arecibo Observatory site.

**3.1.12 Other challenges to overcome:**

- Understanding the coupling of the antenna elements and its effect on the beam shape and array performance.
- The telescope must be secured against fire hazard, which is expected for high power transmission arrays.
- Feed, receiver and transmitter support structure must be designed in such a manner so as to enable regular maintenance. For such large arrays an electrical maintenance cart that can transport one or two people and equipment will be required.
- Environmentally-friendly power generation methods to support the radar transmission should be explored.

## 3.2 New Capabilities

- Increased sky coverage and better sensitivity will allow observations of more targets, including the galactic center, pulsars, other stars, and fast radio bursts, small bodies of the Solar System, etc.

- Increased frequency coverage will significantly enhance the ability to search for techno-signatures, prebiotic molecular emission, redshifted CO observations and pulsars orbiting Sgr A*.
- Greater field of view will allow simultaneous multiple phased beam observations. This capability allows multibeam space weather monitoring observations and is useful for VLBI phase referencing.
- The combination of sensitivity and field of view will provide increased survey speed.
- The NGAT will provide imaging capability when used as an interferometer. The dense sampling of the spatial frequencies will allow use of the array for imaging the Sun, extended structures in the ISM, as well as several cosmological experiments.
- Greater radar transmitter power will increase the number of observable NEOs by a factor of ~15 and allow observations of most of the newly discovered virtual impactors and remove most of their impact solution. It will also allow us to observe smaller and more distant bodies of the Solar System and improve observations of those objects that have been observed before. (Received power increases in proportion to the square of the target's diameter. Received power decreases with the fourth power of the target's distance.)
- Greater radar transmitter power and sky coverage for atmospheric science studies, making it possible to observe along parallel and perpendicular directions to the geomagnetic field.
- The compactness of the antenna array could provide short spacing measurements for other interferometers (eg. ngVLA).
- Radar transmitter power and swift Tx-Rx switching will allow space debris monitoring in geostationary orbit and cislunar space monitoring, while other facilities monitor below low-Earth orbit.
- The array will enable the usage of RFI mitigation techniques like the phased beam 'nulling'.
- The radar transmission beam and reception beam can be shaped by weighting the array elements, which may be useful to study asteroids larger than the size of the phased beam.

## 3.3 Summary of example configurations

| Table 2a: An example configuration of NGAT with 9 m dishes | |
|---|---|
| Equivalent illuminated diameter | 300 m |
| Dish diameter | 9 m |
| Number of dishes | 1,112 |
| Zenith angle coverage | 0 to 48 degrees |
| Frequency coverage | 200 MHz to 30 GHz |
| Beam Width of each dish @ 1.4 GHz | 81 arcmin |
| Resolution @ 1.4 GHz | 2.5 arcmin |
| System temperature (f = 1.4 GHz) | 25 K |
| Peak transmitter power per dish @ 430/220 MHz | 10 kW |
| CW transmitter power per dish @ 2 to 6 GHz | 5 kW |
| Peak transmitter power @ 430/220 MHz | 10 MW |

| Table 2a: An example configuration of NGAT with 9 m dishes (contd.) | |
|---|---|
| CW transmitter power @ 2 to 6 GHz | 5 MW |
| EIRP @ 5 GHz | 800 TW |
| Receivers (4 broadband feeds; instantaneous bandwidth at f < 900 MHz is yet to be determined) | 0.2 - 0.8 GHz (room temp)<br>0.7 - 2.5 GHz (room temp)<br>2.0 - 8.0 GHz (cooled)<br>8.0 - 30.0 GHz (cooled) |
| Signal processing bandwidth | 2.5 GHz |
| Transmitters (2 transmitting feeds + power amplifiers) | 0.2 - 0.8 GHz<br>2.0 - 8.0 GHz |

| Table 2b: An example configuration for NGAT with 15 m dishes | |
|---|---|
| Equivalent illuminated diameter | 300 m |
| Dish diameter | 15 m |
| Number of dishes | 400 |
| ZA coverage | 0 to 48 degrees |
| Frequency coverage | 200 MHz to 30 GHz |
| Beam Width of each dish @ 1.4 GHz | 49 arcmin |
| Resolution @ 1.4 GHz | 2.2 arcmin |
| Tsys (f = 1.4 GHz) | 25 K |
| Peak transmitter power per dish @ 430/220 MHz | 25 kW |
| CW transmitter power per dish @ 2 to 6 GHz | 12.5 kW |
| Peak transmitter power @ 430/220 MHz | 10 MW |
| CW transmitter power @ 2 to 6 GHz | 5 MW |
| EIRP @ 5 GHz | 800 TW |
| Receivers (4 broadband feeds; instantaneous bandwidth at f < 900 MHz is yet to be determined) | 0.2 - 0.8 GHz (room temp)<br>0.7 - 2.5 GHz (room temp)<br>2.0 - 8.0 GHz (cooled)<br>8.0 - 30.0 GHz (cooled) |
| Signal processing bandwidth | 2.5 GHz |
| Transmitters (2 transmitting feeds + power amplifiers) | 0.2 - 0.8 GHz<br>2.0 - 8.0 GHz |

### 3.4 Rough Order of Magnitude Cost estimate

A rough order of magnitude (ROM) cost estimate for NGAT's construction is given in Table 3. A more accurate cost can be estimated only after optimization of the configuration during the design study. For the ROM, we consider an average cost for the two configurations given in Table 2 and assume that the telescope will have 8 segments. Other considerations that have gone in the estimation are mass production rate for carbon composite reflectors and solid state transmitters, cost estimate of the SKA frontend receivers scaled for NGAT requirements, bulk acquisition of conventional power generators and DSA2000 (Hallinan et al. 2019) estimate for backend signal processing, data storage and archival scaled for NGAT application.

| Table 3: NGAT ROM | |
|---|---|
| Design studies | $8M |
| Telescope structure, mount, drive and reflectors | $230M |
| Installation and integration | $37M |
| Receivers (LNA, cryostat, post-amplifier, signal processing circuit, fiber optic transceivers, fiber connection) | $12M |
| Transmitters (waveguide, heliax cable, power amplifiers, signal processing network) | $83M |
| Backend signal processing, data storage and public data archive | $17M |
| Power generation for transmitter (20 MW) | $6M |
| Building and other infrastructure | $2M |
| Sub total | $395M |
| Contingency/Reserve (15%) | $59M |
| Total | $454M |

## 4.0 Summary

The Next Generation Arecibo Telescope (NGAT) satisfies the overwhelming scientific and public desire to rebuild a telescope at the Arecibo Observatory. Although constructed in 1963, the legacy Arecibo telescope (LAT) continued to produce high-impact scientific discoveries, with planetary science, astronomy, and space & atmospheric science results published in scientific journals like *Nature* and *Science* even in the months just prior to its collapse. Rather than simply replacing the now-lost facility, we propose building a new telescope that would significantly advance research in all three areas of science pursued at the Arecibo Observatory. The NGAT will use state-of-the-art radio and transmitter instrumentation and will implement active RFI cancellation techniques. In this paper, we presented the extensive scientific objectives that will be achieved and enhanced by the NGAT concept. Some of the exciting new possibilities with this instrument include searching for pulsars orbiting Sgr A*, observing molecular lines from early Universe, climate change, ISR studies both parallel and perpendicular to the geomagnetic field, space debris characterization, accurate velocity measurements of a larger fraction of near earth objects, space weather forecasts, dark energy and dark matter studies, gravitational wave detection through pulsar timing, etc. These capabilities will vastly increase the user base of the facility and enable cutting-edge science for decades to come.

## Acknowledgements

We are grateful to the Arecibo Observatory users' community for the invaluable support, comments, and discussion that helped us to write this white paper with a short notice. Special thanks to Joanna Rankin from University of Vermont, who led the organization of the AO's user community and Pia Salter for her enthusiastic support to the effort for Arecibo telescope reconstruction. We thank the technical staff of the AO, as well as Dr. Shaffer and the UNC-URSI community for their suggestions and inputs while preparing the white paper. We also greatly appreciate valuable input from Gordon Lacy, Mohammad Islam, and Dean Chalmers at NRC Canada; Peter Dewdney from SKA; Larry D'Addario from Caltech; Anthony van Eyken from SRI International; Alex Kraus from MPIFR, Bonn; Chip Cohen from Fractal Antenna Systems; Tim Bastian from NRAO , Timothy Kennedy from the NASA Johnson Space Center and Catherine Neish from Western University. A huge thanks to B. Serna, SwRI, Texas, for helping with the formatting of the whitepaper.

## Affiliations (for author list)

1. Arecibo Observatory, HC3 Box 53995, Arecibo, PR 00612, USA
2. Physics Department, Western Illinois University, 1 University Circle, Macomb, IL 61455, USA
3. Department of Astronomy, Yale University, New Haven, CT, 06511, USA
4. Southwest Research Institute, San Antonio, TX, 78238 USA
5. NASA Kennedy Space Center, FL 32899, USA
6. Department of Physics and Astronomy, Johns Hopkins University, Baltimore, MD 21218, USA
7. Department of Astronomy and Cornell Center for Astrophysics and Planetary Science, Cornell University, Ithaca, NY 14853, USA
8. George Mason University, Resident at the Naval Research Laboratory, Washington, DC 20375, USA
9. Green Bank Observatory, P.O.Box 2, Green Bank, WV 24944, USA
10. Jet Propulsion Laboratory, California Institute of Technology, 4800 Oak Grove Drive, Pasadena, CA 91109, USA
11. Department of Astronomy, University of California, Berkeley, CA 94720-3411, USA
12. Inter American University, Bayamón, PR 00957, USA
13. Computational Physics Inc. (CPI), North Chelmsford, MA 01863, USA
14. Delft University of Technology, Delft, Netherlands
15. Department of Physical Sciences, University of Puerto Rico, Río Piedras Campus, San Juan, PR 00931, USA
16. Department of Physics and Astronomy, West Virginia University, P.O. Box 6315, Morgantown, WV 26506, USA
17. Department of Physics and Astronomy, University of Georgia, Athens, GA 30602, USA
18. USRA, SOFIA, NASA Ames Research Center, Moffett Field, CA 94035, USA
19. Lunar and Planetary Laboratory, University of Arizona, Tucson, AZ 85721, USA
20. Department of Physics and Astronomy, Texas Tech University, Lubbock, TX 79409, USA
21. Space Telescope Science Institute, 3700 San Martin Drive, Baltimore, MD 21218, USA
22. Florida Space Institute, University of Central Florida, Orlando, FL 32826, USA
23. National Research Council Canada, Herzberg Astronomy and Astrophysics Programs, Dominion Radio Astrophysical Observatory, Penticton, BC V2A 6J9, Canada
24. Department of Physics, New Mexico Tech, Socorro, NM 87801, USA
25. Department of Physics, University of Puerto Rico, Mayagüez, PR 00681, USA

26. INAF-Arcetri Astrophysical Observatory , Florence, Italy
27. Joint Institute for VLBI ERIC, Dwingeloo, Netherlands
28. Consultant, University of Central Florida, Orlando, FL 32826, USA
29. Pennsylvania State University, PA 16802, USA
30. Hillsdale College, MI 49242, USA
31. Purdue University, IN 47907, USA
32. California Institute of Technology, Pasadena CA 91125, USA
33. Electrical and Computer Engineering Department, University of Illinois at Urbana-Champaign, 306N. Wright St., Urbana, IL, 61801, USA
34. Electrical Engineering East, Penn State University Park, PA 16802, USA
35. ASTRON, the Netherland Institute for Radio Astronomy, Dwingeloo, NL
36. Planetary Habitability Laboratory, UPR, Arecibo, PR 00612
37. University of California, Los Angeles, CA 90095
38. Anton Pannekoek Institute for Astronomy, University of Amsterdam, Amsterdam, The Netherlands

# Appendix A: Alternate Design Concepts Considered

## A.1 A Single Dish

The possibility of rebuilding a new single dish at the same location as the LAT was discussed extensively. Most single dish concepts involve a primary reflector with a fixed surface of size at least 305 m and a movable secondary reflector. The secondary could be deformed to improve the aperture illumination, sky coverage, and even to reduce the physical range of the movement of the receivers while pointing the telescope. The secondary could be on a central tower, or suspended by a set of cables, or suspended from an arch or a crane-like mechanism. We discussed the possibility of Cassegrain optics, in which the elevated secondary reflector would be movable but all receivers and transmitters would be near the ground, in an opening near the center of the primary reflector. Preliminary investigations indicate that the major limitation with all these approaches is limited zenith angle coverage with a 305m diameter dish, which might not satisfy some science requirements for ISR studies. Furthermore, there is no technical solution we are aware of that can increase the peak transmitted power above 2 MW, as desired for the next generation ISR and planetary defense applications. However, we note that their design discussions occurred over the three weeks following the collapse of LAT, and viable design options other than the ones presented here are possible.

## A.2 An Array of Individually Pointing Dishes

One possible design we considered is an array of individually steerable dishes. Like the proposed NGAT configuration in Table 2a, the number of antennas required will be approximately 1,100 and the diameter of each dish will be about 9 m. Since each antenna will be pointed individually, conventional servo systems could be used for steering the antenna. The required radar power could be achieved by injecting smaller transmitter power to each antenna and phasing them up, as demonstrated in the KaBOOM experiment for Ka band, but with only three 12 m antennas with spacing of 60 m between each antenna (Geldzahler et al. 2017). The surface brightness sensitivity requirement of ISR studies would require that all the antennas be placed very close to each other. This close spacing of elements is required even if a larger dish size of 15m (see Table 2b) will be used for the array. However, this would mean that the far field pattern of the array would be modified by both geometric shadowing and electromagnetic interaction between antennas in the array. For instance, consider antennas operating at 430 MHz. The near field of a 9 m antenna will be about 20 m at this frequency. This means that for each antenna, all antennas within two antenna diameters will be in its near field, causing them to be electromagnetically coupled. This interaction between the antennas will change constantly as they point to different positions in the sky. This interaction only increases with higher operating frequency. Accurately predicting the radiation pattern of such an array will be a formidable challenge. In the NGAT design, the relative orientations of the dishes in a segment are fixed, which enormously simplifies understanding the electromagnetic coupling in the system.

## Appendix B: A High Frequency Facility with Extended Capabilities and its Application to Space and Atmospheric Sciences

The ionospheric modification program (or the High Frequency facility - HF facility) together with the collocated LAT's ISR was considered a centerpiece of SAS research at AO. The LAT's HF facility was a Cassegrain design using the main 305-m dish as a primary with a light hanging secondary, illuminated by dipole arrays located just above the primary. This facility was partially damaged during the LAT's collapse and will be relocated in the proposed NGAT concept. Below we provide the scientific rationale to relocate the new HF facility to a location near the proposed NGAT.

The HF facility excites irregularities, instabilities, and various radio frequency emissions, from extremely low frequencies through high frequencies, in the ionospheric plasma. The HF facility contributes to improved understanding of (1) the electron acceleration process, (2) the generation of Langmuir/Ion oscillations in the narrow interaction region, (3) electron thermal balance and heat transfer in the ionosphere as heating extends from the illuminated volume (a dual-beam ISR capability permits precise temperature gradient measurements along a field line), (4) the aeronomy of heated electrons and the generation of airglow (experiments with concomitant imaging, photometry, and optical interferometry can provide new information with unprecedented sensitivity), (5) the structuring of field-aligned irregularities at sub-kilometer and meter-scale lengths, and (6) mesospheric structure and dynamics using the system as a mesosphere–lower thermosphere (MLT) radar.

Observations of the LAT's HF facility include the surprising generation of flat structures of electron density that have properties that appear to be partly under the control of the experimenter and partly controlled by ionospheric dynamics (gravity wave propagation). Also, this facility was able to generate very deep high altitude irregularities and study their characteristics in remarkable detail using special HF signals and new 430 MHz analysis techniques.

The LAT's HF had only vertical pointing capability. In the future, it will need to be able to propagate in different directions, including the ability to study the effects of different HF propagation directions with respect to the magnetic field, and different radar viewing angles for a given propagation direction to the field.

The NGAT concept design will preclude using the main facility as the primary reflector for the new (long term) HF facility. However, wire antennas will be constructed near the NGAT. The new antenna array will have the ability to switch between various fixed directions, in order to fulfill many of the needs described above. We pointed out that a radar facility located at AO, looks almost vertical when it propagates with a 45° angle to the field. 45° from vertical makes perpendicular and parallel propagation accessible without extreme increases in range for a fixed altitude. A similar advantage would apply to the HF facility. A wide range of angles will be within the reach of the radar. In addition, *the extended HF facility will be used in NGAT different radar modes to potentially open many new science paths. Some of these include use as a coherent lower atmosphere radar (to study wave interactions in the troposphere and stratosphere), use as a lunar surface radar, use as a solar radar, and imaging ionospheric interactions using an "off-dish" receiver antenna array.*

**It is not possible to predict the breadth of inquiry possible with the new, extended HF system. However, it is inevitable that coupled with the sensitivity of the NGAT, the ionosphere above AO will serve as an unprecedented natural plasma laboratory, which we will exploit with national and international proposal competitions.**

# Appendix C: Additional Science Studies that are enhanced by NGAT

## C.1 Planetary Science

Overall goal: Delay-Doppler radar imaging of near-Earth asteroids and other Solar System objects

- We want the effective area and transmitter power to be as high as possible.
- We want system temperature to be as low as possible, perhaps with a narrow-band receiver that is optimized for the frequency of our radar, like the previous S-band narrow receiver.
- Wider zenith angle coverage will permit observations of more targets.

Better sensitivity and transmit power will improve existing data on:

- Galilean satellites: these have some of the most peculiar radar signatures in the Solar System (Campbell et al. 1977). Comparative studies of these bodies provide a unique opportunity to study different physical and chemical processes in the Solar System, because radar techniques provide unique access to study surface properties at different depths (compared to shorter wavelengths), and the Galilean satellites have widely diverse surfaces from hot to icy, and active to ancient.
- Venus: look for evidence of recent/ongoing tectonic or volcanic activity, impact processes, and aeolian sedimentary transport.
- Mercury: polar ice deposits, volcanic history (especially with longer wavelengths)
- Mars and its moons: Geologic studies of the surface and subsurface. Comparison of radar observations to one of the most surveyed planets in the Solar System help us understand radar observations
- The Moon: volcanic history (lava flows, lava tubes, pyroclastic deposits), impact basin stratigraphy; these will also benefit from longer wavelengths

And possibly allow these new targets and experiments:

- Earth Trojans: For example, 2010 TK7 has a diameter of ~300 m and its oscillation around Earth-Sun L4 brings it to within 0.2 au. This is likely observable with NGAT, depending on the spin rate.
- The largest Jupiter Trojans will be detectable with NGAT, as well as the largest members of asteroid families such as Hildas.
- More main-belt asteroid (MBA) radar astrometry could help to detect orbit perturbations that lead to orbit changes; this could potentially showcase near-Earth object source regions. Until 2020, 138 MBAs have recorded radar observations, while the proposed NGAT design will enable the observation of more than 15,000 MBAs. Radar astrometry of asteroids belonging to collisional families would potentially allow for detection of the non-gravitational Yarkovsky effect and improve their age estimation (e.g. ~80% of the inner main-belt Phocaea family members will be observable). The Yarkovsky effect is also the mechanism that allows an asteroid to get close to orbital resonances and then enter the near-Earth space. The main belt also opens doors to studies of active asteroids: These asteroids pass through physical processes strongly related with the cohesion of the regolith.
- More comets: Only 21 comets have been observed using planetary radar, and increased transmitter power will increase this count. Comets typically have lower radar albedos than asteroids, so on average they require better observation conditions (proximity, size, rota-tion period) than asteroids. Better sensitivity and sky coverage also can help with passive radio observations of OH and other gases in coma.

- Interstellar asteroids and comets: spectroscopic studies enabling the study of the sublimation rate of water ice and the detection of molecules of prebiotic interest on comets in our Solar System and also, those visiting from interstellar space, such as Comet 2I/Borisov.
- Radio occultations of asteroids: diffraction pattern in the Fraunhofer regime
- Look for asteroids' thermal emission at the higher end of AO's frequency range, e.g. C-band (Lovell et al. 2019). Long-wavelength thermal investigations are a key tool to probe the sub-surfaces of asteroids and comets to constrain density and composition, which relate to accretion and collisional history. Opportunities for these observations had been deemed rare due to the lack of sufficient sensitivity, but possibilities for these experiments using NGAT's increased sensitivity and sky coverage should be revisited.
- Assistance to possible future space missions: AO has a long history of spacecraft mission support: OSIRIS-REx benefited from Arecibo's support through radar data based shape modeling, as well as will Double Asteroid Redirection Test (DART), Psyche, and DESTINY$^+$. Mini-RF benefited from Arecibo's S-band radar system illuminating the lunar surface after the Lunar Reconnaissance Orbiter's own radar transmitter failed. NGAT could continue this legacy through support for missions planned beyond 2025, such as Europa Clipper, Jupiter Icy Moons Explorer (JUICE), Dragonfly, and any future missions to other planets.

Faster transmit-receive (Tx-Rx) switch time compared to the old radar system will allow us to observe objects that are closer to the Earth:

- Moon
- Asteroids closer than 4 light-seconds (3.1 lunar distances)
- Cis-lunar space monitoring (out to the Earth-Sun Lagrange points)
- Lunar debris, depending on the size

***Adding additional antenna outside of the sinkhole (bistatic observations with planetary radar)***

Adding a second (smaller) receiving dish near the main facility (compare to the Los Caños antenna; e.g. Stacy 1993); or faster switching between radar transmitter and receiver (on one telescope) will enable:

- Observing targets that are closer than we could previously observe, since we normally took eight seconds to switch from transmit to receive
- Detection and characterization (especially down to mm scale) of space debris in geocentric orbit as done in the 1990s (see Thompson et al. 1992).
- Cis-lunar space monitoring (out to the Earth-Sun Lagrange points) and lunar debris.
- Detection of mini-moons (of Earth), or asteroid Apophis on the day of its closest approach in April 2029.
- (Bistatic) observations of the Moon
- Increased integration time for improved Doppler frequency resolution

## C.2 Solar, Heliospheric, and Space Weather studies

### C.2.1 Solar Wind and Space Weather Studies

The effects of “space weather” at the Earth or elsewhere are determined by the continuous flow of the highly variable solar wind plasma, the interplanetary magnetic field carried by the solar wind into the heliosphere, and the large-scale transient interplanetary disturbances propagating away from the Sun. For example, transient events such as coronal mass ejections (CMEs) and co-rotating interaction regions (CIRs) both cause severe to moderate space weather disturbances at the Earth and other planets. The CMEs are spectacular eruptions of energetic plasma and intense magnetic fields from the surface of the Sun into the heliosphere, whereas CIRs are structures formed by the interaction between slow and fast solar wind streams which have their primary origin in the solar coronal holes which persist for several solar rotations. Additionally, the electromagnetic radiation produced at the time of solar flares/CMEs can also affect the Earth’s atmosphere and ionosphere in several ways. Space weather processes interact with the Earth’s magnetosphere and ionosphere, which can strongly affect many technologies, including satellite operations, telecommunications, navigation systems and power grids. A reliable prediction of space weather is necessary to provide sufficient warning to make effective measures against its adverse effects on human technology. The important point is that the demand for accurate space weather prediction is increasing and observations are required to fully understand the physical principles driving the steady-state solar wind as well as transient flows into the interplanetary space, and their resulting dynamical impacts on the magnetosphere and ionosphere. *NGAT will be a powerful facility that will allow scientists to answer many important questions concerning space weather science with regard to the origin of solar wind and transient events and their evolution in the inner heliosphere. Critically, on a more local scale, this facility will illuminate the impacts of these events on “atmosphere-ionosphere-magnetosphere interactions” (AIMI).*

In the case of steady flow of the solar wind, the exact mechanism of its formation and expansion from the high temperature coronal plasma into the heliosphere is still considered to be a topic of interest. Regarding transient eruptions, the key aspect is that the arrival of a CME at the Earth has the likelihood of causing an intense geomagnetic storm, which includes the impact of the interplanetary shock, followed by the arrival and passage of the plasma cloud and frozen-in-flux magnetic field. There are few techniques available to study the solar wind: (a) the direct sampling of the downstream solar wind plasma, and (b) the remote sensing of the inner heliosphere by radio spectroscopy and radio sounding observations. In-situ measurements are limited to the ecliptic plane along the one-dimensional time scan in the near-Earth space. However, the study of global properties of the solar wind and the efficient tracking of a transient event call for solar wind observations at consecutive locations within the inner heliosphere. The radio remote-sensing technique, e.g., interplanetary scintillation (IPS), provides a three-dimensional view of (i) the velocity of the solar wind, (ii) the level of electron density fluctuations, (ii) the spatial spectrum of density turbulence, (4) the size of the inner scale associated with density fluctuations, and (5) the axial ratio of density irregularities (e.g., Hewish et al. 1964; Coles 1978; Manoharan & Ananthakrishnan 1990; Manoharan 1993, 2012; Tokumaru et al. 1994). Since the Earth-directed CMEs could cause interplanetary disturbances, shocks, and severe geomagnetic storms (e.g., Kumar et al. 2011), IPS observations via NGAT, coupled space-based images (e.g., SOHO, SDO and STEREO), and sophisticated modeling analysis, are of critical importance for the physical understanding of space weather science and forecasting capabilities.

### C.2.2 Highly Efficient Heliospheric Sampling and Tracking of Space Weather Events

Several instruments are available to study IPS (e.g., the Ooty Radio Telescope in India, the Solar Wind Imaging Facility in Japan, and the Mexico Array Telescope). The predominant limitation constraining the usefulness of current IPS measurements and their science output is the sampling resolution of the heliosphere in space and time. The solar wind estimates from such observations limit the detailed inference of the small-scale changes in the solar wind and in the CME kinematics. Particularly, it restricts the precise study of a fast moving CME. Additionally, most of the current IPS arrays operated at meter wavelengths (i.e., in the frequency range of ~100-327 MHz), which can be useful to probe the solar wind at distances >50 solar radii (~0.25 au) from the Sun. The essential fact is that we need to observe a sufficient number of sources in a given time and the statistical noise on the solar wind parameters derived from such observations should be low. For instance, the high sensitivity and large bandwidth of LAT allowed studies of compact radio sources having scintillating flux densities as low as 10 mJy with sufficient signal to noise ratio. The proposed NGAT frequency coverage of 200 MHz to 30 GHz, coupled with excellent sensitivity, is well suited to examine the solar wind all the way from the Sun to near-Earth orbit.

Recently, it was demonstrated that IPS measurements using LAT had the potential to play a significant role in characterizing the properties of the solar wind from near-Sun to the inner heliosphere. A sample plot of scintillation index (m) as a function of observing frequency (ν) is displayed in Fig. 7. This plot includes IPS data sets observed at three observing bands of the legacy Arecibo system, covering the frequency range of ~300 to 3000 MHz. Each data point on the plot corresponds to a bandwidth of 10 MHz. These observations were taken shortly before the telescope's demise, during 12 July to 09 August 2020, and covered a heliocentric distance range from ~40 to 85 solar radii. On some of the days, measurements were taken at multiple bands, quasi-simultaneously within a 1-hour time interval, and these observations are connected by a fitted straight line. This plot shows the potential value of Arecibo to the heliophysics and space weather science communities in probing solar wind at density scales of ~10 to 400 km.

### C.2.3 Tracking Coronal Mass Ejections

We also lack a comprehensive understanding of the physical processes involved in the generation of the exceedingly hot corona and where the plasma is accelerated to form the solar wind. Therefore observations at the near-Sun region (at solar offsets of ~5 to 50 solar radii) are essential to explain the combination of kinetic, magnetic, and turbulent energy to derive the solar wind. Since the micro and macro features of the solar wind carry distinct information about the energetic condition of the coronal plasma in the source region, near-Sun measurements of the solar wind would allow for the inference and diagnosis of the coronal plasma. Additionally, we need to explain the range of turbulence scales in the solar wind and kinetic physical processes involved in the dissipation of energy at small scales as functions of the source region on the Sun and distance from the Sun.

The proposed large sky coverage of NGAT and its multi-beam and multi-frequency facilities will be extremely useful for near-simultaneous monitoring of a grid containing a large number of IPS sources at a given time (refer to the sky coverage Fig. 1 and Fig. 15, in Appendix C.3). Such observations (e.g., in the frequency range of 327 MHz to 10 GHz) in the Sun-Earth distances will be particularly important in deducing the properties of the solar wind at regions inaccessible to spacecraft and to track CMEs (e.g., Yamauchi et al. 1998). Regarding LAT, although its beam switching and the number of sources observed in a given time period were limited, its high sensitivity and the low-level of sampling of the interplanetary medium detected even a weak CME at around 120–140 solar radii from the Sun. This CME was launched from the Sun on 16 July at 04:12 UT, its width was only ~80 deg in the LASCO/SOHO field of view and propagated with a speed of ~350 km/s. Fig. 14 displays the normalized scintillation index measured at the Arecibo as a function of time on 2 days, 17 and 19 July 2020 and an image from LASCO C2 coronagraph. Each day's observations included measurements of several sources and at multiple frequencies (one or more of P, L, and S bands).

Such CME detection and tracking will be much more efficient with the high sensitivity and finer temporal/spatial resolution of the NGAT, which as well as being a good instrument for space weather science studies, will provide advanced predictions of arrival of CMEs at the near-Earth environment. Another important point is that the software being developed for handling the Arecibo IPS data could provide a common interface and analysis tools for IPS data from telescopes across the globe.

### C.2.4 Internal Magnetic Field of CMEs and Faraday Rotation

The knowledge of the strength and structure of the magnetic field associated with a CME is critical because of its key role in deciding the geo effectiveness caused by the CME (e.g., Kooi et al. 2017; Jackson et al. 2020). The plane of the linearly polarized radio source signal rotates, while passing through the magnetized plasma and this effect being known as Faraday Rotation (FR). The degree of rotation is quantified by the Rotation Measure (RM). A CME carries its internal magnetic field from the Sun. The RM values estimated by a grid of compact sources passing through the CME cloud allow us to probe the spatial structure of the magnetic field. Such observations at different solar offsets from the Sun would be important to study the evolution of the magnetic structures within the CME. Polarization studies with LAT were well established. However, the sky coverage and multiple beams of the NGAT will provide new insights of magnetic configuration/evolution via RM measurements of CMEs at solar offsets below 20 solar radii and a simple modeling can be employed to derive the magnetic field configuration which is actually the critical parameter for the space weather science and forecasting.

### C.2.5 Solar Radio Studies

The solar magnetic field is the main cause for all kinds of solar activities. A quantitative knowledge of the solar/coronal magnetic field is essential to understanding solar phenomena above the photosphere, including the structure and evolution of solar active regions, release of magnetic energy, acceleration of charged particles, flares, CMEs, coronal heating, formation of the solar wind and, eventually, space weather and its impact on Earth. In the case of the quiet Sun, the radio emission is dominated by free-free opacity, which is proportional to the square of the electron density, and whose polarization is proportional to the longitudinal magnetic field. For instance, the radio spectrum of moving sources of synchrotron emission (moving type IV bursts) at frequencies >200 MHz, combined with the routinely available space-based solar images and ground-based radio images from the dedicated heliograph (e.g., Nancay Radio Heliograph), have provided the useful diagnostic of magnetic field strength of CMEs (e.g., Bastian et al. 2001; Dulk & Marsh 1982; Maia et al. 2007; Simoes & Costa 2006). The magnetic field strength estimated from various radio observations at 100 to 450 MHz range between 0.3 and 5 G at solar offsets below 4 solar radii. The spread of these field strengths is possibly due to the different electron energy ranges as well as the spectral slopes employed in the analysis (Carley et al. 2017). It is clear from these studies that moving type IV bursts can be used as a useful diagnostic of CME magnetic field strength. Thus at frequencies above 200 MHz, the studies of the CME site can provide a more accurate value for CME magnetic field strength, as well as the characteristics of the distribution of non-thermal electrons causing the radio emission. Such a combined effort with the array of antennas of the NGAT will provide the valuable inputs to the understanding of basic physics of CME onset (magnetic reconnection) and the 2-D distribution of CME associated magnetic field, which is the essential component to understanding the space weather effects.

### C.2.6 Solar Wind and Space Weather Impacts on AIMI System

The solar wind energy transfer to the AIMI system is controlled by the mechanism of the magnetic reconnection between southward components of the interplanetary magnetic field (IMF) and the Earth's northward magnetopause field (Dungey, 1961). The coupling process results in varying geomagnetic responses during different phases of the solar cycle due to varying solar and interplanetary causes, and associated IMF structures.

Around solar maximum, the main features present on the Sun are sunspots and active regions. These structures are related to sporadic coronal mass ejections (CMEs) and associated flares, and their interplanetary counterparts or interplanetary CMEs (ICMEs) (Burlaga et al. 1981; Klein and Burlaga, 1982). The ICMEs are known to be the main causes of major geomagnetic storms during solar maximum (Tsurutani et al. 1988; Gosling et al. 1990). On the other hand, during the declining and minimum solar cycle phases, flares and ICMEs become less frequent and coronal holes become the dominant solar features leading to geomagnetic activity. The coronal holes are open magnetic field regions, from which emanate high-speed solar wind streams (HSSs).

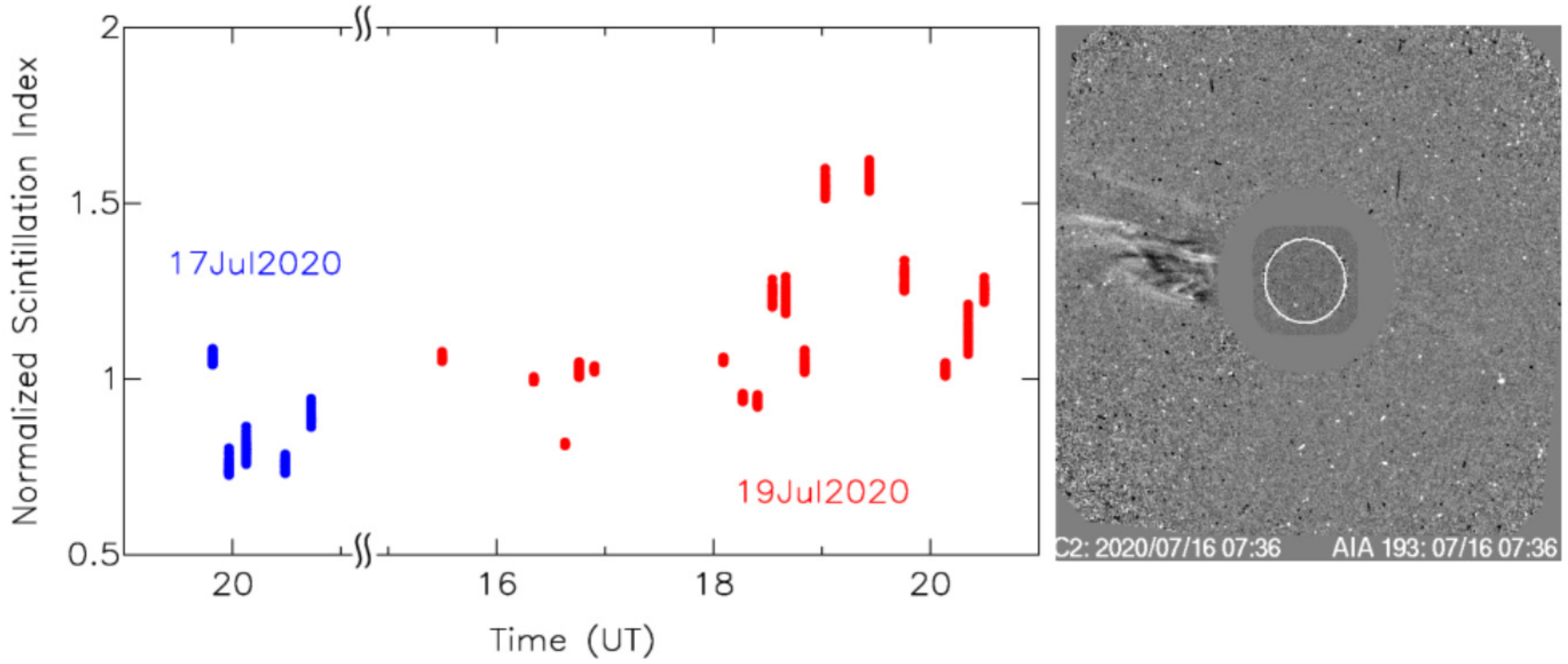

**Figure 14.** Normalized scintillation index observed with LAT on 17 and 19 July 2020. The CME crossing at around solar offset of ~120–140 solar radii is seen on 19 July ~19 UT. Each day's data sets include observations of multiple sources, as well as different observing frequencies. The image on the right: the white-light image of the CME from the LASCO C2 coronagraph on 16 July, at 07:36 UT is shown (image courtesy, http://sidc.oma.be/cactus/catalog) (Manoharan, Perillat, et al. in preparation).

CMEs and HSSs can cause different responses of the AIMI system as a whole, which makes key the development of a comprehensive program of research to improve understanding of the mechanisms that drive the Sun's activity and the fundamental physical processes underlying near-Earth plasma dynamics; to determine the physical interactions of Earth's atmospheric layers in the context of the connected Sun-Earth system; and to greatly enhance the capability to provide realistic and specific forecasts of Earth's space environment that will better serve the needs of society, as pointed out by the Decadal Survey for Solar and Space Physics. The new NGAT facility and the cluster of supporting devices will enormously improve our knowledge of the impacts of the Sun on the AIMI system. For the first time, a single institution (AO) will be able to track the plasma clouds released from the Sun into the interplanetary medium and their impact on the Earth's atmosphere, virtually from the troposphere up to the protonosphere/plasmasphere. For instance, phenomena such as geomagnetic storms and sub-storms and High-Intensity Long-Duration Continuous AE Activity (HILDCAAs) will have their intricate physics unraveled, and the dynamic and chemical responses of the Earth's atmosphere will be investigated.

### C.2.7 In-situ Data Comparison and Cometary Plasma Tail Investigations

The Parker Probe will reach a heliocentric distance in the range of ~25 – 30 solar radii and NGAT IPS in coincidence with *in-situ* locations of the Parker Solar Probe will provide a comparison of IPS measurements with the historically important in-situ data. The combining of the IPS observations of the solar wind from a grid of radio sources at different solar offsets in ensemble modeling of CME propagation can provide improvement of the current state of the art in CME forecasting.

The scintillations caused by the plasma tails of comets are of great utility for investigating the properties of "comet - solar wind" interaction. Such observations with the LAT system provided sensitive measurements of the tail plasma density structures. NGAT will provide much improved sensitivity, plus better sampling of the plasma tail of comets, to evaluate its spatial density structure, as well as studying its associated magnetic field as a function of the comet's heliocentric distance.

## C.3 Pulsar studies

### C.3.1 Pulsar searches

Large-scale Galactic surveys at radio wavelengths have uncovered more than 2800 pulsars over the past 50 years, including MSPs, exotic binaries such as the double pulsar system (Lyne et al. 2004) and the triple system (Archibald et al. 2018), transitional MSPs (Archibald et al. 2009), pulsars with puzzling behavior with their emission and rotation (Kramer et al. 2006a, Lyne et al. 2010, Perera et al. 2015, Perera et al. 2016), and rotating radio transients (RRAT – McLaughlin et al. 2006). The PALFA (Cordes et al. 2006) and AO327 (Deneva et al. 2013) surveys at the LAT have discovered more than 250 pulsars until the closure of the telescope, including MSPs, RRATs, and two Fast Radio Bursts (FRBs) with the first FRB repeater (e.g. Crawford et al. 2012, Lazarus et al. 2015, Lyne et al. 2017, Parent et al. 2019, Deneva et al. 2016, Martinez et al. 2019, Spitler et al. 2014, Spitler et al. 2016). The globular clusters are likely hosting central intermediate mass BHs, and discovering MSPs close to their centers is useful to study the dynamics and probe the central sources (Perera et al. 2017). With a large declination coverage and a better sensitivity, NGAT will be able to detect many more pulsars and FRBs in the Arecibo sky than before (see Table 4 and Fig. 15). The declination coverage is an important factor to cover a large fraction of the sky and increase the productivity of a survey. *The proposed NGAT will be able to observe more than 70%* *of the Galactic plane, including the Galactic center (GC), which was not visible with LAT, leading to a factor of more than 2.3 improvement in the declination coverage, and capable of observing more than twice the number of known pulsars compared to legacy Arecibo telescope (*see Table 4 and Fig. 15). A large fraction of known pulsars (more than 75%) are located within |b| < $10^0$ as they are born in the Galactic plane. Thus, the capability of sampling a larger fraction of the Galactic plane is a great advantage in pulsar search with large-scale surveys with NGAT.

**Table 4: Pulsars visible based on the sky coverage of NGAT. The second and third columns show the declination and the sky coverage, respectively. The fourth column presents the number of known pulsars that can be observed (MSPs with spin period < 20 ms and normal pulsars) with the two designs based on the sky coverage. The fifth column gives the number of simulated pulsars detectable in an all-Arecibo sky survey with a phased-array feed system at 1.4 GHz. The sixth column gives the number of simulated MSPs that could be timed with precision required for PTA experiments.**

| Design | Dec. coverage (deg) | Sky coverage | Known pulsars (MSPs/Normal) | Pulsars detectable with all-sky survey (MSPs/Normal)† | PTA-quality MSPs† |
|---|---|---|---|---|---|
| Legacy Arecibo | [-1, +37.5] | 31% | 104/661 | 207/1306 | 57 |
| NGAT | [-30, +66] | 71% | 240/1449 | 981/5477 | 226 |

†Pulsar timing simulations described in detail at http://www.aoc.nrao.edu/~tcohen/research/popsynth.shtml

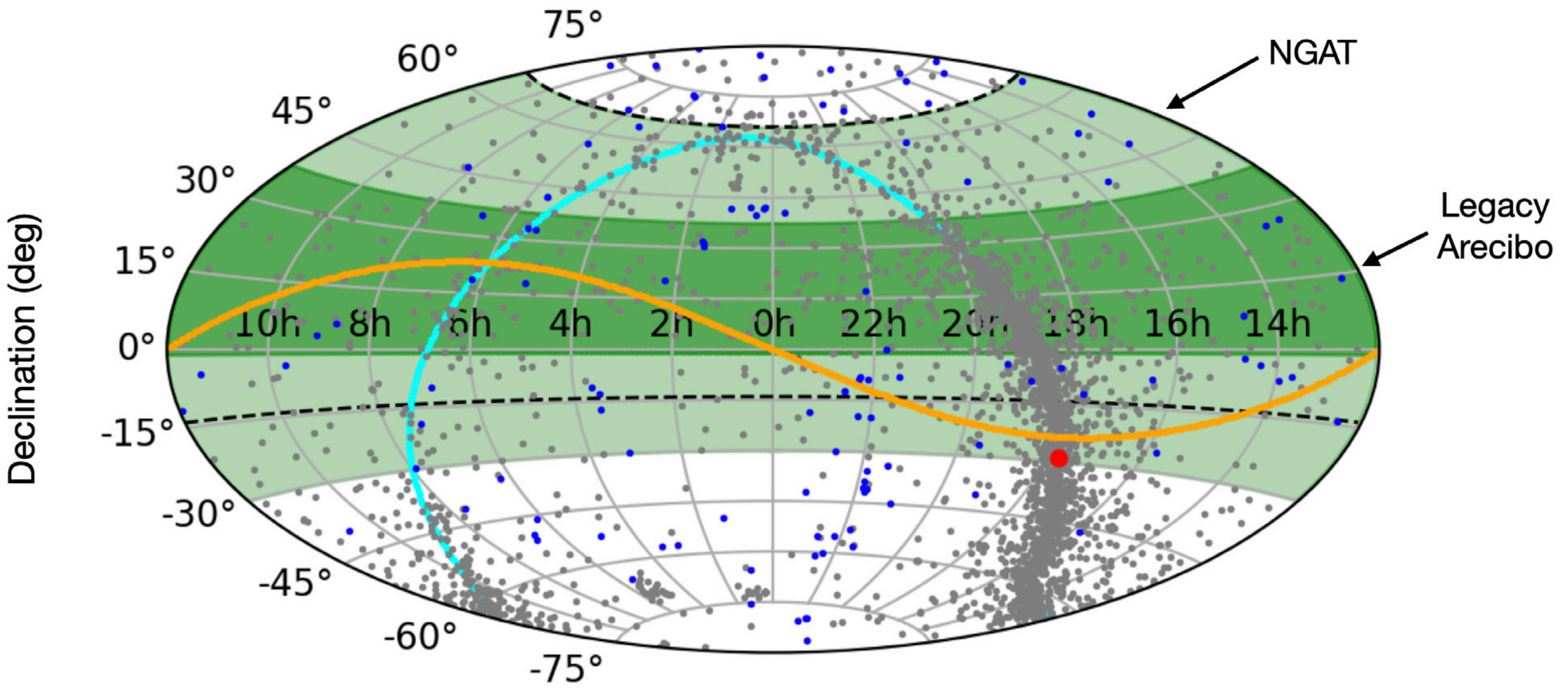

**Figure 15**: Sky coverage for NGAT design. The coordinates are in Right Ascension and Declination. The different shaded areas represent the sky coverages for LAT (dec = [-1,+37.5]$^0$) and NGAT (dec = [-30,+66]$^0$). The FAST telescope declination range [-14,+66]$^0$ is denoted by black dashed lines. The known pulsars (gray circles) and FRBs (blue circles) are over-plotted, and the Galactic center is denoted by the red circle. NGAT design is capable of probing a larger sky coverage than before. The Galactic (cyan) and ecliptic (orange) planes are over-plotted.

A simulation of pulsar population analysis determines that *NGAT will be able to detect more than 6400 pulsars at 1.4 GHz, which is a factor of 4.3 more pulsars compared to LAT* (see Table 4). Furthermore, the simulation indicates that NGAT will be able to detect 226 MSPs that are suitable for PTAs, which is a factor of 4 improvement than before. The wide FoV of the new design will provide improved survey capabilities and speed for pulsars and FRBs. The ultra-wideband receivers will improve the efficiency of follow-up observations of new pulsar discoveries and also their timing. Further, these receivers will also help to characterize the ionized interstellar medium using pulsars (Stinebring et al. 2019). With the high sensitivity, wide FoV, and large collecting area, NGAT will enable us to discover many pulsars in the Galaxy including MSPs, RRATS, and the faint ones that other telescopes around the world cannot detect.

### C.3.2. Binary pulsars

The binary neutron star systems are excellent laboratories to understand the gravity and the neutron star matter. The first double neutron star was discovered with LAT (Hulse & Taylor 1975), and its timing revealed the existence of GWs, leading to the authors receiving the 1993 Physics Nobel prize. The timing of double neutron star systems is used to test GR and the double pulsar system provides the stringent GR test in the strong field regime (Kramer et al. 2006b). The triple systems are important to understand the gravity further, and also improve the tests of universality of free fall (Archibald et al. 2018). Timing double neutron star systems allow us to measure their neutron star masses accurately. Finding massive neutron stars is highly important to improve our understanding of their interiors, equation of states, moment of inertia, and their formation and evolution. The recent discovery of the most massive neutron star (2.14 $M_{\odot}$) known to date resulted in improvements to the current equation of state models for neutron star matter (Cromartie et al. 2020). The legacy Arecibo was recently used to discover the most asymmetric-mass double neutron star system (Ferdman et al. 2020). Understanding and characterizing this double neutron star population is vital in gravitational wave astronomy as they are among the main targets in the LIGO experiment (Abbott et al. 2017). With the high sensitivity, NGAT will be able to discover and time more binary pulsars, including the heaviest neutron star, possibly an exotic binary such as a pulsar-black hole system, leading to understanding the formation and evolution of neutron stars and stellar mass black holes and to probe the space-time around a black hole, and study and test gravity.

### C.3.3. Pulsar Emission Mechanism and individual pulses

Studying individual pulse emission behavior at high time resolution and increased sensitivity will lead to further insights into the pulsar emission physics. Individual pulse studies of pulsars have been limited due to weak emission nature in general. Right from the start of pulsar science, LAT contributed significantly to these studies; e.g., finding the period of the Crab pulsar, detecting high-energy giant pulses, emission microstructures, drifting sub-pulses, abrupt emission switch-off known as nulling, changing between different emission modes (Lovelace et al. 1968, Hankins et al. 2003, Craft et al. 1968, Drake & Craft 1968, Backer 1970, Backer 1970b). Early work by Biggs et al. (1992) indicated that there is a correlation between nulling fraction and pulsar period. This result is so far contested. Some emission models suggest that in old pulsars, approaching the so-called "death line", the emission is stalling and leading to nulling. *With enormous sensitivity and larger sky coverage, NGAT will observe various emission features in many pulsars which have not been detected yet with current telescopes due to their sensitivity limitation.* Pulsar polarization provides independent studies to understand emission physics. Extracting the geometry information from pulsar polarization measurements may elucidate the emission region, emission altitude, and other characteristics (Rankin 1983a,b, Weltevrede et al. 2010). The increased sensitivity, combining with long tracking time and zenith coverage, NGAT will be advantageous for observing individual pulses, their polarimetry and better calibration. Furthermore, the broad frequency coverage offered by NGAT will allow multi frequency individual pulse studies. Pulse-to-pulse amplitude and phase variations that affect timing arise at an individual pulse level. Therefore, in particular, it is crucial to understand and model the individual pulse emission behavior of MSPs to help probe fundamental physics such as GW detection with PTAs, and to test gravity theories with binary pulsars. Therefore, studying the whole range of different individual pulse phenomena with NGAT will lead to an understanding of their (common) emission mechanism which will answer the question of what makes pulsars shine and facilitate tests of fundamental physics with pulsars.

## C.4. VLBI studies

VLBI offers a very high spatial resolution that is necessary to investigate, for example, never explored regions close to SMBH, to get accurate distances and proper motion measurements of galactic and extragalactic sources. When a VLBI station is equipped with a high-sensitivity telescope like the proposed NGAT, one can achieve lower noise levels, approaching a few microJy/beam. Several areas that will benefit immensely from such a high sensitive telescope include but not limited to the following:

**Strong Field Tests of General Relativity:** This key science goal of several international collaborations (and future telescopes) is projected to be achieved via; (a) timing relativistic systems such as pulsar-neutron star (and potentially pulsar-black hole) binaries; and (b) monitoring an array of millisecond pulsars (pulsar timing array), to directly detect gravitational waves. However, accurate distances and proper motions for the observed systems are essential to correct for the acceleration that affect intrinsic system parameters such as spin and orbital period derivatives. These are the parameters used to test alternative theories of gravity. We note that the current limitation on the test of GR using the Hulse-Taylor pulsar, B1913+16, is due to the uncertainty in its estimated distance, derived using the pulsar's dispersion measure and a model for the distribution of the ionized interstellar medium, that has a precision of only ~30% (Weisberg et al. 2008). Trigonometric parallax measurements using phase-referencing VLBI techniques can determine model-independent distances to much higher precision, supplying vital input to such studies. Future pulsar surveys (well into the "SKA era") will hopefully discover many relativistic binaries whose pulsars will be an order of magnitude weaker than current examples. NGAT's sensitivity will be indispensable for such measurements (as demonstrated for the micro-Jansky stars in the Pleiades cluster, Melis et al. 2014).

**Stellar Physics:** High-sensitivity VLBI is useful in the detection of weak sources (exoplanets, distant pulsars, SNR, star clusters) and accurate measurement of basic astrometric parameters, viz position, proper motion, and trigonometric parallax. This will lead to accurate and model independent distance measurements and luminosity estimates of stellar objects. It will be possible to localize the repeating FRB within their host galaxies. Proper motion measurements help determine the distances of GRBs, and the estimation of angular expansion and deceleration rates. These facilitate the study of jet emission and the evaluation of surrounding environments.

**Extragalactic Continuum Polarization VLBI:** Continuum polarization of compact extragalactic radio sources is usually at most a few percent of their total intensity, and hence requires the presence of the most sensitive possible telescopes in VLBI arrays planning to investigate it.

However, broadband imaging of the Stokes Q and U parameters allow derivation of the Rotation Measure over the jet of a source, allowing recovery of the intrinsic angle of polarization of the emission, and hence derivation of the magnetic field in the emitting region. By making repeated observations of the Stokes-Parameter images as the source expands, a study can be made of the overlying magneto-ionic material in the nuclear region of the host galaxy via secular changes in the measured distribution of Rotation Measure.

NGAT will continue to be part of the VLBI network and find answers to numerous outstanding questions (Venturi et al. 2020). Launching a global array of sensitive telescopes like NGAT, SKA1-MID, and FAST will provide a huge increase in sensitivity opening up a whole new set of scientific opportunities.

## C.5 New high-frequency explorations in Radio Astronomy

A combination of high telescope gain at frequencies above a few GHz, wide-band receivers and backend capabilities will allow sensitive studies of a plethora of spectral transitions that serve a wide range of scientific purposes. We describe in Table 5, a sample of the interesting molecular transitions in the 1-30 GHz range, followed by a brief description of their scientific applications.

### C.5.1 Probing the Nature of Early and Late-type Stellar Evolution from the Local to the Distant Universe

**Table 5: A sample of the interesting molecular transitions in the 1-30 GHz range.**

| Frequency (GHz) | Molecule |
|---|---|
| 1.612, 1.665, 1.667, 1.720, 4.765, 6.035, 6.049, 7.8, 8.2 | OH (hydroxyl) |
| 3.263, 3.335, 3.349 | CH (methylidine) |
| 4.829, 14.488 | $H_2CO$ (formaldehyde) |
| 6.668, 12.18 | $CH_3OH$ (methanol) |
| 5.640, 7.896, 9.024, 10.15, 10.65 | $HC_7N$ (cianotriacetylene) |
| 2.662, 5.324, 7.988, 10.65 | $HC_5N$ (cianodiacetylene) |
| 12.16 | OCS (carbonyl sulfide) |
| 22.23 | $H_2O$ (water) |
| 23.69, 23.72, 24.14, 24.53 | $NH_3$ (ammonia) |

The 1-30 GHz portion of the spectrum abounds with molecular lines and masers that provide insight into the processes of stellar formation at various stages. Observations of the linear and circular polarization of different maser species allow for the determination of magnetic field properties, such as morphology and strength, and enable exploration of the role magnetic fields play in the collapse of molecular clouds. For example, the methanol ($CH_3OH$) 6.7 GHz transition produces one of the strongest and the most commonly observed masers in massive star-forming regions and can be used to probe different parts of protostars, such as accretion discs and outflows. Masers from other molecular species (OH, $H_2CO$, and also the 12.2 GHz transition of $CH_3OH$; e.g., Al-Marzouk et al. 2012), are sometimes exhibited in periodic stellar flares (e.g., Goedhart et al. 2007; Szymczak et al. 2018), and may trace accretion events in binary systems, proto-stellar pulsations, and periodic enhancements of ionized wind shocks in young eccentric binaries, among other possibilities (e.g., Araya et al. 2010). The K-doublet transitions at 2 cm and 6 cm of $H_2CO$ can be used as an effective densitometer (Ginsburg et al. 2011) and provide an effective probe to obtain the physical characteristics of molecular gas.

In addition, higher excitation transitions of OH (at 4.8 GHz), for example, have been shown to trace expanding gas in pre-planetary nebulae (Strack et al. 2019), a discovery made at LAT and the first detection of this particular OH transition in a late-type stellar object. The increased sensitivity and sky coverage of NGAT along the Galactic place will widely open the opportunities for similar novel discoveries. These capabilities are also conducive to exploring unconventional tracers of stellar formation. For example, the CH 3.3 GHz lines are common tracers of molecular gas in both the atomic-molecular transition regions and dark clouds of the ISM (Sakai et al. 2012; as discussed in C.5.4) and likewise might be useful to trace outflows in high-mass star forming regions.

The capabilities of NGAT ensure such studies are not limited to our own Milky Way. An exciting facet of NGAT's capabilities is that it could more than double the collecting area of the upcoming ngVLA if included as part of the array. Just as LAT did in the High-Sensitivity Array, a combination that could function as a high resolution follow up opportunity following new NGAT detections. This powerful combination could enable the detection of megamasers across various cosmic distances. As a stand-alone telescope, NGAT will enable searches for highly red-shifted megamasers such as the $H_2O$ line near 22 GHz[13] and $H_2CO$ line near 4.8 GHz[14] that would provide a unique opportunity to investigate the molecular environment of active galaxies at cosmological distances, including starbursts and AGNs. In addition, extending earlier surveys of the methanol 6.7 GHz maser line (as done for part of the sky visible with LAT by Pandian et al. 2007) could be done both along the plane of our galaxy and in nearby (e.g., M33[15] and M31) and distant galaxies.

Finally, there has long been interest in the OH 18 cm lines which, in addition to their prevalence, offer the possibility to measure magnetic fields via Zeeman splitting, as demonstrated by Robishaw et al. (2008) who made with LAT the first extragalactic detection of the Zeeman effect in an emission line. Sensitivity is key for Zeeman observations which provide direct measurements of the magnetic fields of their host galaxies. For example, the sensitivity of the LAT (as part of the HSA) enabled McBride et al. (2013) to map the magnetic field of distant starburst galaxy Arp 220. Thus, widespread searches of Zeeman splitting in OH megamasers with NGAT could sample the magnetic fields of ultra-luminous infrared and starburst galaxies. In fact, NGAT is the only instrument that, if used in conjunction with other components of the HSA, could provide the additional sensitivity needed to enable the simultaneous study of the magnetic fields in these galaxies. If these capabilities are not restored, such studies can no longer be done.

While NGAT's sensitivity is expected to be slightly larger than China's FAST telescope, they may have similar capabilities for detecting masers at < 3 GHz. However, the required sensitivity for the higher frequency maser searches can only be provided by NGAT since FAST is frequency-limited. In addition, because of their geographic locations, only NGAT can be used to more than double the collecting area for ngVLA and VLBI in general for high resolution studies, much less maser searches. The next generation of maser science in the Milky Way and other galaxies is very bright, and NGAT will be a powerful force for investigating the physics of stellar evolution and various additional science products that come from maser studies over galactic and extragalactic scales.

[13] The $H_2O$ 22 GHz line would be redshifted to ~11 GHz at z=l. Existing receivers were already capable of searching this redshift regime, so extending to higher redshifts with the capabilities of NGAT is quite reasonable.

[14] Extragalactic detections of this maser line as well as OH megamasers were made with the legacy telescope (Baan et al. 1992, Baan et al. 1993), and the enhanced capabilities of NGAT allow similar studies to further redshifts

[15] A methanol search in M33 was done with the Arecibo legacy telescope, the most sensitive at the time, and yielded a non-detection (Goldsmith et al. 2008). In addition, a search of M31 conducted with the VLA resulted in one positive detection (Sjouwerman et al. 2010). The increased sensitivity of NGAT will justify a new search, and provide an opportunity to determine whether the 6.7 GHz maser line is as prevalent in our galactic neighbors as it is in our own.

### C.5.2 Pulsars and Radio Transients at High Frequencies

Pulsar observations at relatively high frequencies, especially for distant and/or high dispersion measure (DM) pulsars, may benefit from minimized variable scattering and time delays due to the turbulent ionized interstellar medium. For example, the high-DM Arecibo pulsar J1903+0327 would be better observed at a much higher frequency to improve its timing precision (Lam et al. 2018). High frequencies (> 9 GHz) are essential to probe the pulsars in the Galactic center due to large DMs and extreme scattering effects in the region (see Section 2.12). High frequencies also provide optimal conditions to search for radio transients in the Galactic center (e.g. Chiti et al. 2016, Hyman et al. 2002, Hyman et al. 2009). The radio emission from magnetars and pulsars that show inverted/flat spectra may be detectable at the high observing frequencies. Generally, pulsars show narrow pulse profiles at high frequencies, explained by high frequencies occurring close to the surface of the star. Therefore, observations of pulsars at high frequencies may offer insights to the pulsar magnetospheres and emission properties, and also discover and better characterize new classes of radio sources through transient events.

### C.5.3 Zeeman Measurements to detect Galactic Magnetism

The ultrawide frequency capabilities of the NGAT system could enable the simultaneous detection of more than 100 hydrogen radio recombination lines (RRLs) in the 1-30 GHz range alone. The detection of weak Zeeman splitting of RRLs from HII regions and photodissociation regions would be possible with this system by averaging a large number of RFI-free transitions, enabling the study of magnetic fields in these regions. In addition, the helium and carbon lines are also enticing candidates for Zeeman measurements (Greve and Pauls, 1980), particularly since carbon RRLs generally trace photodissociation regions, and, when combined with the suite of hydrogen lines, they would present upward of 300 possible lines for these measurements. Such observations are not possible currently due to enormous observing time requirements with existing narrowband systems and would be revolutionary not only with respect to measuring the Zeeman effect on RRLs but in various other astrophysical regimes as well.

## C.6 A Comprehensive Snapshot of the Galactic Plane

The interplay between gas and stars dictates the evolution of spiral galaxies, both large and small. By the mid-1970s, CO(1-0) surveys of the Milky Way had established that the coldest, densest portion of the interstellar medium was mostly contained in large entities called Giant Molecular Clouds (GMCs) which were the sites of copious star formation. It became clear fairly quickly that the molecular clouds originated from the surrounding cold atomic medium (known as the Cold Neutral Medium or CNM), but the exact mechanism for how these objects form is still debated. Equally troubling are the details of how these large clouds are supported against large-scale gravitational collapse. Magnetic fields and turbulence are clearly involved, but the exact nature of the supporting mechanisms has, thus far, avoided elucidation.

Perhaps more disturbingly, gamma-ray and infrared data revealed that a substantial amount of molecular gas may have been missed by the large-scale CO mapping (Grenier et al. 2005). This gas is now known as 'CO-dark gas' and many studies have tried to address how much of the ISM is tied up in this material (e.g., Planck Collaboration XIX, 2011; Donate and Magnani 2017). If the CO-dark gas is not amenable to CO mapping, there are other spectroscopic ways to detect it and map its distribution. Spectroscopy is critical here because it provides velocity information whereas the gamma-ray and infrared maps may reveal its presence but say nothing about the kinematics.

The velocity structure of this low-density, extended molecular gas holds the key to unraveling its relationship with the atomic hydrogen medium from which it condensed. This gas is likely to be gravitationally unbound and thus carries the primordial kinematic signature from its formation processes. Several studies have shown that the OH 1665 and 1667 MHz and the CH 3335 MHz hyperfine, ground state, main lines are capable tracers of CO-dark molecular gas (for OH: Wannier et al. 1993; Barriault et al. 2010, Cotten et al. 2012; Li et al. 2018; Donate et al. 2019; Busch et al. 2019; for CH; Magnani and Onello 1993, Liszt and Lucas 1996; Xu and Li 2016).

In an effort to resolve the above issues, the NGAT will be ideal for conducting a large-scale Galactic survey using the OH 18 cm lines[16] and the CH 9 cm lines to detect all of the Galaxy's molecular gas and determine its relationship to the cold atomic hydrogen gas (i.e., the CNM). This project would complete the great CO plane surveys (Dame et al. 2001; Solomon et al. 1985) by including the lowest density, outermost portions of the CO-traced molecular clouds. In addition, simultaneous observations of the 21-cm neutral hydrogen line would be possible, and could be done in full polarization mode such that one could measure Zeeman splitting in both the HI and OH lines. This will allow for more accurate cloud properties and modeling and allow us to better study the interface regions between the atomic and molecular gas as well as how the atomic-molecular transition is affected by magnetism. Finally, because of the wideband capabilities, one could observe numerous hydrogen RRLs simultaneously to include the inventory of ionized gas and capture all phases of the ISM.

The NGAT will provide enough resolution to resolve GMCs out to 10 kpc and the large declination coverage will permit complete mapping of three of the Galactic quadrants. *Such a study would provide a truly comprehensive picture of the most fundamental processes in the ISM, and the NGAT will provide the wide FOV, largely unparalleled sensitivity, wideband receiving capabilities, and RFI mitigation necessary to make it possible and efficient.* These large-scale projects would also complete our view of GMCs and, because of the kinematic information, finally allow us to study the relationship between molecular gas and the CNM on a Galactic scale.

## C.7 Near-Field HI 21 cm Line Cosmology

According to the ΛCDM paradigm, the local Universe should be teeming with dark matter mini-halos with masses $< 10^9$ $M_{\odot}$. While the ΛCDM model is extremely successful in describing the overall evolution of cosmic structure, observational results fail to detect its predicted large numbers of "satellites." While the number of known satellites in the Local Group has increased dramatically in recent years, the number with small circular velocities remains too low. Interactions between satellites and the hot coronae of large galaxies can explain the dearth of stars and gas in small halos, but explaining the kinematic mismatch between the smallest galaxies and their host halos is harder. In fact, the occupation of low mass ($< 10^9$ $M_{\odot}$) halos by visible galaxies is poorly constrained, and the observed luminosity and velocity functions fall well below the predictions of ΛCDM.

Converging the paradigm with our local Universe requires a complete multiwavelength census of low mass galaxy populations and their general characteristics. A cornerstone of these studies rests on observing the HI content of the low mass populations, as it provides not only information on the HI mass but also dynamical information including recessional velocities and velocity widths. Studies like the Arecibo Legacy Fast ALFA (ALFALFA; Haynes et al. 2018) survey, the AGN and Galaxy Evolution Survey (AGES; Auld et al. 2006), and Arecibo Pisces-Perseus Supercluster Survey (APPSS; O'Donoghue et al. 2019) have delivered a wealth of data to that effect. In conjunction with optical studies such as the Sloan Digital Sky Survey (SDSS) and the Widefield Infrared Explorer (WISE) survey these studies have allowed the characterization (stellar, gas, dust and dark matter content) of diverse populations of galaxies both near and far as well as illuminated their distributions within the large scale Cosmic Web while simultaneously challenging our understanding of the local Universe in many ways (Haynes et al. 2018, Minchin 2020).

[16] To ensure an accurate assessment of the OH content, one should observe it in both emission and absorption, as the low excitation temperatures and their proximity to the background can have a large effect on their emission line intensities. See, for an example, Tang et al. 2017. However, there is a good chance sightlines toward the outer Galaxy could be observed in emission only, as the OH emission, albeit weak, is ubiquitous and observable, and the profiles rarely switch into absorption (Allen et al. 2012, Allen et al. 2015, Busch et al. 2019).

For example, some of the nearby sources in ALFALFA have extremely faint optical counterparts, e.g., Leo P (Skillman et al. 2013) and AGC 198691 (Hirschauer et al. 2016). Such low metallicity dwarf galaxies constitute 40% of the known population in the local universe, and represent some of the targets of the next generation of HI galaxy surveys. NGAT's tremendous sensitivity, large (~1.0 deg) FoV, and high resolution (~2.5') at 1.4 GHz will enable both fast and highly sensitive surveys of such galaxy populations, ushering in a new generation of 21-cm line cosmological studies.

## C.8 Detection of Galactic Cold Dark Matter and Testing the Standard Model of Particle Physics

The question of the existence of the hypothetical axion is one of the remaining mysteries in the Standard Model of Particle Physics, primarily because its existence provides the most well-known solution to the strong CP problem in Quantum Chromodynamics (Peccei & Quinn 1977; Weinberg 1978; Wilczek 1978).

Within the context of cosmology, axions may address the matter-antimatter asymmetry in the Early Universe (Co & Harigaya 2019), and are one of the leading candidates for constituents of cold dark matter (Preskill et al. 1983; Dine & Fischler 1983; Abbott & Sikivie 1983). The frequency range covered by axion-related signals (discussed below) is accessible to radio telescopes that meet the high sensitivity, high resolution requirements, ushering in for the first time the possibility of detecting this particle.

Axions are thought to be converted into photons by strong magnetic fields. The transition is expected to occur when an axion is located at a region within a magnetosphere at which the plasma frequency is equivalent to the axion mass, inciting a resonance which allows the particle to convert into a photon (Hook et al. 2018). Hence, magnetars or similar sources of intense cosmic magnetism provide test beds for detecting axions and illuminating the aforementioned long-standing questions. To date, one study has put constraints on the axion mass using radio data, but higher sensitivity, high resolution, and RFI-free observations will be needed to cover the entire parameter space (Darling 2020).

The expected frequency range of photons generated in this axion-photon conversion (based on expected axion mass) is 240 MHz - 24 GHz (Hook et al. 2018), which is entirely covered by the frequency capabilities of NGAT. In cases where confusion of the axion-photon conversion (see Darling, 2020) signal is not a serious issue, NGAT's high sensitivity and spatial and spectral resolution, especially at the higher frequencies, would be sufficient to conduct such searches alone. The very detection of their existence would be groundbreaking on multiple fronts, and the capabilities of NGAT will poise the telescope to contribute to constraining the axion mass or detecting the par-ticle which has eluded the realm of observational possibilities for the past 50 years.

## C.9. Space and Atmospheric Sciences

**AO is strategically located in the Caribbean, a region with the largest traffic of vessels in the world that connect the Atlantic and Pacific oceans through the Panama Canal**. Over 14,000 vessels cross the Caribbean Sea every year transporting more than 470,000,000 tons of products (numbers of 2019). Also, the Caribbean Sea hosts a large number of offshore oil and gas exploration vessels and hundreds of floating production, storage, and offloading units (PSOU). All of these *are geo-positioned by transionospheric radio signals (GPS)*. On the other hand, *a reliable forecast of the upper atmosphere is crucial for geo-positioning accuracy.* **Building the NGAT at AO is critical to supply data inputs for forecasting models and mitigate the effects of the space weather impacts on the Caribbean region, as well as to serve as a tool to better understand the local and global climate change.** Moreover, the *NGAT data taken at AO is critical to technologies that employ the bottomside ionosphere as a reflecting layer in order to extend the range of radio communications or sensing, such as over-the-horizon radar (OTHR).*

**NGAT together with lidars and passive optical observations (AOL and ROF) can support spacecraft missions in terms of temporal and spatial resolution as well as calibration.** *For instance, the AO is uniquely located to provide simultaneous measurements of ionospheric features along with the GOLD NASA mission.* This global v/s local dynamics can be linked to provide new insights into (a) How global dynamics influences coupling of different atmospheric regions, (b) what are the timescales involved in the disturbance propagation during a transient or a geomagnetically disturbed event? (c) how do different atmospheric regions respond to such events?

Below we highlight additional SAS science objectives the NGAT concept will enable to continue and improve.

### C.9.1 Ion-Neutral Interactions

The complexities of the ionospheric variability and its link to the neutrals can be studied using simultaneous ISR and lidar measurements. The unique combination of calcium ion lidar and the LAT's ISR facilitated the validation of mesospheric chemistry models for the first time (Raizada et al. 2011). A good correlation between sporadic E and calcium ions layers further suggested that calcium ions can act as a proxy for major metal ions in the ionosphere that cannot be studied using ground based lidars (Tepley et al. 2003; Raizada et al. 2012, Raizada et al. 2020). *New observations with the improved NGAT will provide insights into ion-electron-neutral interactions in the collision dominated region, and will advance our knowledge about instabilities in the neutral and ionized atmosphere. Such studies will further benefit from additional HF experiments at 3 MHz that will allow creation of artificial irregularities in the lower ionosphere.*

Another interesting and unique investigation that resulted from the simultaneous observations from the LAT's ISR and lidar is shown in Fig. 16. This study demonstrated for the first time that the tidal ion layers, which are an ubiquitous phenomena at AO location are not related to the thermospheric neutral metal layers as observed by the resonance lidars at this location (Raizada et al. 2015). *This study revealed that the occurrence of neutral metal layers above the differential ablation zone of meteoroids is still unresolved and probably is a latitude dependent event. This mystery can be resolved by the improved NGAT together with multi-instrument data sets, as the AO's cluster of optical and radio sounders.*

Other investigations include Sporadic-E ($E_s$), valley region structures, quasi-periodic echoes,

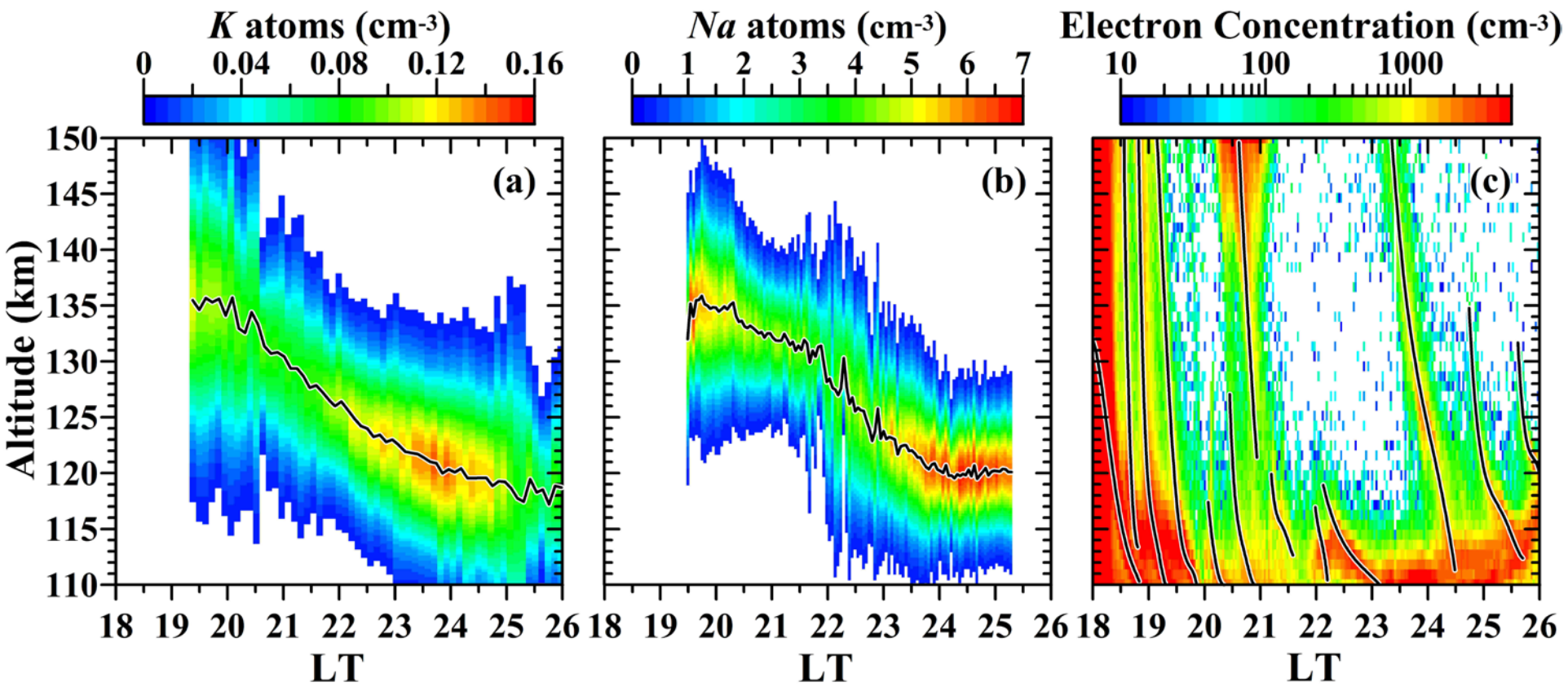

**Figure 16:** Potassium and sodium lidar data (left and middle panels), and electron concentrations obtained using ISR data (right panel) illustrate different descent rates of the neutral and ionized layers at thermospheric altitudes. (*Adapted from Raizada et al. 2015*).

influences of dynamical instabilities like Kelvin-Helmholtz on the ionosphere, which can be best studied using combined ion and neutral observations. In this regard, multi-metal probing of the meteoric metal layers are very important as they provide new insights into mesospheric chemistry (Raizada et al. 2015a, b; Zhou et al. 2005). A recent investigation has revealed differences in the layer morphology at Antarctica and has opened up new questions regarding the factors causing variability in the neutral layer distribution (Chu et al. 2020). Thus, it is important to continue neutral metal measurements at AO using different resonance lidars, since these will provide new insights into neutral layer characteristics at a low latitude site. Such observations in conjunction with the NGAT improved radar system will result in the profound understanding of the factors responsible for such differences.

*The new proposed beam steering capabilities of the NGAT radar system promise investigations of two-dimensional characteristics of $E_s$ in both space-time dimensions. This in conjunction with high-resolution wind measurements with lidars can provide shed new light on the role of advection, shears and evolution of the $E_s$ layers.*

Further daytime temperature lidar capabilities in conjunction with the NGAT's ISR will provide accurate estimation of ion-neutral collision frequency in the lower E region. In addition, high resolution wind measurements using a Doppler lidar system together with the NGAT's ISR will provide new information about the role of dynamical instabilities in the triggering E region irregularities, their evolution and influence on F region coupling.

Above the E-region ionospheric, the ion-neutral interactions also can be used to retrieve the neutral atomic oxygen. Resonant charge exchange between O and $O^+$, the principal constituents in the terrestrial atmosphere between 200 and 500 km, is the dominant mechanism governing momentum and energy exchange between the ionosphere and the thermosphere. The efficacy of this ion-neutral coupling offers a unique means of ground based remotely sensing of thermospheric neutral O density [O] (Josh et al. 2018).

Due to the geomagnetic geometry of AO, the plasma drift velocity arises from an imbalance between ion diffusion velocity and momentum transfer due to collisions with neutrals moving along the field line. Thus, through the most important source of momentum exchange of $O^+$ ($O{-}O^+$) charge exchange collisions on the molecular diffusion term of plasma equilibrium, it is possible to retrieve the [O] combining ISR plasma drift velocities, ISR electron density and the Fabry-Perot interferometer (FPI) neutral winds (Fig. 17). The continuity of FPI observations at AO and ROF, associated with a better temporal and spatial resolution from the NGAT's ISR will provide a better altitudinal coverage to retrieve the neutral O.

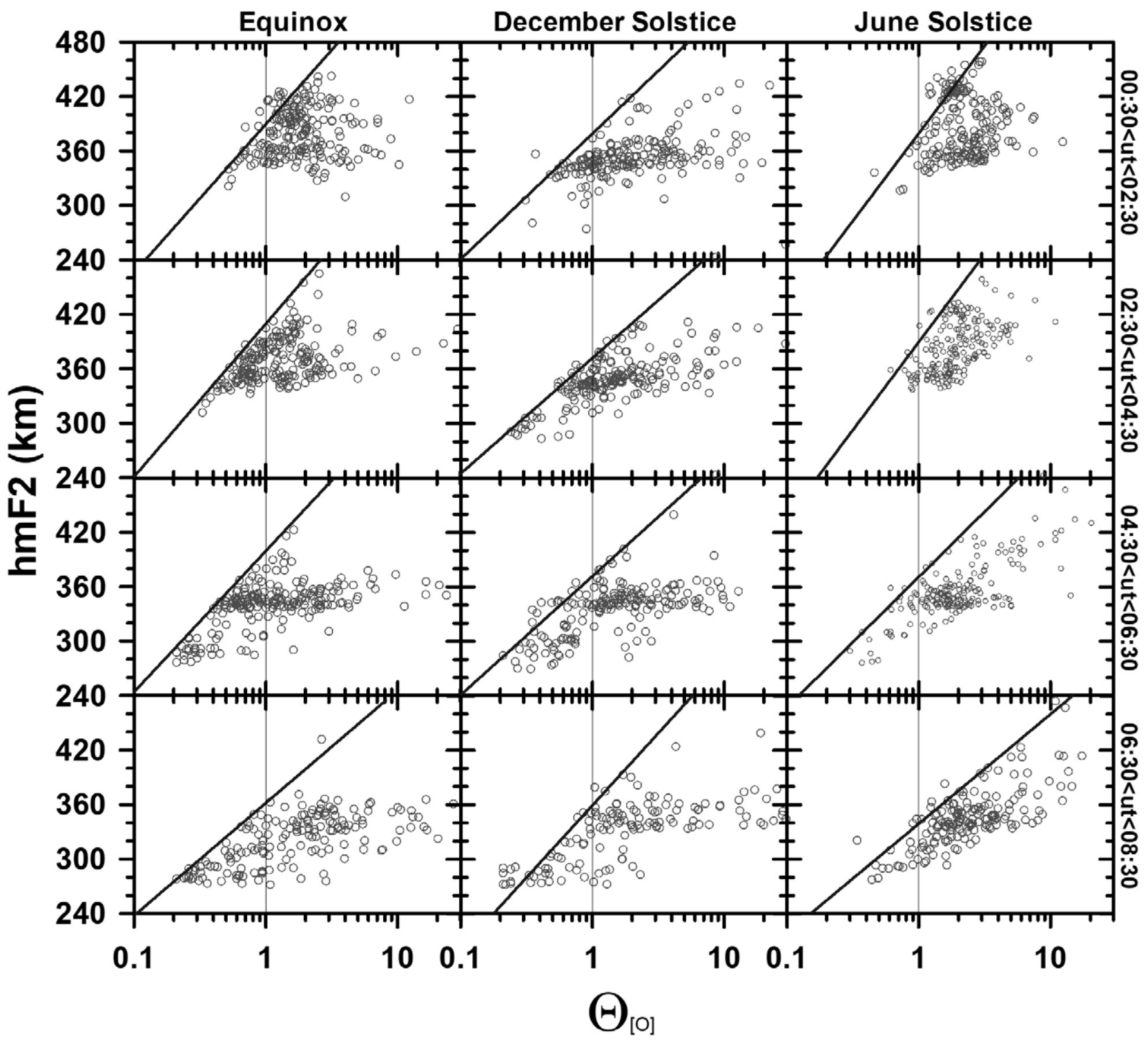

**Figure 17:** Ratio between MSIS model and the neutral atomic oxygen density obtained by this ion-neutral coupling method with altitude using 30 years of ISR data and an empirical model developed using AO Fabry-Perot data (Brum et al. 2009).

### C.9.2 Sudden Stratospheric Warming Events (SSW)

Longitudinal chains of ISRs are crucial for understanding the role of stratospheric warming events on the upper atmosphere regions. AO's ISR is uniquely located to fill the gap in the geomagnetically mid latitude location, while being a geographically low latitude site. *The new NGAT's ISR will provide high altitudinal resolution and make possible the fine structures studies generated by SSW.*

This effort is further complemented by the state-of-the-art Rayleigh and Resonance Lidars systems, as well as the cluster of passive optical instruments and radio receivers from AOL and ROF.

### C.9.3 Plasmaspheric studies and modeling

The AO strategic location at a relatively geographic low-latitude and a geomagnetic mid-latitude, together with the proposed NGAT will allow us an unrivaled ability to study outflow from the ionosphere into the inner plasmasphere and also the ability to detect precipitation of scattered ions and electrons. Furthermore, the NGAT's ISR, optical instruments, and the HF facility will perform active experiments that address issues related to the plasmasphere plasma refilling. Observations indicate that plasma refilling (following erosion of the "outer" plasmasphere) proceeds very rapidly: e.g. flux tubes at L~6.6 refill within 24-48 hours as observed by in-situ satellite instrumentation close to the equatorial plane. This rate of refilling is far in excess of refilling rates predicted by theoretical first-principle models. Additionally, the plasmasphere is known to play an important role in scattering more energetic ions and electrons in the Earth's radiation belts. Such scattering can cause these energetic particles to enter the loss cone and precipitate into the upper atmosphere. However, the precise nature of this interaction (e.g. the conditions under which it is maximized, the rate of scattering, the temporal and spatial extent of the scattering) are not known in any detail. *The high altitude coverage of the NGAT's ISR will provide information of the plasma population at heights of plasmasphere/protosnosphere and thus, making possible the understanding of plasma refilling at geomagnetic middle latitude.*

### C.9.4 Inter-hemispheric flux of particles and its impacts on the Caribbean Sector

The conjugate magnetic points of AO and ROF are moving northward from the vicinity of Bahia Blanca (Argentina) towards Rio Grande state (southern Brazil). At the same time, the South Atlantic Magnetic Anomaly (SAMA) is moving westward, and will hit the AO and ROF conjugate points in a few years (Fig. 18). We expect an interhemispheric flux of plasma towards the Caribbean though the second adiabatic invariant (Ganguly et al. 2014a and b). *The development of NGAT at AO (and the maintenance of ROF) is crucial to understand the impacts of the SAMA in the local upper atmosphere as well as the impacts on the trans-ionospheric radio signal, which is responsible for geo-positioning service and radio communication.*

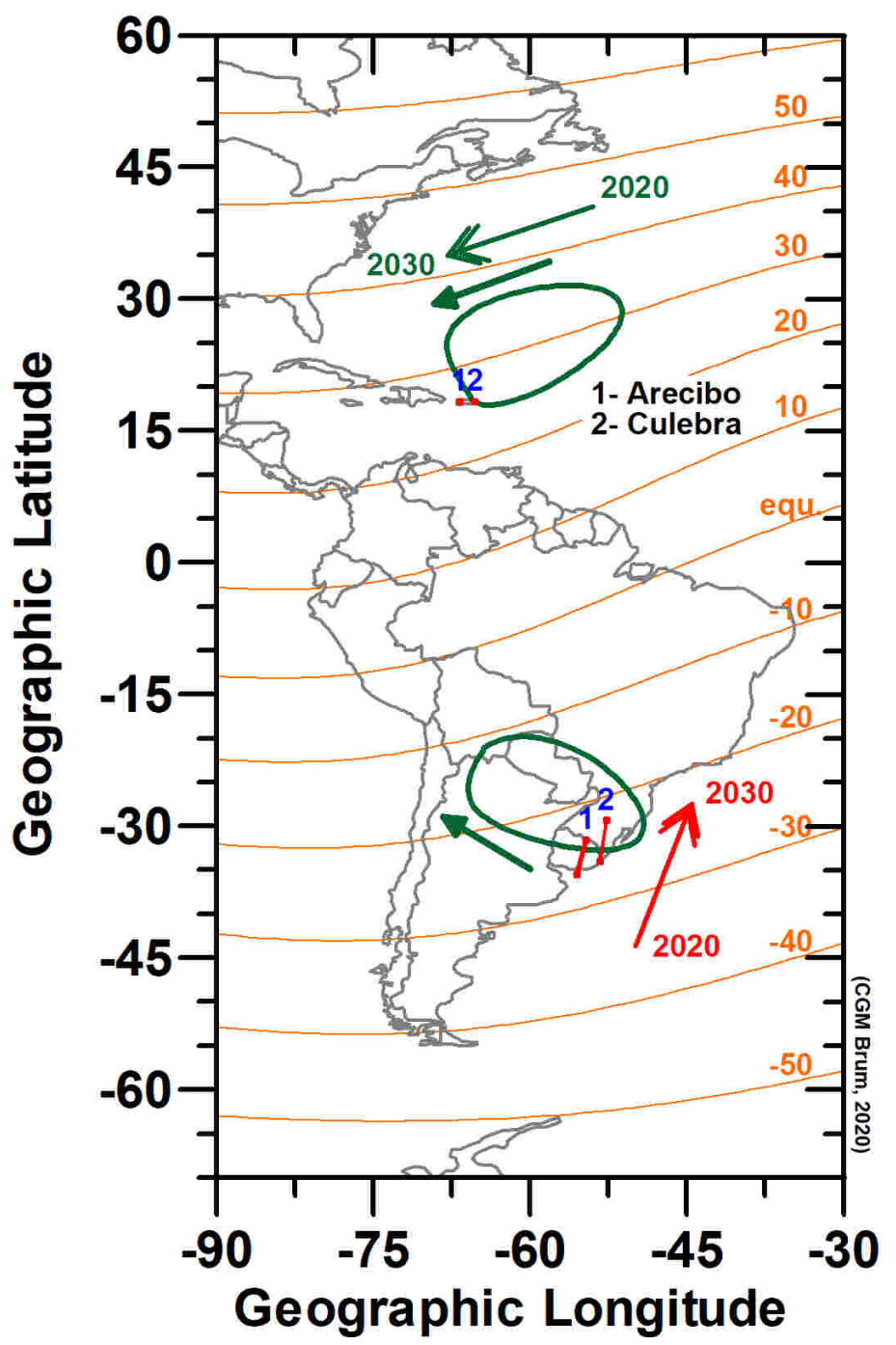

**Figure 18:** AO and ROF geographic location in respect to the American Sector (red dots 1 and 2 on the northern hemisphere) and its respective conjugate points for the years 2020 and 2030 (red dots on the southern hemisphere). The green southern "oval" is the SAMA based on the magnetic total intensity of 25000nT, while the northern one is the SAMA conjugate point at 350km of altitude (magnetic total intensity was retrieved from IGRF13).

### C.9.5 Atmosphere-ionosphere-magnetosphere interactions (AIMI)

ISR can study AIMI in different ways. For example, the topside ionospheric response to an interplanetary high-speed solar wind stream (HSS) or to external solar wind forcing was never explicitly studied before. At AO, scientists have analyzed the HSS impacts on the topside ionosphere and their finds reveal an expansion of the topside ionosphere and retraction of the protonosphere using the LAT's ISR (Hajra et al. 2017). The noontime and midnight peak electron density was ~7.5 and ~2.7 times higher in average, respectively, during the HSS event compared to the non-HSS interval while the $O^+$ ion concentration exhibited an overall enhancement, deep penetration of the $H^+$ ions below $O^+/H^+$ transition height ($h_T$) were observed during the day and the night. The $h_T$ increased by ~200 km during the day and by ~100 km at night. At the $h_T$, the peak ionospheric electron and ion temperatures increased by ~200-500 K during day and by ~50-70 K at night. *What is the source of this ionospheric heating? At the present time we do not understand this. Building the NGAT's ISR at AO is crucial to answer this open scientific question because this is a location where magnetospheric particle precipitation is not expected.*

### C.9.6 Vertical Coupling of the Earth's Atmospheric Layers

Understanding the Earth's atmosphere and ionosphere as an integrated, coupled system is one of the biggest challenges of the aeronomy community. The continuous influence of meteorological (below) and space weather (above) drivers on this system makes it even more challenging to identify and avoid ambiguities in interpreting the observations.

Though the Geospace inputs to the Earth's atmosphere are more explored, the potential tropospheric forcings are not (e. g., Immel et al. 2018). For instance, there are multiple unresolved questions of how atmospheric gravity waves (AGW) couple the different atmospheric regions and trigger the energy and momentum transfer between neutrals and plasma in the ionosphere-thermosphere (IT) system (Trinh et al. 2018), and on the physical origin of Medium Scale Traveling Ionospheric Disturbances (MSTID).

The AO's lidar systems (see section D.1.2) and the passive optical instruments from AOL and ROF (see Section D. 1. 5) have been important resources for untangling the role of wave propagation in the atmospheric coupling and in the generation of ionosphere irregularities. These facilities continue operating after the collapse of the LAT's ISR. Thus, *building the NGAT at AO's location will provide the community a powerful and unique set of tools to investigate the Earth's atmospheric coupling from the troposphere to the thermosphere and IT system via wave propagation. Unique experimental results will be provided to model predictions on the dissipation of AGW in the lower thermosphere.*

### *C.9.6.1 Science Driver for the NGAT 220 MHz Coherent Radar*

AGW are generated in the troposphere and propagate into the middle- and upper- atmosphere depending on their phase velocity and horizontal wavelengths (Fritts and Alexander, 2003). These waves transport energy and momentum from the lower atmosphere to the middle and upper atmosphere (Hines, 1960), affecting the middle atmosphere's mean flow and radiative balance due to heating and forcing by dissipative or breaking AGW producing turbulence and mixing (Gardner, 1994). Given the importance and effects on global circulation, AGW research remains one of the highlighted topics in the scientific community.

*Adding the 220 MHz frequency on the NGAT's radar system is in consideration to improve AO's capability and enable multi-layer measurements beyond the altitude range covered by other facilities.* This effort will contribute to a better understanding of the Earth's atmosphere vertical coupling via AGW. The 220 MHz frequency lies in the inertial subrange of turbulence within the lower-middle atmosphere, as shown in Fig. 19 by Hocking (1985). The red line indicates the Bragg wavelength of 0.7m corresponding to 220 MHz.

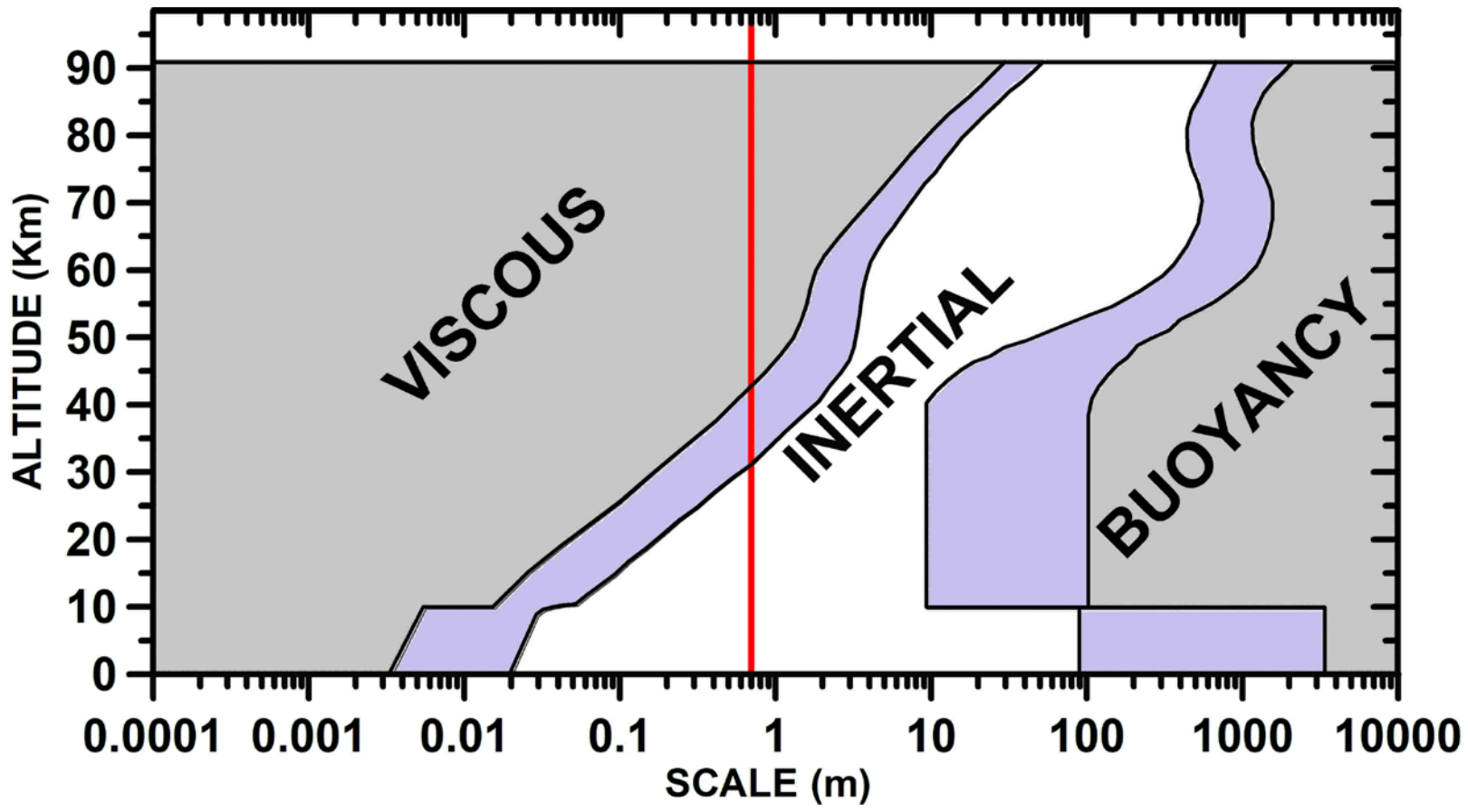

**Figure 19:** Typical inner and inertial range/buoyancy range transition scales for turbulence in the atmosphere (modified from Hocking,1985) .

The 220 MHz coherent radar will help to better understand the lower and middle atmospheric structure and dynamics in extended height coverage than 430 MHz (Bragg wavelength of 0.35 m), whereas the upper limit for turbulence investigation of 430 MHz is below 30 km. A power spectra of 212MHz radar is shown in Fig. 20 (a) by Devi et al. (2019). Echoes of turbulence origin are used for quantification of turbulence by Jaiswal et al. (2019), using ARIES-ST radar at frequency of 206.5MHz. Fig. 20 (b) shows the energy dissipation rate of PMWE turbulence in the mesosphere estimated from EISCAT-VHF radar at 224 MHz. It can be noted that, in the mesosphere, the UHF of 430 MHz at AO is not capable of obtaining echoes from mesospheric turbulence, whereas these can be observed even with a low-power VHF radars which was demonstrated by Rottger et al. (1981).

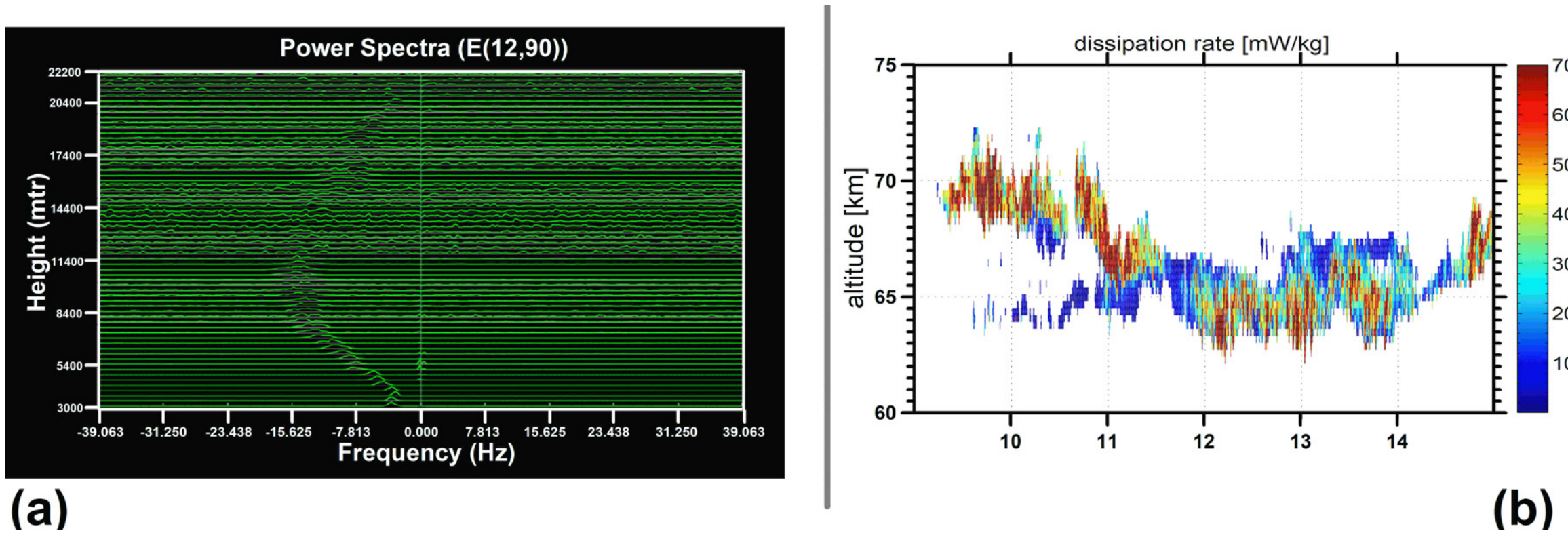

**Figure 20:** (a) Doppler Power Spectra (from 3 km to 22 km) from the Gauhati-ST Radar of 212 MHz (26N, 91E); (b) Energy dissipation rates inside the PMWE layer (Strelnikova and Rapp, 2013).

Thus, VHF radars in the range of 206-224MHz are capable of measuring the echoes of turbulence originating from the troposphere, stratosphere and mesosphere. *These echoes could also be used for quantification of turbulence and instability processes in order to understand the energetics and dynamics of the lower- and middle-atmosphere*.

### *C.9.6.2 Wave energy in the F-region thermosphere*

It becomes increasingly evident that AGW propagate into the thermosphere from below and thereafter, often induce secondary waves within the thermosphere (Vadas and Becker, 2018). Besides, it is well documented that tidal and planetary scale waves modulate ionospheric densities and temperatures. The Midnight Temperature Maximum (MTM) at AO (Martinis et al. 2013) is common, and appears to be generated by upward propagation of the terdiurnal tide component. These are important mechanisms for horizontal and vertical momentum transfer into the thermosphere. For example, Sudden Stratospheric Warming (SSW) events generate ionospheric structure through upward propagating AGW from the source. Previous work from AO (Gong et al. 2016) have even cataloged an MTM event that appears to be correlated with an SSW event.

*The optical instruments from AOL and ROF will continue to be important resources for unraveling the role of tides and AGW in the generation of quiescent mid-latitude ionosphere-thermosphere-mesosphere system irregularities.* Bi-static observations of the same wave event from AOL and ROF provide accurate altitude triangulation as a powerful new diagnostic for this research. *The NGAT radar frequencies (220 MHz and 430 MHz) will leverage these studies by providing information of the vertical atmospheric coupling from the lower- to middle- and upper- atmosphere and ionosphere-thermosphere (IT) system.*

### *C.9.6.3 Climatology, morphology and equatorward propagation of medium scale traveling ionosphere disturbances (MSTIDs)*

MSTIDs are the most common ionospheric irregularity phenomenon at middle latitudes, with defining wavelengths of approximately 150-250 km and periods of 10 – 60 min. Propagation direction is characteristically to the southwest in the northern hemisphere, and to the northeast in the southern hemisphere, implying the influence of a polarization electric field in their generation. Morphology and climatology of MSTID events at AO and ROF have been described by Martinis et al. (2010) and recently by Terra et al. (2020), respectively.

The seed of nighttime midlatitude MSTIDs is generally attributed to the perturbation of the neutral wind field by AGW activity (e.g. Hysell et al. 2018). Those winds induce electric fields as ion-neutral collisions create polarization, and inhomogeneous conductivity manifests as the ionospheric wave disturbance). Vertical Pedersen currents driven by differences in zonal neutral winds and zonal plasma drift in turn drive small scale instabilities that may trigger mid-latitude spread-F (Hysell et al. 2018).

*Modeling of MSTID origins, morphology and climatology, and evaluating the roles that MSTIDs play in different ionospheric irregularities (e.g. midlatitude spread-F, and equatorial plasma bubble) formation, are now supported by the wind measurements from AOL and ROF serving as the northern post in a distributed FPI chain across the magnetic equator.* That chain now formally includes similar FPIs and imagers at Manaus and Cachoeira Paulista, Brazil. APL collaborates with CPI and AO in this research initiative.

*The information retrieved from the proposed NGAT will leverage the existing research at AO with new inputs to improve the current knowledge of the occurrence and generation of MSTIDs.* NGAT will provide measurements with an exceptional altitudinal and temporal high resolution of ionospheric parameters essential for a better understanding of the MSTID phenomena.

#### *C.9.6.4 Tropospheric Forcing on the upper atmosphere during Extreme Weather Systems*

Extreme Weather Systems (EWS) – which include tropical depressions, tropical storms, and hurricanes – are among the Climate Change research areas. These EWS travel from Africa's coast towards the Caribbean and U.S. regions every year, affecting thousands of people. A recent study by Kossin et al. (2020) showed that *the probability of storms reaching major hurricane levels increased 8% per decade in the last 40 years, being the most significant changes observed in the North Atlantic sector. However, the physics of the IT system's response to this neutral phenomenon is not entirely understood, and the observational infrastructure needed for scientific closure on these issues is minimal.*

On the other hand, EWS are a primary tropospheric forcing system and a prominent source of AGW. These waves can reach well into very high altitudes and propagate in all directions from the convective source (Vadas and Fritts 2004). However, there is neither a consistent theory on how these oscillations are transferred to the ionosphere nor how the ionosphere responds to AGW generated by EWS (Zakharov et al. 2019). The literature shows separate lines of investigation: 1) studies evaluating the impact of these tropospheric forcing on altitudes of the D-region (e. g. NaitAmor et al. 2018), and 2) studies focusing on that impact on the E- and F-regions, without the inclusion of the D-region (e. g. Bishop et al. 2006). This segregation is mostly due to the observational difficulty probing the Earth's atmosphere with enough altitude and spatial coverage. *The strategic location of AO in the Caribbean Sea and its fusion of optical and radio instruments can fill up these observational gaps.*

The all-sky imager systems (ASIs) co-allocated at AO and ROF cover hundreds of kilometers away from Puerto Rico and can detect waves generated by EWS along their trajectory. The ASIs also provide the direction of the wave propagation and allow performing common-volume observations. *The improved NGAT radar operating at 220 MHz and 430 MHz will provide an optimal altitude coverage of the atmosphere (from the troposphere to beyond 2000 km), being the most powerful ground based tool available nowadays to study the wave propagation from their source in the troposphere through the IT system.*

Furthermore, *the continuous operations of LAT's ISR from the 1960's to 2020 guarantee a long-term database that, together with the future NGAT observations, are crucial to retrieve new proxies to improve weather forecast and also uncover the effects of climate changes on Extreme Weather Systems (EWS).*

#### *C.9.6.5 Aerosol and Coupling Processes in the Lower Atmosphere*

*Aerosols can have a major impact on climate and play an important role in the energy budget of the Earth's lower atmosphere. Puerto Rico is a strategically important location for precise and height-resolved aerosol observations.* It fills the observational gap between Barbados in the south-east, the main continental US in the north, and Mexico in the west. The recently awarded Culebra Aerosol Research Lidar (CARLA) will be installed at the ROF in Culebra Island by 2021. CARLA will deliver information about aerosol properties over time and altitude. Only the knowledge of all three variables: aerosol properties, time, and altitude enable the study of the dynamics of aerosol layers. This includes dust coming across the Atlantic Ocean from the African continent that is named Saharan Air Layer. The Saharan Air Layer (SAL) plays an important role for the lower atmosphere because of its influence on cloud and hurricane formation, weather, and the health of our population, among others. The purpose of the CARLA project is twofold: (a) study and provide data to the scientific community of aerosol properties and SAL over the island of Culebra, and (b) increase opportunities for the educational system by outreach activities for Puerto Rico's K-12 schools as well as for STEM students to perform hands-on research experience.

Furthermore, the lower stratosphere is a source region of waves propagating upward through the atmosphere. Stratospheric aerosols can be used as a tracer to study these waves and can help to understand coupling processes in the lower atmosphere.

*The proposed NGAT operating at 220 MHz can contribute to the study of aerosols at the main site of AO. Thus, the two distinct locations (AO and ROF) give further insides of the transport of aerosol over Puerto Rico.* This research has a direct impact on the public as aerosols and the SAL influence hurricane formation, weather, and the health of the Puerto Rican population, among others.

# Appendix D: Other science activities at the Arecibo Observatory that interlock with the NGAT

## D.1 Space and Atmospheric Sciences

The AO's scientific leadership in Space and Atmospheric Sciences (SAS) during the past 57 years was especially compelling due to the extraordinary capabilities of the LAT's ISR and the nested observational resources that supported the radar investigations including several radio receivers (e. g. GPS, Riometers, HF receiver, VLF/LF receiver, among others), the Arecibo Optical Laboratory (AOL), the Lidar facility, and most recently HF facility and the Remote Optical Facility (ROF) at the island of Culebra. Below we highlight some current (and under study) investigations carried out at the SAS that interlock with the proposed NGAT.

### D.1.1 Ion Transport Processes using Lidars and ISR

The processes associated with transfer of energy from high, low or equatorial latitudes are still unresolved. Satellites provide excellent spatial coverage but have restrictions in providing insights into temporal evolution of structures at a high resolution. Such processes can be best understood by combining ground based instruments and satellite mission data sets. Having the legacy ISR at AO along with direct measurements of meteoric metal ions using resonance lidars have provided new insights into the F-region valley dynamics that appears to be unique to AO location (Raizada et al. 2020a, b). Future endeavors should also focus on the development of new solid state transmitters to be used in Calcium ion measurements using a lidar as it opens up new dimensions and promises new discoveries with narrow-bandwidth of such lasers.

### D.1.2 Climate Studies and Forcing of the Ionosphere from below using Lidars

The combination of Rayleigh and potassium resonance lidar systems can deliver temperatures from 30 km to 105 km. Adding a Raman channel will further extend the temperatures down to tropospheric altitudes. With this, the forcing of the ionosphere from below can be studied by delivering coupling and wave propagation from the lower stratosphere and mesosphere into the ionosphere. Knowledge about different atmospheric parameters are crucial for weather forecast. Current limitations of the models arise due to undersampling of these parameters. Lidars at AO can fulfill these gaps and make significant contributions in this field. Research accomplished using the state of the art lidars that are capable of autonomous measurements of temperatures from troposphere to mesosphere will be extremely beneficial to society. Continued observations of temperatures are required for Madden-Julian Oscillation, vertical coupling between different regions of the atmosphere, stratospheric warming events, and can shed new light on the lower atmosphere as a driver for the ionospheric variability.

### D.1.3 Geocoronal hydrogen: Secular change and storm response

AO has monitored the geocoronal Balmer-alpha emission since Meriwether et al. (1980) measured a decrease in the emission line width with increasing shadow height at night. This "gravitational cooling" is due to the thermal escape of outbound H atoms exceeding the escape velocity – which decreases with altitude. An increase of [H] in the upper thermosphere predicted by Roble and Dickinson (1989) (and a decrease in F2 region temperature) remains a compelling bellwether of global atmospheric change to validate. Brightness data gathered from 1983-2001 analyzed by Kerr et al. (2001) showed a weak but inconclusive increase in H column abundance during the period. Using new calibrated measurements and data acquired from 2001-2020, that investigation should continue. Using a 10-fold improvement in FPI existent sensitivity and an improved instrument design, new H-alpha line profile analysis is a timely research initiative. *An outstanding question is the detail of the altitude profile and velocity distribution of H atoms between the 500 km exobase and the plasmapause, and how this distribution is throttled by charge exchange with protons and O+. The H distribution in this region is the source of cold ions in the magnetosphere.*

### D.1.4 Horizontal winds as a function of altitude: New wind measurements above the exobase

The 1083-nm airglow arises from solar resonance by metastable helium atoms, which are primarily created by photoelectron impact in local or conjugate twilights. Another contributor to the population comes from He+ recombination. Metastable He from either source forms a narrow layer near 700 km, so the emission (which approaches 100R in the twilight airglow) offers an excellent opportunity to measure winds above the exobase – which has never been accomplished by direct Doppler techniques. The behavior of the horizontal wind field with altitude above the F2 peak is not known, but the structure of the topside ionosphere and the local H escape flux depend on those neutral dynamics.

### D.1.5 The AO Remote Optical Facility (ROF) in Culebra Island

The AO ROF is located in Culebra, a small island in the east of Puerto Rico's archipelago (approximately 150 km from AO, Fig. 21). *ROF is a strategic site for optical observations due to its low light pollution and geography that results in less cloud coverage than the main island, and also ideal for radio receiver instruments due to the small population and a large portion of the island dedicated to natural resources (low levels of RFI).* The ROF was established in November 2015, and moved to a dark site on that island in May 2018. The ROF is equipped with a solar power system and backup as an alternative source of energy, which allows the facility to resume operations immediately following Extreme Weather Systems (EWS) strike if the shutdown is necessary.

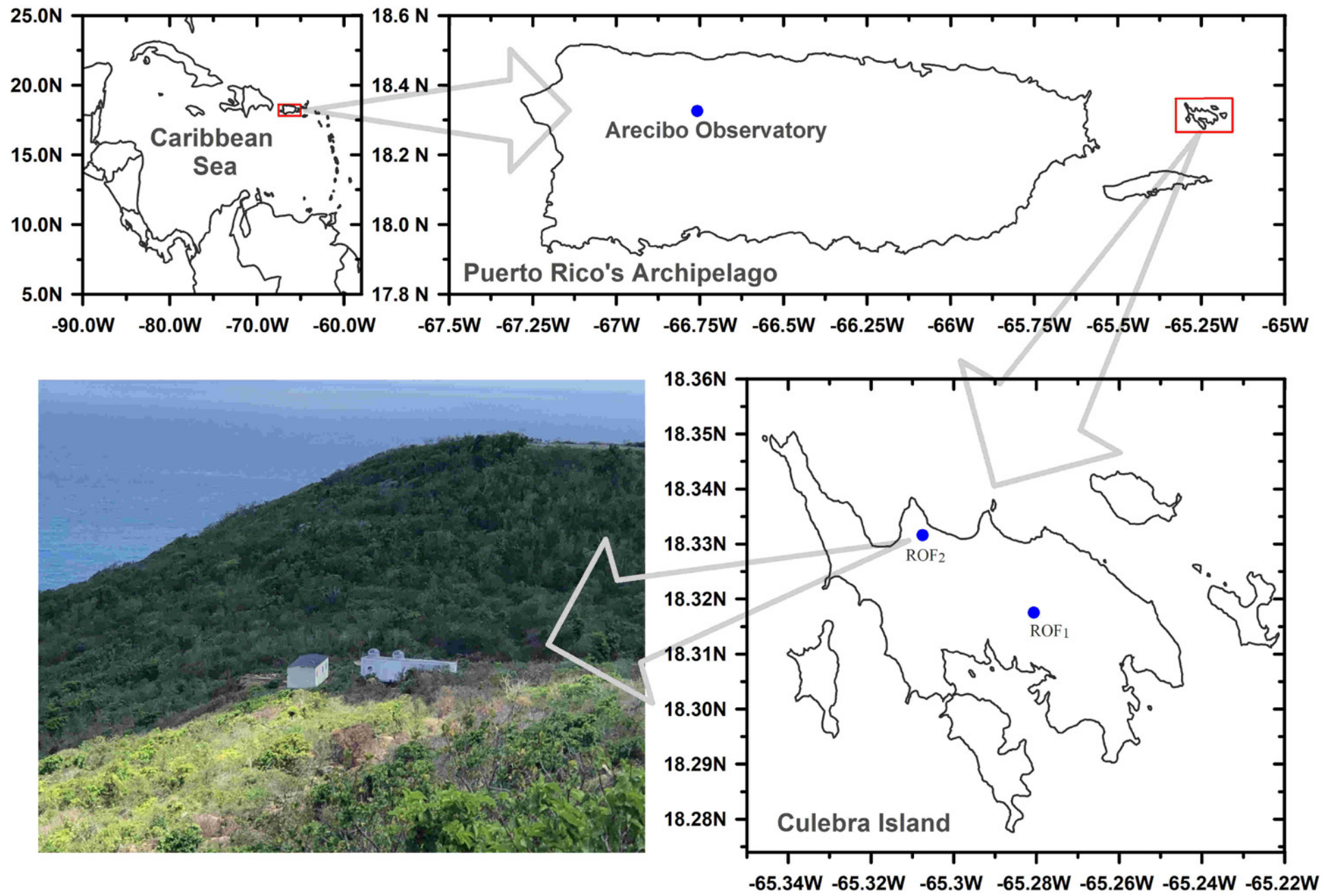

**Figure 21:** ROF location in the Caribbean Sea/Puerto Rico sector. The $ROF_2$ and $ROF_1$ denote the current and previous locations of ROF, respectively. The container with the domes on top in the left bottom pane hosts the optical and radio instrumentation and a control room, while the other one is lodging for scientists and technicians (Terra et al. 2020).

The ROF hosts a riometer, a GPS, a High Frequency Receiver, an all-sky imager system with five slots filter wheel, a 630.0-nm Fabry-Perot interferometer, and a cloud sensor. ROF will also receive a photometer, the CARLA – Culebra Aerosol Research Lidar, and a VHF all-sky Meteor Radar in the second semester of 2021.

*The ROF promises expanded sampling of upper atmospheric dynamics as a second potentially clear-sky site near AO, in addition to the purpose of novel common volume experiments with the onsite AOL.*

#### *D.1.5.1 High Doppler resolution measurements of vertical motion in the thermosphere*

Since 2012 the Fabry-Perot interferometers (FPI)s at AOL have been upgraded with imaging detectors, automated or remote internet-aware calibration and operation, and daily updates of processed thermospheric wind and temperature data. The 40-fold increase in sensitivity has thus been accompanied by an explosive expansion of temporal coverage. More than 450 nights of usable wind and temperature data were acquired in the two years 2015-2016 at Arecibo, compared to 473 nights of Arecibo data assimilated by Drob et al. (2008) in the 20-year period 1980-1999. (This large database with unmatched data quality, now covering 2012-2020, is ripe for application of modern machine learning algorithms). Wind measurements with <1 $ms^{-1}$ statistical errors have made measurements of vertical winds possible, and the first climatological empirical model of vertical motions in the nighttime midlatitude thermosphere is currently being developed (Kerr et al. 2017). *Using simultaneous common volume measurements from the FPI at ROF, the vertical winds measurements from the single AOL site will be validated by unambiguous isolation of the vertical neutral wind vector.*

#### *D.1.5.2 Field line diffusion of HF produced electrons as a function of energy*

HF induced electron heating can produce photoelectrons with sufficient energy to excite the OI $3p^3P$ state, at 11 eV above the ground $2p^3P$ state, which decays to the $3s^3S$ state producing the 8446Å emission. By observing this emission in the HF heated volume above Arecibo from the AOL, with simultaneous measurements up the field line from ROF in Culebra, this experiment seeks to quantify the generation and field-aligned diffusion rate of >10 eV electrons in the F-region. *This project requires the sustenance of the AOL and the ROF, and a timely reconstruction of the HF facility.*

#### *D.1.5.3 First Caribbean meteor radar and its application to enhance the atmospheric probing*

ROF will receive an all-sky VHF meteor radar in the second semester of 2021. *This will be the first meteor radar in the Caribbean and Central America and will add information from latitudes rarely considered in the global statistics of meteor flux*. The data will also provide 3-D background mesospheric winds and neutral temperatures from the meteor trails with a large observation window, as the radar is active 24 hours/day regardless of weather conditions. *This feature will leverage the AO and its ROF observational capability by bringing an advantageous accounting for the intermittency of wave sources. These parameters will be used to investigate the momentum and energy fluxes and drag due to breaking/dissipating gravity waves in Puerto Rico's vicinity, which is an invaluable opportunity to deeply understand the coupling of lower and upper atmospheric*

*regions.*

## D.2 12m Telescope

The AO has a 12-m diameter dish antenna, which will be one of the radio astronomy facilities used by the scientific staff during the construction of the proposed NGAT. The 12-m antenna was commissioned in 2011. It is fully steerable and equipped with room temperature receivers at S and X bands to record dual-polarization signals. We currently plan to equip the telescope with a broadband (2 to 14 GHz) cryogenic receiver. A feasibility study to have a 400 - 800 MHz prime focus receiver is underway. The main science drivers for the 12-m antenna include, but not limited to,

- Participation in European VLBI, VLBA and geodetic VLBI networks.
- As a single-dish it is extremely useful to monitor interplanetary scintillation of selected compact radio sources and to track coronal mass ejections between the near-Sun region to inner heliosphere and to predict the arrival of space weather events at the Earth's magnetosphere
- Regular monitoring of strong pulsars
- Participate with a network of telescopes using the 400 - 800 MHz receiver for FRB search and localization.
- Educational and training programs for the university and motivated higher level K-12 students.

## D.3 e-CALLISTO spectrometer

The e-CALLISTO spectrometer has proven to be a valuable tool for monitoring solar activity and for space weather research. It provides dynamic spectra of type II, III and IV radio bursts in the frequency range of 45 - 900 MHz. This instrument is under commissioning and the deployment will also support education and training of space weather observers.

## Appendix E: Study of planetary subsurfaces with 40-60 MHz radar observations

A direct measurement of planetary bodies' interiors requires a radar system using lower frequencies than those commonly used in ground-based planetary radar observations, as lower frequencies allow the radio waves to propagate through the regolith with less absorption. The lower boundary is set by the ionosphere, which becomes increasingly opaque at frequencies below 20 MHz. Spacecraft missions from Rosetta's COmet Nucleus Sounding Experiment by Radiowave Transmission (CONSERT) to the future mission Europa Clipper's Radar for Europa Assessment and Sounding: Ocean to Near-surface (REASON) have used or planned radar systems in the frequency range of 60-90 MHz for direct measurements of the targets' subsurfaces. In fact, asteroid tomography at the frequency range 40-60 MHz could provide the first direct measurement of the internal structure of an asteroid (Haynes et al. 2020). Arecibo could collaborate with a spacecraft that has an instrument such as CONSERT, REASON, or MASCOT (Herique et al. 2019). Future missions to near-Earth destinations, such as the ESA Comet Interceptor, could also plan to collaborate with Arecibo for small body bistatic radar tomography experiments. Potential targets could be 99942 Apophis, which will pass the Earth at less than 40,000 km in April 2029, or similar close passes of (137108) 1999 AN10 on 7 July 2027 at 400,000 km, or (153814) 2001 WN5 on 26 June 2028 at 250,000 km. **High-power radar systems in this frequency range could also enable improved subsurface studies of the Moon and the terrestrial planets, and could support planned robotic missions and human exploration of the Solar System.**

Such a radar system is not available at the Arecibo Observatory at this moment, but its science interest and implementation have been an open discussion at AO since early 2020. Such a radar system could work in a facility independent of NGAT, perhaps in coordination with the new, extended HF system (see Appendix B). Thus, further development of this idea should be subject to discussion in the science, operations, and management AO teams.

## Appendix F: Acronyms

| | |
|---|---|
| AGN | Active galactic nuclei |
| AGW | Atmospheric gravity waves |
| AIMI | Atmosphere-Ionosphere-Magnetosphere Interaction |
| AMiBA | Array for Microwave Background Anisotropy |
| AO | Arecibo Observatory |
| AOL | Arecibo Optical Laboratory |
| AOSA | Arecibo Observatory Space Academy |
| APL | Applied Physics Laboratory |
| APPSS | Arecibo Pisces-Perseus Supercluster Survey |
| ATCA | Australian Telescope Compact Array |
| ATLAS | Asteroid Terrestrial-impact Last Alert System |
| BH | Black hole |
| CARLA | Culebra Aerosol Research Lidar |
| CIR | Corotating Interaction Region |
| CME | Coronal Mass Ejection |
| CNM | Cold Neutral Medium |
| CPI | Computational Physics Inc. |
| DART | Double Asteroid Redirection Test |
| DESTINY+ | Demonstration and Experiment of Space Technology for INterplanetary voYage Phaethon fLyby dUSt science |
| DM | Dispersion measure |
| ECME | Electron Cyclotron Maser Emission |
| EPOXI | Extrasolar Planet Observations and Characterization (EPOCh) and Deep Impact Extended Investigation (DIXI) extended mission |
| ESA | European Space Agency |
| ESF | Equatorial spread-F |
| EVN | European VLBI Network |
| EWS | Extreme Weather Systems |
| FAST | Five-hundred-meter Aperture Spherical radio Telescope |
| FoV | Field of view |
| FPI | Fabry-Perot interferometer |
| FR | Faraday Rotation |
| FRB | Fast Radio Burst |
| GC | Galactic center |
| GMC | Giant Molecular Cloud |
| GPS | Global Positioning System |
| GR | General relativity |
| GSSR | Goldstone Solar System Radar |
| GW | Gravitational Waves |
| HF | High Frequency |
| HILDCAA | High-Intensity Long-Duration Continuous AE Activity |
| HSA | High Sensitivity Array |
| HSS | High-speed Solar wind Stream |
| IGRF | International Geomagnetic Reference Field model |
| IMF | Interplanetary Magnetic Field |
| IPS | Interplanetary Scintillation |
| ISM | Interstellar Medium |
| ISR | Incoherent Scatter Radar |
| JAXA | Japan Aerospace Exploration Agency |
| JUICE | JUpiter ICy moons Explorer |
| LASCO | Large Angle and Spectrometric Coronagraph |
| LAT | Legacy Arecibo Telescope |
| LEO | Low Earth Orbit |
| LF | Low Frequency |
| Lidar | Light Detection and Ranging |
| MBA | Main-Belt Asteroid |
| MLT | Mesosphere–Lower Thermosphere |
| MSP | Millisecond Pulsar |
| MSTID | Medium Scale Travelling Ionospheric Disturbances |
| MTM | Midnight Temperature Maximum |
| NANOGrav | North American Nanohertz Observatory for Gravitational waves |
| NASA | National Aeronautics and Space Administration |
| NEO | Near-Earth Object |
| NEOSM | NEO Surveillance Mission |
| NGAT | Next Generation Arecibo Telescope |
| NS | Neutron Star |
| OSIRIS-REx | Origins, Spectral Interpretation, Resource Identification, Security, Regolith Explorer |
| OTHR | Over-the-Horizon Radar |
| PanSTARRS | Panoramic Survey Telescope and Rapid Response System |
| PMWE | Polar Mesospheric Winter Echoes |
| PDF | Probability density function |
| PSOU | Production, Storage, and Offloading Units |
| PTA | Pulsar Timing Array |
| RM | Rotation Measure |
| RFI | Radio frequency interference |
| ROF | Remote Optical Facility |
| RRAT | Radio Rotating Transient |
| RRL | Radio Recombination Line |
| SAL | Saharan Air Layer |
| SAMA | South Atlantic Magnetic Anomaly |
| SAS | Space and Atmospheric Sciences |
| SDA | Space Domain Awareness |
| SDO | Solar Dynamics Observatory |
| SDSS | Sloan Digital Sky Survey |
| Sgr A* | Sagittarius A* |
| SIMPLEx | Small Innovative Missions for Planetary Exploration |
| SMBH | Supermassive Black Hole |
| SMBHB | Supermassive Black Hole Binary |
| SNR | Supernova remnant |
| SOHO | Solar and Heliospheric Observatory |
| SSA | Space Situational Awareness |
| SSW | Sudden Stratospheric Warming |
| STAR | STEM Teaching at Arecibo Academy |
| STEM | Science, technology, engineering, and mathematics |
| STEREO | Solar-Terrestrial Relations Observatory |
| TESS | Transiting Exoplanet Survey Satellite |
| TRAPPIST | Transiting Planets and Planetesimals Small Telescope |
| VHF | Very High Frequency |
| VI | Virtual impactor |
| VLBI | Very long baseline interferometer |
| VLF | Very Low Frequency |
| WISE | Wide-field Infrared Survey Explorer |